\documentclass[aps,pre,reprint,longbibliography]{revtex4-2}

\usepackage{here} 
\usepackage{amsmath}
\usepackage{amssymb}
\usepackage{graphicx}
\usepackage{placeins} 
\usepackage{hyperref}
\hypersetup{
colorlinks = true,
linkcolor = blue,
citecolor = blue
}

\begin{document}

\title{Energy Conversion, Fluctuation Suppression, and Information Transfer in the Thermodynamic Performance of Kinesin}


\author{Riku Kato$^1$}
\author{Takayuki Ariga$^{2,3}$}
\author{Tomohiro Tanogami$^1$}

\affiliation{$^1$Department of Earth and Space Science, The University of Osaka, Japan\\
$^2$Graduate School of Frontier Biosciences, The University of Osaka, Japan\\
$^3$Department of Biological Sciences, The University of Osaka, Japan}
\date{\today}

\begin{abstract}

Kinesin is a molecular motor that transports intracellular cargoes along microtubules.
Recent studies have quantified kinesin performance using various efficiencies within the framework of stochastic thermodynamics; however, quantitative comparisons remain difficult because different models and assumptions have been employed.
As a result, it remains unclear which aspect of kinesin performance, if any, is thermodynamically optimized.
Here, we systematically compare multiple thermodynamic efficiencies within a single kinesin--cargo model.
To this end, we construct a thermodynamically consistent two-state kinesin–cargo model that retains both the discrete stepping of kinesin and its coupling to the cargo.
Assuming a separation of time scales between the motor and the cargo, we derive analytical expressions for the thermodynamic efficiency, the information-thermodynamic efficiency, the thermodynamic uncertainty relation (TUR) efficiency, and the bipartite TUR efficiency, and compare them with numerical simulation results.
We find that these efficiencies generally remain low, suggesting that kinesin is not optimized for maximizing the thermodynamic efficiencies considered here.
Our results suggest that thermodynamic efficiencies alone may not fully characterize kinesin performance and motivate further investigation of complementary kinetic perspectives for assessing molecular motor function.

\end{abstract}

\maketitle


\section{Introduction}
Kinesin-1 (hereafter referred to as kinesin) is a molecular motor that typically functions in the cytoplasm to transport various intracellular cargoes, such as vesicles and organelles~\cite{vale2003molecular, hirokawa2009kinesin}. 
A single kinesin moves along a microtubule filament in discrete 8 $\mathrm{nm}$ steps by alternately advancing its two heads in a hand-over-hand manner~\cite{svoboda1993direct, kaseda2003alternate, asbury2003kinesin, yildiz2004kinesin}.
This stable and unidirectional motion is achieved by regulating the affinity of the two heads for the microtubule according to their respective nucleotide-binding states~\cite{cross2000conformational, mori2007kinesin}, using free energy supplied by ATP hydrolysis~\cite{schnitzer1997kinesin, hua1997coupling}. 
Although the molecular mechanisms underlying kinesin motility have been progressively clarified~\cite{dogan2015kinesin, isojima2016direct}, it remains unclear which aspect of kinesin's performance is optimized from a thermodynamic perspective.
Clarifying this issue is important for understanding the design principles of kinesin.

Recent advances in nonequilibrium statistical mechanics have extended thermodynamics to small systems in which thermal fluctuations are significant~\cite{seifert2025stochastic,peliti2021stochastic,shiraishi2023introduction}.
This theoretical framework, known as stochastic thermodynamics, has been applied to various molecular motors, including kinesin.
For example, one of the authors experimentally and numerically studied the energetics of single-molecule kinesin \textit{in vitro} and found that approximately 80\% of the input free energy from ATP hydrolysis is dissipated within kinesin, suggesting a low thermodynamic efficiency of approximately 20\%~\cite{ariga2018nonequilibrium}. 
Hwang \textit{et al.}~\cite{hwang2018energetic} evaluated both the thermodynamic efficiency and a transport efficiency based on the thermodynamic uncertainty relation (TUR)~\cite{barato2015thermodynamic, gingrich2016dissipation}, which quantifies the trade-off between current fluctuations and thermodynamic cost. 
Their analytical calculations yielded a thermodynamic efficiency of $12\%$ and a transport efficiency of $28\%$ under cellular conditions.
Moreover, Leighton and Sivak introduced an information-thermodynamic efficiency that incorporates information transfer between the motor and the cargo, and estimated that this efficiency can be as high as $70$--$90\%$ under cellular conditions~\cite{leighton2023inferring}.

These previous studies suggest that no single efficiency can fully characterize kinesin performance; a low efficiency in one process may coexist with a high efficiency in another.
Consequently, a systematic comparison of multiple efficiencies is essential for elucidating the design principles of kinesin.
However, such a comparison remains difficult because these efficiencies have been evaluated using different models and assumptions.
For example, although both Refs.~\cite{ariga2018nonequilibrium,hwang2018energetic} employ Markov jump models, their levels of coarse-graining differ.
Specifically, Ref.~\cite{ariga2018nonequilibrium} constructed a two-state model describing transitions between the internal states of kinesin, whereas Ref.~\cite{hwang2018energetic} adopted a standard six-state model.
Furthermore, Ref.~\cite{leighton2023inferring} employs a Langevin-type model for the kinesin--cargo system that does not explicitly account for kinesin's internal states.
These methodological differences make it difficult to determine whether the reported differences in efficiency reflect intrinsic biophysical features of kinesin or artifacts of modeling choices.
This motivates a unified framework in which several efficiencies can be evaluated on an equal footing while retaining the minimal ingredients required to describe the chemomechanical cycle of kinesin.

\begin{figure*}[t]
    \centering
    \includegraphics[width=\textwidth]{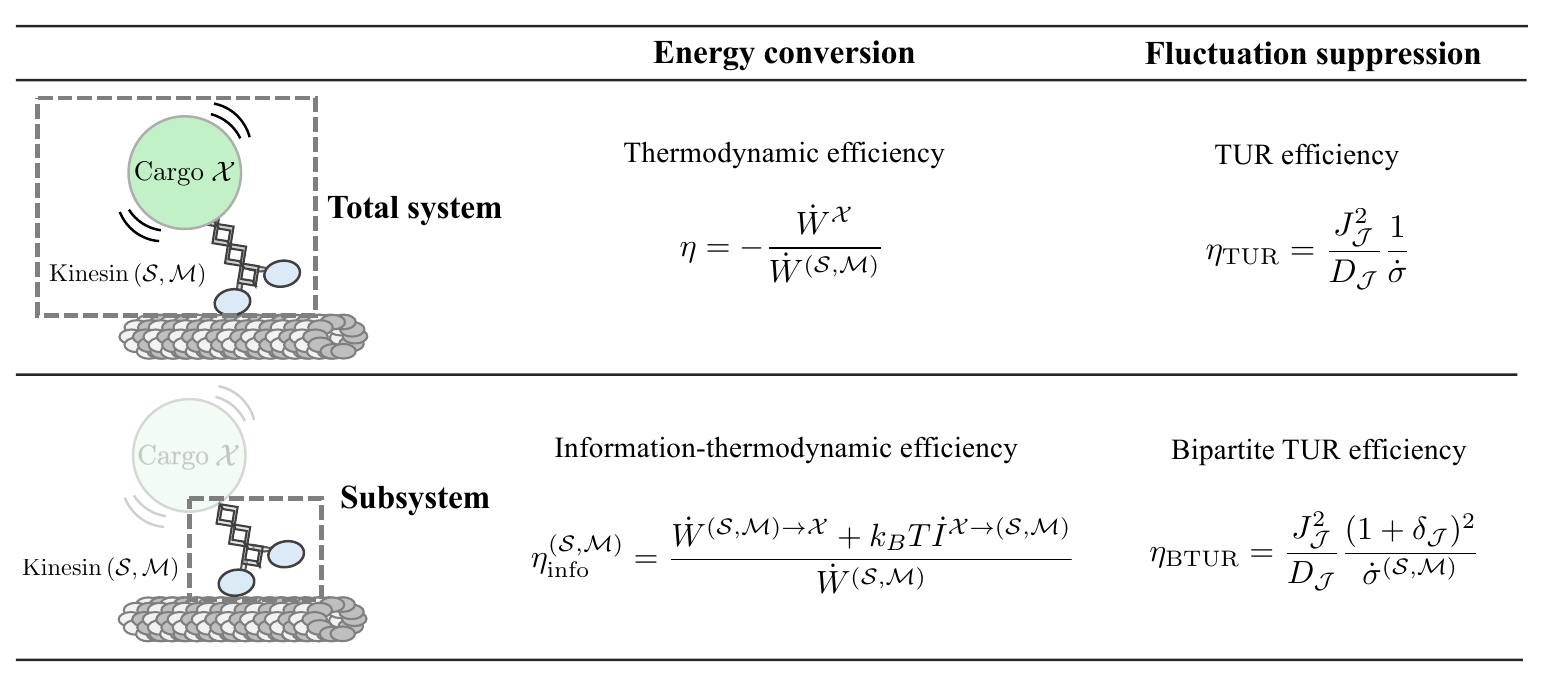}
    \caption{
    Schematic classification of the four efficiencies considered in this work.
    The efficiencies are classified according to whether they characterize energy conversion or fluctuation suppression, and according to whether they are defined for the total system or for a subsystem.
    The pair $(\mathcal{S},\mathcal{M})$ represents the kinesin subsystem—comprising the internal state $\mathcal{S}$ and the spatial degree of freedom $\mathcal{M}$—while $\mathcal{X}$ represents the cargo subsystem, specifically its spatial degree of freedom.
    Precise definitions are given in Sec.~\ref{sec: Basic properties}.
    }
    \label{fig:efficiencies}
\end{figure*}

In this paper, we aim to provide such a systematic comparison within a single thermodynamically consistent model.
Specifically, we construct a thermodynamically consistent modification of the two-state kinesin--cargo model proposed in Ref.~\cite{ariga2018nonequilibrium}, while keeping the model closely connected to experimentally observable quantities.
The model retains explicit internal states of kinesin, discrete stepping, and elastic coupling to the cargo.
For this model, we analytically calculate several thermodynamic efficiencies by performing a perturbative expansion based on the time-scale separation between the motor and cargo degrees of freedom.
We then compare these analytical results with numerical simulations of the kinesin--cargo dynamics.

Specifically, we investigate four types of efficiencies: the thermodynamic efficiency, the information-thermodynamic efficiency, the TUR efficiency, and the bipartite TUR efficiency~\cite{tanogami2023universal}.
As summarized in Fig.~\ref{fig:efficiencies}, the first two efficiencies quantify energy conversion, whereas the latter two quantify the fluctuation suppression at a given thermodynamic cost.
In each pair, one efficiency is defined for the total kinesin--cargo system, while the other is defined for the kinesin subsystem.
The precise definitions of these quantities are given in Sec.~\ref{sec: Basic properties}.
We analytically and numerically evaluate these efficiencies within a parameter regime corresponding to the \textit{in vitro} experimental setup of Ref.~\cite{ariga2018nonequilibrium}.
To complement this, we also examine an \textit{in vivo}-like condition to mimic the cellular environment, although its details are deferred to Sec.~\ref{sec:Results on various efficiencies}.
Our main results for the four efficiencies are summarized as follows:

\begin{itemize}
    \item[(i)] \textit{Thermodynamic efficiency.}---For the total kinesin--cargo system, the thermodynamic efficiency is defined as [cf.~Eq.~(\ref{eq:setup: def. thermodynamics efficiency})]
    \begin{align}
        \eta
        := \frac{-\dot{W}^{\mathcal{X}}}
        {\dot{W}^{(\mathcal{S},\mathcal{M})}},
    \end{align}
    where \(\dot{W}^{(\mathcal{S},\mathcal{M})}\) denotes the input power supplied by ATP hydrolysis to the kinesin, and \(-\dot{W}^{\mathcal{X}}\) denotes the output power extracted from the cargo.
    Here, the pair $(\mathcal{S},\mathcal{M})$ represents the kinesin subsystem—comprising the internal state $\mathcal{S}$ and the spatial degree of freedom $\mathcal{M}$—while $\mathcal{X}$ represents the cargo subsystem, specifically its spatial degree of freedom.
    This efficiency quantifies the conversion of chemical free energy input into mechanical output work. 
    The analytical prediction obtained from a perturbative expansion agrees well with the numerical simulations and is consistent with the value under the \textit{in vitro} condition reported in Ref.~\cite{ariga2018nonequilibrium}.
    Under the \textit{in vivo}-like condition, the thermodynamic efficiency reaches a maximum of \(0.18\), which is lower than the corresponding value obtained in the \textit{in vitro} condition.

    \item[(ii)] \textit{Information-thermodynamic efficiency.}---For the kinesin subsystem \((\mathcal{S},\mathcal{M})\), the information-thermodynamic efficiency is defined as [cf.~Eq.~(\ref{eq:setup: Definition of Info. Efficiency})]
    \begin{align}
        \eta_{\mathrm{info}}^{(\mathcal{S},\mathcal{M})}
        := \frac{
        \dot{W}^{(\mathcal{S},\mathcal{M})\to \mathcal{X}}
        + k_{\mathrm{B}}T\dot{I}^{\mathcal{X}\to(\mathcal{S},\mathcal{M})}
        }{\dot{W}^{(\mathcal{S},\mathcal{M})}},
    \end{align}
    where \(\dot{W}^{(\mathcal{S},\mathcal{M})\to \mathcal{X}}\) denotes the power delivered from the kinesin \((\mathcal{S},\mathcal{M})\) to the cargo \(\mathcal{X}\), and \(k_{\mathrm{B}}T\dot{I}^{\mathcal{X}\to(\mathcal{S},\mathcal{M})}\) denotes the energetic contribution associated with the information flow from \(\mathcal{X}\) to \((\mathcal{S},\mathcal{M})\).
    This efficiency characterizes energy transduction within the kinesin subsystem while accounting for kinesin--cargo information transfer.
    We show that, under the assumption of time-scale separation, the information-thermodynamic efficiency reduces to the standard thermodynamic efficiency $\eta$.
    Furthermore, under no-load conditions ($F=0$), it nearly vanishes under the \textit{in vitro} condition, and remains at most approximately $0.3$ even under the \textit{in vivo}-like condition.
    This contrasts with the estimate $\eta_{\mathrm{info}}^{(\mathcal{S},\mathcal{M})}=0.7$--$0.9$ reported in Ref.~\cite{leighton2023inferring}.

    \item[(iii)] \textit{TUR efficiency.}---For the total kinesin--cargo system, the TUR efficiency is defined as [cf.~Eq.~(\ref{eq:setup: Definition of TUR Efficiency})]
    \begin{align}
        \eta_{\mathrm{TUR}}
        := \frac{J_{\mathcal{J}}^{2}}{D_{\mathcal{J}}}
        \frac{1}{\dot{\sigma}},
    \end{align}
    where \(J_{\mathcal{J}}\) and \(D_{\mathcal{J}}\) denote the mean and fluctuation of a selected current \(\mathcal{J}\), respectively, and \(\dot{\sigma}\) is the total entropy production rate.
    This efficiency measures how effectively the thermodynamic cost is utilized to suppress fluctuations of the selected current.
    Under the \textit{in vitro} condition, the maximum TUR efficiency is approximately \(0.25\).
    Under the \textit{in vivo}-like condition, the TUR efficiency is consistent with the previous estimate in Ref.~\cite{hwang2018energetic}, yielding \(\eta_{\mathrm{TUR}} \simeq 0.28\) at an external load of \(F=-1\,\mathrm{pN}\).

    \item[(iv)] \textit{Bipartite TUR efficiency.}---Finally, we consider the bipartite TUR efficiency [cf.~Eq.~(\ref{eq:setup: bipartite TUR Efficiency})]
    \begin{align}
        \eta_{\mathrm{BTUR}}
        := \frac{J_{\mathcal{J}}^{2}}{D_{\mathcal{J}}}
        \frac{(1+\delta_{\mathcal{J}})^2}
        {\dot{\sigma}^{(\mathcal{S},\mathcal{M})}},
    \end{align}
    where \(\dot{\sigma}^{(\mathcal{S},\mathcal{M})}\) is the partial entropy production rate associated with the kinesin subsystem \((\mathcal{S},\mathcal{M})\), and \(\delta_{\mathcal{J}}\) represents a correction term arising from the interaction between the kinesin and the cargo~\cite{tanogami2023universal}.
    This efficiency is a subsystem version of the TUR efficiency for bipartite systems.
    Under the \textit{in vitro} condition, the bipartite TUR efficiency is nearly identical to the standard TUR efficiency.
    Under the \textit{in vivo}-like condition, its value is lower than that under the \textit{in vitro} condition.
\end{itemize}
Taken together, these results suggest that kinesin may not be optimized for any of the thermodynamic efficiencies considered here.
In particular, because the information-thermodynamic efficiency---which was presumed to be high in Ref.~\cite{leighton2023inferring}---is also low in our analysis, kinesin does not appear to be optimized for kinesin--cargo information transfer within the present modeling framework.
These observations do not rule out functional optimization of kinesin; rather, they suggest that thermodynamic efficiencies alone may be insufficient to characterize kinesin performance.
Complementary kinetic viewpoints based on activity (frenesy) may therefore be essential~\cite{baiesi2018life,maes2020frenesy}.

The remainder of this paper is organized as follows.
In Sec.~\ref{sec:model}, we describe the thermodynamically consistent kinesin--cargo model used in this study.
In Sec.~\ref{sec: Basic properties}, we define the thermodynamic and information-theoretic quantities for our model, as well as the four efficiencies considered in this paper.
In Sec.~\ref{sec:Perturbation expansion}, we present the details of the perturbation expansion based on the time-scale separation between the motor and the cargo.
In Sec.~\ref{sec:Results on various efficiencies}, we present both the analytical and numerical results for each efficiency.
Finally, we provide concluding remarks in Sec.~\ref{sec:conclusion}.
The details of the analytical calculations and parameter estimations, as well as the estimation results for the various efficiencies under other conditions, are provided in the Appendices.
\begin{figure*}[t]
  \centering
  \includegraphics[width=\textwidth]{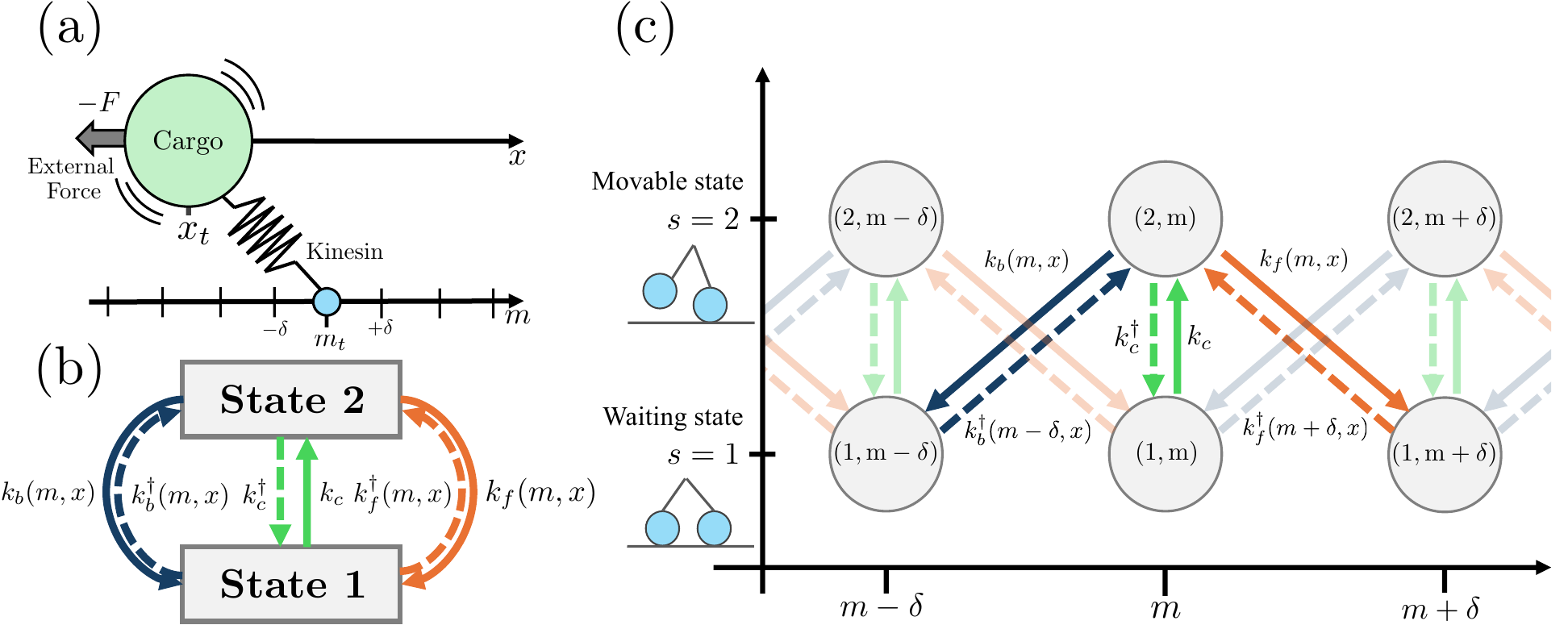}
  \caption{Schematic of (a) the kinesin--cargo complex,
  (b) the two-state kinesin stepper model,
  and (c) the transition diagram of kinesin governed by the transition matrix.
  The state \(s=1\) represents a waiting state where both kinesin heads are bound to the microtubule, whereas the state \(s=2\) represents a state where one of the heads is detached.}
  \label{fig:model}
\end{figure*}

\section{Model
\label{sec:model}}
We consider a coarse-grained description of the kinesin--cargo complex immersed in an equilibrium environment at temperature $T$ (see Fig.~\ref{fig:model}(a)).
This model can be regarded as a thermodynamically consistent modification of the model proposed in Ref.~\cite{ariga2018nonequilibrium}.
The state of the kinesin is described by two variables $(s,m)$, where $s\in\{1,2\}$ denotes the internal state of the kinesin and $m\in\delta\mathbb{Z}$ denotes the position of the kinesin, which takes the discrete value $\delta n$ ($n\in\mathbb{Z}$) with the step size $\delta=8\,\mathrm{nm}$~\cite{svoboda1993direct}.
Similarly, we denote by $x\in\mathbb{R}$ the position of the cargo pulled by the kinesin.
Then, the state of the kinesin--cargo complex is represented by the three variables $(s,m,x)$.
Below, we use the notation $(\mathcal{S},\mathcal{M})$ and $\mathcal{X}$ to represent the labels of each subsystem (the kinesin and the cargo), and also for the corresponding random variables taking values $(s,m)$ and $x$.

Let $p_t(s,m,x)$ be the joint probability density of the state $(s, m, x)$ at time $t$.
The time evolution of the joint probability density for the entire system is described by the following master equation:
\begin{align}
  \partial_t p_t(s, m, x) = \sum_{s', m'} W^{ss'}_{mm'}(x) &p_t(s', m', x) - \partial_x J_t^{\mathcal{X}}(s, m, x).
  \label{eq:setup: Entire Master eq.} 
\end{align}
Here, the first and second terms on the right-hand side of Eq.~(\ref{eq:setup: Entire Master eq.}) represent the time evolution associated with the kinesin and the cargo, respectively.
Specifically, $J_t^{\mathcal{X}} (s, m, x)$ in the second term denotes the probability current associated with the cargo defined by
\begin{align}
  J_t^\mathcal{X} (s, m, x) := &\frac{1}{\gamma} \bigg( - \partial_x U(m, x) + F \bigg) p_t(s, m, x) \notag \\
  &- \frac{k_{\mathrm{B}} T}{\gamma} \partial_x p_t (s, m, x),
  \label{eq:setup: Probability current of cargo}
\end{align}
where $\gamma$ denotes the friction coefficient, $U(m, x) := \kappa(m - x)^2/2$ denotes the interaction potential between the kinesin and the cargo with a spring constant $\kappa$, and $F$ denotes a constant external force, which represents, e.g., the force exerted on the cargo within a cell or the force applied by optical tweezers.
We note that the time evolution of the cargo described by Eq.~(\ref{eq:setup: Probability current of cargo}) is equivalent to the following overdamped Langevin equation~\cite{risken1996fokker},
\begin{align}
  \gamma \dot{x}_t = - \partial_x U(m_t, x_t) + F +  \sqrt{2 \gamma k_{\mathrm{B}} T}\xi_t.
  \label{eq:setup: Equation of cargo}
\end{align}
Here $\xi_t$ is a zero-mean white Gaussian noise that satisfies the fluctuation-dissipation relation of the second kind $\langle \xi_t \xi_{t'}\rangle = \delta (t - t')$, where $k_{\mathrm{B}}$ denotes the Boltzmann constant.

For the first term on the right-hand side of Eq.~(\ref{eq:setup: Entire Master eq.}), $W_{mm'}^{ss'}(x)$ denotes the transition rate matrix element for the kinesin from the state $(s',m')$ to $(s,m)$ at a fixed cargo position $x$ (see Figs.~\ref{fig:model}(b) and (c)).
For $(s,m)\neq (s',m')$, the transition rate matrix elements are defined as
\begin{align}
    W^{ss'}_{mm'}(x) =
        \begin{cases}
            k_f(m-\delta,x)
            &\text{for } \ (2,m-\delta) \to (1,m), \\[3pt]
            k_b(m+\delta,x)
            &\text{for } \ (2,m+\delta) \to (1,m), \\[3pt]
            k_c
            &\text{for } \ (1,m) \to (2,m), \\[3pt]
            k_f^\dagger(m+\delta,x)
            &\text{for } \ (1,m+\delta) \to (2,m), \\[3pt]
            k_b^\dagger(m-\delta,x)
            &\text{for } \ (1,m-\delta) \to (2,m), \\[3pt]
            k_c^\dagger
            &\text{for } \ (2,m) \to (1,m),
        \end{cases}
\end{align}
for any $m\in\delta\mathbb{Z}$, where the notation
\[
    (s',m') \to (s,m)
\]
in the conditions specifies the corresponding transition from the state $(s',m')$ to $(s,m)$ at a fixed cargo position $x$.
For $(s, m)=(s', m')$, the transition rate matrix elements are defined to satisfy the normalization condition $\sum_{s,m} W^{ss'}_{mm'}(x)=0$, i.e.,
\begin{align}
W^{ss}_{mm}(x) &:= -\sum_{(s',m')(\neq(s,m))} W^{s's}_{m'm}(x)\notag\\
  &=
  \begin{cases}
  - (k_f^\dag (m, x) + k_b^\dag (m, x) + k_c)&\quad\text{if}\quad s=1, \\
  - (k_f (m, x) + k_b (m, x) + k_c^\dag)  &\quad\text{if}\quad s=2.
  \end{cases}
\end{align}
Here, $k_f(m, x)$ denotes the forward transition rate from state $(2, m, x)$ to $(1, m + \delta, x)$, which depends on the positions of the kinesin and the cargo $(m,x)$.
Similarly, $k_b(m,x)$ denotes the backward transition rate from state $(2, m, x)$ to $(1, m - \delta, x)$, while $k_c$ denotes the constant transition rate from state $(1, m, x)$ to $(2, m, x)$, which is independent of $m$ and $x$. 
In other words, $k_f(m,x)$ and $k_b(m,x)$ correspond to the mechanical steps, while $k_c$ corresponds to the internal chemical process.
The daggered transition rates, $k^\dag_\alpha$ $(\alpha \in \{f, b, c\})$, denote the reverse transition rates corresponding to $k_\alpha$.

To ensure the thermodynamic consistency of the model, we impose the local detailed balance condition~\cite{maes2021local,peliti2021stochastic}, which states that the logarithm of the ratio between $k_\alpha$ and $k^\dag_\alpha$ is identified as the entropy produced in the environment:
\begin{align}
  \frac{k_f(m, x)}{k_f^\dag(m + \delta, x)}
  &= \exp \bigg[
    \frac{U(m, x) - U(m + \delta, x) + \Delta\mu_{\mathrm{mech}}}{k_{\mathrm{B}} T}
  \bigg],
  \label{eq:setup: LDB of the kinesin (kf)}\\
  \frac{k_b(m, x)}{k_b^\dag(m - \delta, x)}
  &= \exp \bigg[
    \frac{U(m, x) - U(m - \delta, x) + \Delta\mu_{\mathrm{mech}}}{k_{\mathrm{B}} T}
  \bigg],
  \label{eq:setup: LDB of the kinesin (kb)}\\
  \frac{k_c}{k_c^\dag}
  &= \exp \bigg[
    \frac{\Delta \mu_{\mathrm{chem}}}{k_{\mathrm{B}} T}
  \bigg].
  \label{eq:setup: LDB of the kinesin (kc)}
\end{align}
Here, $\Delta\mu_{\mathrm{mech}} > 0$ and $\Delta \mu_{\mathrm{chem}} > 0$ represent the free energy changes associated with the mechanical forward/backward transition $(2, m, x) \to (1, m \pm \delta, x)$ and the chemical transition $(1, m, x) \to (2, m, x)$, respectively. 
Their sum, $\Delta \mu = \Delta\mu_{\mathrm{mech}} + \Delta \mu_{\mathrm{chem}}$, corresponds to the total input free energy obtained from the hydrolysis of a single ATP molecule. 
This free energy is determined by the chemical potential difference of the hydrolysis reaction given by $\Delta \mu = \Delta \mu^0 + k_{\mathrm{B}}T \ln \{[\mathrm{ATP}]/([\mathrm{ADP}][\mathrm{P_i}])\}$, where $[\mathrm{ATP}]$, $[\mathrm{ADP}]$, and $[\mathrm{P_i}]$ are the concentrations of adenosine triphosphate (ATP), adenosine diphosphate (ADP), and inorganic phosphate ($\mathrm{P_i}$), respectively. 
The term $\Delta \mu^0$ is the standard chemical potential at reference concentrations.
We note that the specific form of $k_\alpha(m,x)$ and $k_\alpha^\dag (m,x)$ ($\alpha\in\{f,b\}$) cannot be determined only from the local detailed balance condition. 
Although the specific choice of transition rates may significantly affect physics in the nonequilibrium case~\cite{tasaki2004remark}, here we adopt the following specific functional forms, following Ref.~\cite{kawaguchi2014nonequilibrium}:
\begin{align}
  k_f(m, x) &:= \frac{1}{\tau_f} e^{\frac{\theta_f}{k_{\mathrm{B}} T} [U(m, x) - U(m + \delta, x) + \Delta\mu_{\mathrm{mech}} ]}, \label{eq:setup:form of kf}\\
  k_f^\dag(m, x) &:= \frac{1}{\tau_f} e^{\frac{\theta_f - 1}{k_{\mathrm{B}} T} [U(m - \delta, x) - U(m, x) + \Delta\mu_{\mathrm{mech}}] }, \label{eq:setup:form of kf dag}\\
  k_b (m, x) &:= \frac{1}{\tau_b} e^{\frac{\theta_b}{k_{\mathrm{B}} T} [U(m, x) - U(m - \delta, x) + \Delta\mu_{\mathrm{mech}}] }, \label{eq:setup:form of kb}\\
  k_b^\dag(m, x) &:= \frac{1}{\tau_b} e^{\frac{\theta_b - 1}{k_{\mathrm{B}} T} [U(m + \delta, x) - U(m, x) + \Delta\mu_{\mathrm{mech}}] },
  \label{eq:setup: Specific transition rate of kinesin (kb dag)}
\end{align}
where the parameter $\tau_\alpha$ denotes the characteristic time scale for the forward/backward transition, and $\theta_\alpha \in [0, 1]$ is an asymmetry parameter that determines the dependence of these rates on $x$ and $m$, while respecting the local detailed balance condition. 
Note that $k_\alpha(m,x)$ depends only on the relative displacement $(x-m)$: $k_\alpha (m, x) = k_\alpha (0, x - m)$ and  $k^\dag_\alpha (m, x) = k^\dag_\alpha (0, x - m)$.
We remark that the functional form of the transition rates used in this study [Eqs.~(\ref{eq:setup:form of kf})-(\ref{eq:setup: Specific transition rate of kinesin (kb dag)})] is different from that used in previous studies~\cite{taniguchi2005entropy, nishiyama2002chemomechanical,ariga2018nonequilibrium,ariga2021noise}.
In the previous studies, the transition rates $k_f$ and $k_b$ were described by Bell's equation~\cite{bell1978models}, which is independent of $m$ and $x$ and therefore does not satisfy the local detailed balance condition of the form in Eqs.~(\ref{eq:setup: LDB of the kinesin (kf)})-(\ref{eq:setup: LDB of the kinesin (kc)}).
For a more detailed comparison between Eqs.~(\ref{eq:setup:form of kf})-(\ref{eq:setup: Specific transition rate of kinesin (kb dag)}) and Bell's equation, see Appendix~\ref{sec:appendix: comparison with bell's equation}.

Although our model includes many parameters, they can all be estimated using the experimental data obtained in Ref.~\cite{ariga2018nonequilibrium}.
Specifically, the spring constant $\kappa$ and the friction coefficient $\gamma$ can be estimated from the experimentally measured power spectral density of the probe position~\cite{ariga2018nonequilibrium}.
The remaining six parameters $\tau_f, \tau_b, \theta_f, \theta_b, k_c, \Delta \mu_{\mathrm{chem}}$ can be determined by fitting the theoretical prediction of the force--velocity relationship derived from the model to the experimental data.
For the details of the fitting procedure and the specific values of the estimated parameters, see Appendix~\ref{sec:appendix:parameter estimation}.

For later convenience, we describe the effective dynamics for the kinesin and the cargo, respectively.
Let $p_t(s,m):=\int_{-\infty}^{\infty}dx\,p_t(s,m,x)$ be the marginal probability distribution for the kinesin.
Its time evolution can be derived from Eq. \eqref{eq:setup: Entire Master eq.} as 
\begin{align}
  \partial_t p_t(s, m) &= \sum_{s', m'}  \overline{W}_{mm'} ^{ss'} p_t(s', m') 
  \label{eq:setup: Time evolution eq. of (S,M)}.
\end{align}
Here, $\overline{W}_{mm'}^{ss'}$ is the effective transition matrix defined as
\begin{align}
  \overline{W}_{mm'} ^{ss'} := \int_{-\infty}^{\infty} dx W_{mm'} ^{ss'} (x) p_t (x | s', m'),
  \label{eq:setup: effective transition rate matrix}
\end{align}
where $p_t (x | s', m'):=p_t(s',m',x)/p_t(s',m')$ is the conditional probability density for $x$ conditioned on $(s',m')$.
Similarly, the time evolution equation for the marginal probability density for the cargo, $p_t(x):=\sum_{s}\sum_{m}p_t(s,m,x)$, is given by
\begin{align}
  \partial_t p_t(x) &= - \partial_x \overline{J}^\mathcal{X}_t (x)
  \label{eq:setup: Time evolution eq. of X}.
\end{align}
Here, $\overline{J}^\mathcal{X}_t (x)$ is the effective probability current, defined as
\begin{align}
  \overline{J}^\mathcal{X}_t (x) := \sum_{s} \sum_{m} J^\mathcal{X}_t (s, m, x).
  \label{eq:setup: effective probability current}
\end{align}


\section{Basic properties\label{sec: Basic properties}}
In this section, we define the thermodynamic and information-theoretic quantities used to quantify the performance of the kinesin--cargo system.
In Sec.~\ref{subsec: Thermodynamic properties}, we introduce the thermodynamic quantities of the model and present the corresponding first and second laws of thermodynamics.
In Sec.~\ref{subsec: Information-thermodynamic properties}, we define the information-theoretic quantities and formulate the first and second laws of information thermodynamics.
In Sec.~\ref{Various definitions of efficiency}, we define the four efficiencies analyzed in this paper: the thermodynamic efficiency, the information-thermodynamic efficiency, the TUR efficiency, and the bipartite TUR efficiency.

\subsection{Thermodynamic properties\label{subsec: Thermodynamic properties}}

\begin{figure}
    \centering
    \includegraphics[width=\linewidth]{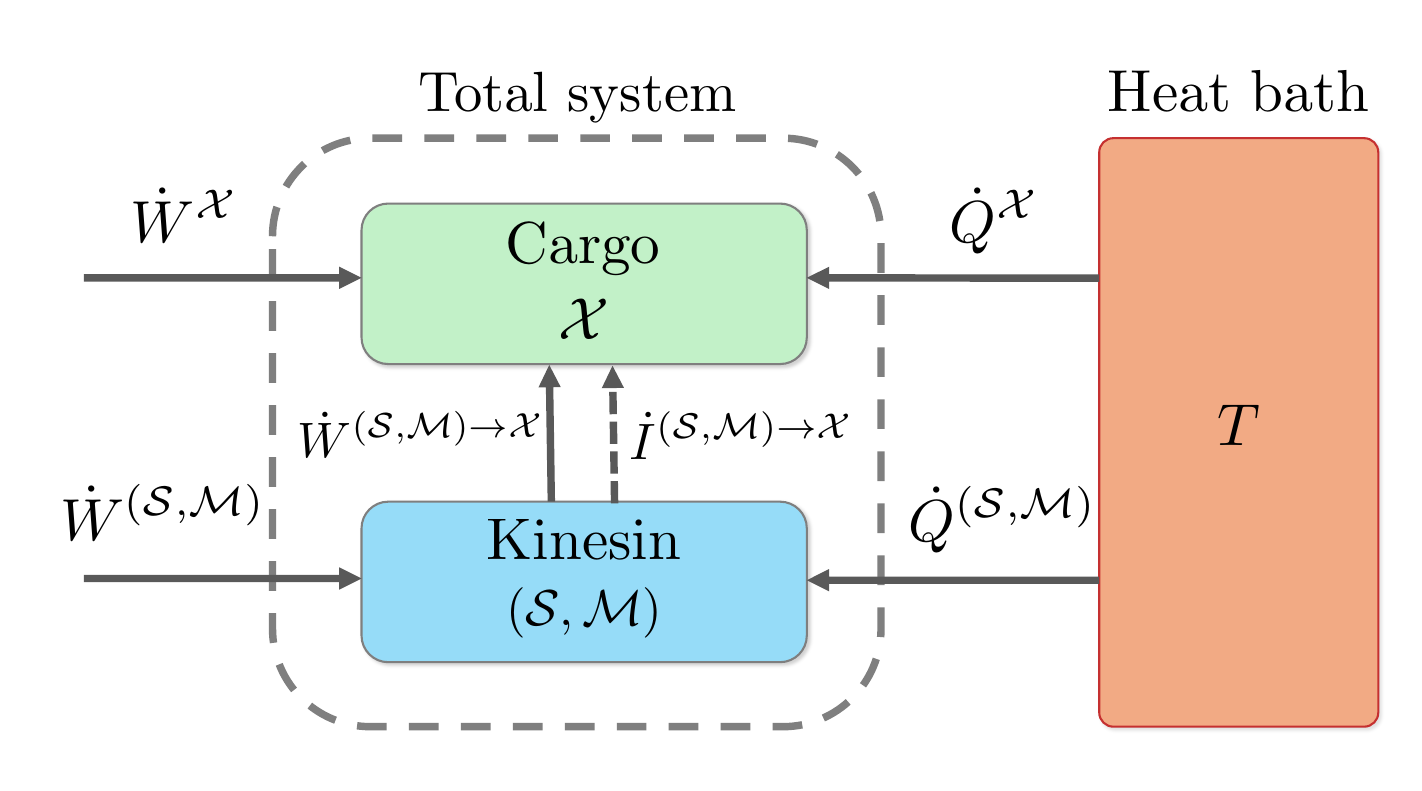}
    \caption{Schematic of the relevant thermodynamic and information-theoretic quantities in our model.
    Note that arrows indicate positive flows under our sign convention.
    }
    \label{fig:basic properties}
\end{figure}

\subsubsection{Definitions of thermodynamic quantities}
Here, we introduce the relevant thermodynamic quantities for our study. 
A schematic overview of the corresponding physical quantities in our model is shown in Fig.~\ref{fig:basic properties}.
Throughout this paper, we treat entropy as dimensionless by dividing it by the Boltzmann constant~$k_\mathrm{B}$.

The system entropy of the kinesin--cargo complex is identified with the Shannon entropy $S[\mathcal{S}_t, \mathcal{M}_t, \mathcal{X}_t]$ for the joint probability density $p_t(s, m, x)$, defined by
\begin{align}
  S[\mathcal{S}_t, \mathcal{M}_t, \mathcal{X}_t] := - \int_{-\infty}^{\infty} dx \sum_{s,m} p_t(s, m, x) \ln p_t(s,m,x).
  \label{eq:setup: The entropy of the composite system}
\end{align}
Here, we use the notation $S[\mathcal{S}_t, \mathcal{M}_t, \mathcal{X}_t]$ to indicate the relevant random variables $\mathcal{S}_t$, $\mathcal{M}_t$, and $\mathcal{X}_t$, although $S[\mathcal{S}_t, \mathcal{M}_t, \mathcal{X}_t]$ is not a function of these random variables but a functional of $p_t(s, m, x)$.
The average rate of change of the system entropy is given by the time derivative of this quantity,
\begin{align}
  &d_t S[\mathcal{S}_t, \mathcal{M}_t, \mathcal{X}_t]  \notag \\
  &= -  \int_{-\infty}^{\infty} dx \sum_{s', s, m', m} W^{ss'}_{mm'} (x) p_t(s', m', x) \ln \frac{p_t(s, m, x)}{p_t(s', m', x)} \notag \\
  &\hspace{30pt} - \int_{-\infty}^{\infty}dx \sum_{s, m} J_t^\mathcal{X} (s, m, x) \partial_x \ln p_t (s, m, x).
  \label{eq:setup: The entropy "Change" of the composite system}
\end{align}

The average rate of entropy produced in the environment can be decomposed into two parts associated with the kinesin and the cargo.
The average entropy change in the environment due to transitions of the kinesin is identified from the local detailed balance as~\cite{peliti2021stochastic}
\begin{align}
  &\dot{S}^{(\mathcal{S}, \mathcal{M})}_{\mathrm{env}} \notag \\
  &:= \int_{-\infty}^{\infty} dx \sum_{s, s', m, m'} W_{mm'}^{ss'}(x) p_t (s',m',x) \ln \frac{W_{mm'}^{ss'} (x)}{W_{m'm}^{s's} (x)}.
  \label{eq:setup: the average entropy increase rate in the environment due to transitions of the kinesin}
\end{align}
The average entropy increase rate in the environment associated with the cargo can be identified according to Sekimoto's argument~\cite{sekimoto2010stochastic}.
Because the model satisfies the fluctuation-dissipation relation of the second kind, the thermal environment is assumed to remain in equilibrium at temperature $T$.
Then, by noting that $-\gamma\dot{x}_t+\sqrt{2 \gamma k_{\mathrm{B}} T}\xi_t$ can be interpreted as a force exerted by the environment on the cargo, the entropy change in the environment is identified as the work done by the cargo on the environment per unit time divided by $k_{\mathrm{B}}T$:
\begin{align}
  \dot{S}^\mathcal{X}_{\mathrm{env}} &:= \frac{1}{k_{\mathrm{B}} T} \langle \dot{x}_t \circ (\gamma \dot{x}_t - \sqrt{2 \gamma k_{\mathrm{B}} T} \xi_t) \rangle \notag \\
  &= \frac{1}{k_{\mathrm{B}} T} \int_{-\infty}^{\infty} dx \sum_{s,m} J_t ^\mathcal{X}(s, m, x) (- \partial_x U(m, x) + F),
  \label{eq:setup: the average entropy increase rate in the environment due to the time evolution of the cargo}
\end{align}
where the symbol $\circ$ denotes the multiplication in the sense of Stratonovich~\cite{gardiner1985handbook}.
Note that these quantities are related to the average rates of heat absorbed from the environment by the kinesin, $\dot{Q}^{(\mathcal{S}, \mathcal{M})}$, and the cargo, $\dot{Q}^\mathcal{X}$, through the following relations,
\begin{align}
  \dot{Q}^{(\mathcal{S}, \mathcal{M})} &= - k_{\mathrm{B}} T \dot{S}^{(\mathcal{S}, \mathcal{M})}_{\mathrm{env}}, \label{eq:setup:(S,M) relation of entropy and heat} \\
  \dot{Q}^\mathcal{X} &= - k_{\mathrm{B}} T \dot{S}^\mathcal{X}_{\mathrm{env}}.
  \label{eq:setup:X relation of entropy and heat}
\end{align}

The average rate of work performed on the kinesin--cargo complex can also be decomposed into two parts associated with each subsystem.
To concisely describe the average rate of work performed on the kinesin, we denote the free-energy change in the environment associated with the transition $(s',m')\rightarrow(s,m)$ by the symbol $\Delta\mu_{mm'}^{ss'}$.
Specifically, it is given by
\begin{align}
    \Delta\mu_{mm'}^{ss'} :=
        \begin{cases}
            \Delta\mu_{\mathrm{mech}}
            &\text{for } \ (2,m-\delta) \to (1,m), \\[3pt]
            \Delta\mu_{\mathrm{mech}}
            &\text{for } \ (2,m+\delta) \to (1,m), \\[3pt]
            \Delta \mu_{\mathrm{chem}}
            &\text{for } \ (1,m) \to (2,m), \\[3pt]
            -\Delta\mu_{\mathrm{mech}}
            &\text{for } \ (1,m+\delta) \to (2,m), \\[3pt]
            -\Delta\mu_{\mathrm{mech}}
            &\text{for } \ (1,m-\delta) \to (2,m), \\[3pt]
            -\Delta \mu_{\mathrm{chem}}
            &\text{for } \ (2,m) \to (1,m),
        \end{cases}
\end{align}
and $\Delta\mu_{mm'}^{ss'}=0$ for $(s,m)=(s',m')$.
Then, the average rate of work performed on the kinesin can be identified as
\begin{align}
  &\dot{W}^{(\mathcal{S}, \mathcal{M})} \notag\\
  &= \int_{-\infty}^{\infty} dx \sum_{s, s', m, m'} W_{mm'}^{ss'}(x) p_t (s',m',x) \Delta\mu_{mm'}^{ss'},
  \label{eq:setup: work of (S,M)}
\end{align}
which can be interpreted as the input power supplied to the kinesin by ATP hydrolysis.
Similarly, the average rate of work performed on the cargo can be identified as the input or output power exerted by the constant external force $F$ on the cargo, which reads
\begin{align}
  \dot{W}^\mathcal{X} &= F \langle \dot{x}_t \rangle \notag \\
  &= \int_{-\infty}^{\infty} dx \sum_{s,m} F J_t^\mathcal{X} (s, m, x).
  \label{eq:setup: Definition of the work of the cargo (W X)}
\end{align}

\subsubsection{First law for the total system}
By relating the change in the interaction energy to the heat and work exchanged with the environment, we can formulate the first law of thermodynamics for this system. 
The average rate of change of the interaction energy $U$ is given by
\begin{align}
  d_t \langle U \rangle &= \int_{-\infty}^{\infty} dx \sum_{s, m} \partial_t p_t (s,m,x) U(m, x).
  \label{eq:setup: average rate of change of the interaction energy}
\end{align}
By substituting Eq.~(\ref{eq:setup: Entire Master eq.}) into Eq.~(\ref{eq:setup: average rate of change of the interaction energy}) and using Eqs.~(\ref{eq:setup:(S,M) relation of entropy and heat}), (\ref{eq:setup:X relation of entropy and heat}), (\ref{eq:setup: work of (S,M)}), and (\ref{eq:setup: Definition of the work of the cargo (W X)}), we obtain
\begin{align}
  d_t \langle U \rangle =\dot{W}^{(\mathcal{S}, \mathcal{M})} + \dot{W}^\mathcal{X} + \dot{Q}^{(\mathcal{S}, \mathcal{M})} + \dot{Q}^{\mathcal{X}}.
  \label{eq:setup: the first law of thermodynamics (entire system)}
\end{align}
If we identify the interaction potential as the internal energy of the kinesin--cargo complex, this relation can be regarded as the first law of thermodynamics for this model.

\subsubsection{Second law for the total system}
The total entropy production rate, which we denote by $\dot{\sigma}$, is identified as the sum of the average rate of change of the system entropy [Eq.~\eqref{eq:setup: The entropy "Change" of the composite system}] and the average entropy change in the environment [Eqs.~\eqref{eq:setup: the average entropy increase rate in the environment due to transitions of the kinesin} and~\eqref{eq:setup: the average entropy increase rate in the environment due to the time evolution of the cargo}]:
\begin{align}
  \dot{\sigma} &= d_t S[\mathcal{S}_t, \mathcal{M}_t, \mathcal{X}_t] + \dot{S}^{(\mathcal{S}, \mathcal{M})}_{\mathrm{env}} + \dot{S}^\mathcal{X}_{\mathrm{env}} \notag \\
  &= \int_{-\infty}^{\infty} dx \sum_{s, s',m , m'} W^{ss'}_{mm'} (x) p_t(s', m', x) \notag \\
  &\hspace{20pt} \times \ln \frac{W^{ss'}_{mm'} (x)p_t(s', m', x)}{W^{s's}_{m'm} (x) p_t(s, m, x)} \notag \\
  &\hspace{30pt} + \frac{\gamma}{k_{\mathrm{B}} T} \int_{-\infty}^{\infty} dx \sum_{s,m}\frac{(J_t^\mathcal{X} (s, m, x))^2}{p_t (s, m, x)} \geq 0,
  \label{eq:setup: the second law of thermodynamics}
\end{align}
where the non-negativity of the first term on the right-hand side follows from the relation $\ln a\le a-1$ ($a\ge0$).
The non-negativity of the total entropy production rate is a manifestation of the second law of thermodynamics.
Within the framework of stochastic thermodynamics, which is constructed based on the thermodynamic consistency, Eq.~(\ref{eq:setup: the second law of thermodynamics}) is also called the second law of stochastic thermodynamics~\cite{peliti2021stochastic}.

\subsection{Information-thermodynamic properties\label{subsec: Information-thermodynamic properties}}

\subsubsection{Definitions of information-theoretic quantities}
Since we are interested in the information transfer between the kinesin and the cargo, we introduce the mutual information between $(\mathcal{S}_t, \mathcal{M}_t)$ and $\mathcal{X}_t$ defined by
\begin{align}
  I[(\mathcal{S}_t, \mathcal{M}_t)\colon\!\mathcal{X}_t] := \int_{-\infty}^{\infty} dx \sum_{s, m} p_t (s, m, x) \ln \frac{p_t (s, m, x)}{p_t (s, m) p_t(x)}.
  \label{eq:setup: definition of MI}
\end{align}
Here, we used the notation $I[(\mathcal{S}_t, \mathcal{M}_t)\colon\!\mathcal{X}_t]$ to explicitly indicate the relevant random variables, as in the Shannon entropy.
The mutual information is nonnegative and is equal to zero if and only if $(\mathcal{S}_t, \mathcal{M}_t)$ and $\mathcal{X}_t$ are statistically independent.
Intuitively, $I[(\mathcal{S}_t, \mathcal{M}_t)\colon\!\mathcal{X}_t]$ quantifies the mutual dependence between the kinesin and the cargo.

The decomposition of the time derivative of the mutual information yields the information flow~\cite{allahverdyan2009thermodynamic,horowitz2014thermodynamics,parrondo2015thermodynamics}. 
Specifically, the information flow from the cargo $\mathcal{X}$ to the kinesin $(\mathcal{S}, \mathcal{M})$ is defined as
\begin{align}
  &\dot{I}^{\mathcal{X} \to (\mathcal{S}, \mathcal{M})} \notag\\
  &:=\lim_{\Delta t \to 0} \frac{I[(\mathcal{S}_{t + \Delta t} , \mathcal{M}_{t + \Delta t})\colon\!\mathcal{X}_t] - I[(\mathcal{S}_t, \mathcal{M}_t)\colon\!\mathcal{X}_t]}{\Delta t} \notag \\
  &= \int_{-\infty}^{\infty}dx \sum_{s, s', m, m'} W^{ss'}_{mm'} (x) p_t (s', m', x) \ln \frac{p_t (s, m, x)}{p_t (s, m) p_t(x)}.
  \label{eq:setup: Info Flow of I(S,M)}
\end{align}
Similarly, the information flow from the kinesin $(\mathcal{S}, \mathcal{M})$ to the cargo $\mathcal{X}$ is defined as 
\begin{align}
  \dot{I}^{(\mathcal{S}, \mathcal{M}) \to \mathcal{X}} &:= \lim_{\Delta t \to 0} \frac{I[(\mathcal{S}_{t} , \mathcal{M}_{t})\colon\!\mathcal{X}_{t + \Delta t}] - I[(\mathcal{S}_t, \mathcal{M}_t)\colon\!\mathcal{X}_t]}{\Delta t} \notag \\
  &= \int_{-\infty}^{\infty} dx \sum_{s, m} J_t^\mathcal{X} (s, m, x) \partial_x \ln \frac{p_t (s, m, x)}{p_t (s, m) p_t(x)}.
  \label{eq:setup: Info Flow of I(X)}
\end{align}
The sum of these two information flows equals the time derivative of the mutual information:
\begin{align}
  d_t I[(\mathcal{S}_t, \mathcal{M}_t)\colon\!\mathcal{X}_t] &= \dot{I}^{\mathcal{X} \to (\mathcal{S}, \mathcal{M})} + \dot{I}^{(\mathcal{S}, \mathcal{M}) \to \mathcal{X}}. 
\label{eq:setup: sum of two info flow}
\end{align}
Note that in the steady state where $d_t I[(\mathcal{S}_t, \mathcal{M}_t)\colon\!\mathcal{X}_t]=0$, which will be examined in Sec.~\ref{sec:Perturbation expansion}, the two information flows have opposite signs: $\dot{I}^{\mathcal{X} \to (\mathcal{S}, \mathcal{M})} = - \dot{I}^{(\mathcal{S}, \mathcal{M}) \to \mathcal{X}}$.

The information flow $\dot{I}^{\mathcal{X} \to (\mathcal{S}, \mathcal{M})}$ quantifies the rate at which the kinesin gains information about the cargo. 
In particular, its sign indicates the direction in which the information is transmitted.
For example, if $\dot{I}^{\mathcal{X} \to (\mathcal{S}, \mathcal{M})} > 0$, the mutual information increases due to infinitesimal time evolution of the kinesin.
In this sense, the information of the cargo is propagated to the kinesin.
If $\dot{I}^{\mathcal{X} \to (\mathcal{S}, \mathcal{M})} < 0$, in contrast, $(\mathcal{S}_{t + \Delta t} , \mathcal{M}_{t + \Delta t})$ is less correlated with $\mathcal{X}_t$ than $(\mathcal{S}_{t} , \mathcal{M}_{t})$, and the information is destroyed or transferred from the kinesin to the cargo.
Thus, we can quantify the direction and magnitude of the information transmitted between the subsystems using the information flow.

Here, we remark on the bipartite property of our system.
In general, two stochastic processes $\mathcal{Z}_{1,t}$ and $\mathcal{Z}_{2,t}$ are said to satisfy the bipartite condition if the transition probability $p(z_{1,t+\Delta t},z_{2,t+\Delta t}|z_{1,t},z_{2,t})$ satisfies
\begin{align}
&p(z_{1,t+\Delta t},z_{2,t+\Delta t}|z_{1,t},z_{2,t})\notag\\
&=p(z_{1,t+\Delta t}|z_{1,t},z_{2,t})p(z_{2,t+\Delta t}|z_{1,t},z_{2,t})
\end{align}
for $\Delta t\rightarrow0^+$.
This property means that the two stochastic processes do not jump simultaneously in the case of Markov jump processes and that the noises acting on $\mathcal{Z}_{1,t}$ and $\mathcal{Z}_{2,t}$ are uncorrelated in the case of diffusion processes.
Whether our model satisfies the bipartite condition depends on which two stochastic processes we consider.
For example, the two stochastic processes $(\mathcal{S}_t,\mathcal{M}_t)$ and $\mathcal{X}_t$ satisfy the bipartite condition although they are affecting each other through the interaction potential $U(m,x)$.
In contrast, $\mathcal{S}_t$ and $\mathcal{M}_t$ do not satisfy the bipartite condition because there are transitions where both $\mathcal{S}_t$ and $\mathcal{M}_t$ change simultaneously.
We remark that the generalization of the information flow to the non-bipartite case is non-trivial and the relation (\ref{eq:setup: sum of two info flow}) no longer holds in that case~\cite{chetrite2019information}.

\subsubsection{First law for the subsystem}
Under the bipartite condition, we can formulate the first law of thermodynamics for each subsystem, the kinesin $(\mathcal{S}, \mathcal{M})$ and the cargo $\mathcal{X}$.
First, we note that the average rate of change of the interaction energy [Eq.~(\ref{eq:setup: average rate of change of the interaction energy})] can be decomposed into contributions due to the respective dynamics of each subsystem:
\begin{align}
    d_t \langle U \rangle = \dot{W}^{(\mathcal{S}, \mathcal{M}) \to \mathcal{X}} + \dot{W}^{\mathcal{X} \to (\mathcal{S}, \mathcal{M})},
    \label{eq:setup: dU decomposition}
\end{align}
where $\dot{W}^{(\mathcal{S}, \mathcal{M}) \to \mathcal{X}}$ and $\dot{W}^{\mathcal{X} \to (\mathcal{S}, \mathcal{M})}$ are defined as
\begin{align}
  \dot{W}^{(\mathcal{S}, \mathcal{M}) \to \mathcal{X}} &= \int_{-\infty}^{\infty} dx \sum_{s, s', m, m'} W^{ss'}_{mm'} (x) p_t(s', m', x) \notag \\
  &\hspace{40pt} \times \Big( U(m, x) - U(m', x) \Big),
  \label{eq:setup: W (S,M) to X} \\
  \dot{W}^{\mathcal{X} \to (\mathcal{S}, \mathcal{M})} &= \langle \dot{x}_t \circ \partial_x U(m, x)\rangle \notag \\
  &= \int_{-\infty}^{\infty} dx \sum_{s, m} J_t^{\mathcal{X}}(s, m, x) \partial_x U(m, x).
  \label{eq:setup: W X to (S,M)}
\end{align}
Here, we used the notation $\dot{W}^{(\mathcal{S}, \mathcal{M}) \to \mathcal{X}}$ to indicate that it can be identified as the power delivered from the kinesin $(\mathcal{S}, \mathcal{M})$ to the cargo $\mathcal{X}$, if we regard $(s,m)$ as an externally manipulated control parameter driving $\mathcal{X}$~\cite{ehrich2023energy}. 
Similarly, $\dot{W}^{\mathcal{X} \to (\mathcal{S}, \mathcal{M})}$ can be interpreted as the power delivered from $\mathcal{X}$ to $(\mathcal{S}, \mathcal{M})$. 
In the steady state with $d_t \langle U \rangle = 0$, we find that $\dot{W}^{(\mathcal{S}, \mathcal{M}) \to \mathcal{X}} = -\dot{W}^{\mathcal{X} \to (\mathcal{S}, \mathcal{M})}$.
By using Eqs.~(\ref{eq:setup:(S,M) relation of entropy and heat}), (\ref{eq:setup:X relation of entropy and heat}), (\ref{eq:setup: work of (S,M)}), and (\ref{eq:setup: Definition of the work of the cargo (W X)}), the first law for the kinesin--cargo complex Eq.~\eqref{eq:setup: the first law of thermodynamics (entire system)} can be decomposed as
\begin{align}
  \dot{W}^{(\mathcal{S}, \mathcal{M}) \to \mathcal{X}} &= \dot{W}^{(\mathcal{S}, \mathcal{M})} + \dot{Q}^{(\mathcal{S}, \mathcal{M})}, \label{eq:setup: the first law of the kinesin}\\
  \dot{W}^{\mathcal{X} \to (\mathcal{S}, \mathcal{M})} &= \dot{W}^\mathcal{X} + \dot{Q}^\mathcal{X}.
  \label{eq:setup: the first law of the cargo}
\end{align}
Equations \eqref{eq:setup: the first law of the kinesin} and \eqref{eq:setup: the first law of the cargo} represent the energy balance for the kinesin and the cargo, respectively. In other words, these relations can be interpreted as the first law of thermodynamics for each subsystem~\cite{ehrich2023energy,leighton2023inferring}.

\subsubsection{Second law for the subsystem}
Based on the bipartite condition, we can also formulate the second law of thermodynamics for each subsystem.
From Eq.~\eqref{eq:setup: the second law of thermodynamics}, we decompose the total entropy production rate into two nonnegative contributions $\dot{\sigma}^{(\mathcal{S}, \mathcal{M})}$ and $\dot{\sigma}^\mathcal{X}$ as follows:
\begin{align}
  \dot{\sigma} = \dot{\sigma}^{(\mathcal{S}, \mathcal{M})} + \dot{\sigma}^{\mathcal{X}},
\end{align}
where
\begin{align}
  \dot{\sigma}^{(\mathcal{S}, \mathcal{M})} &:= \int_{-\infty}^{\infty} dx \sum_{s, s',m , m'} W^{ss'}_{mm'} (x) p_t(s', m', x) \notag \\
  &\hspace{70pt} \times \ln \frac{W^{ss'}_{mm'} (x)p_t(s', m', x)}{W^{s's}_{m'm} (x) p_t(s, m, x)}\ge0,
  \label{eq:setup: entropy production rate of (S,M)}\\
  \dot{\sigma}^\mathcal{X} &:= \frac{\gamma}{k_{\mathrm{B}} T} \int_{-\infty}^{\infty} dx \sum_{s, m} \frac{(J_t^\mathcal{X} (s, m, x))^2}{p_t (s, m, x)}\ge0,
  \label{eq:setup: entropy production rate of X}
\end{align}
denote the \textit{partial entropy production rates}~\cite{shiraishi2015fluctuation} due to the time evolution of the kinesin $(\mathcal{S}, \mathcal{M})$ and the cargo $\mathcal{X}$, respectively.

The partial entropy production rates can be further decomposed into thermodynamic and information-theoretic parts. 
To see this, we introduce the system entropy for each subsystem, $(\mathcal{S}, \mathcal{M})$ and $\mathcal{X}$: 
\begin{align}
  S[\mathcal{S}_t, \mathcal{M}_t] &:= - \sum_{s, m} p_t (s, m) \ln p_t(s,m),
  \label{eq:setup: (S,M) Shannon entropy}\\
  S[\mathcal{X}_t] &:= - \int_{-\infty}^{\infty} dx p_t (x) \ln p_t(x).
  \label{eq:setup: X Shannon entropy}
\end{align}
From Eqs.~\eqref{eq:setup: Time evolution eq. of (S,M)} and \eqref{eq:setup: Time evolution eq. of X}, the time derivative of these quantities can be calculated as 
\begin{align}
  d_t S[\mathcal{S}_t, \mathcal{M}_t] &= - \sum_{s, s', m, m'}  \overline{W}^{ss'}_{mm'} p_t(s', m') \ln \frac{p_t (s,m)}{p_t(s',m')}, \\
  d_t S[\mathcal{X}_t] &= - \int_{-\infty}^{\infty} dx \overline{J}^\mathcal{X}_t (x) \partial_x \ln p_t(x).
\end{align}
Let $\dot{S}^{(\mathcal{S}, \mathcal{M})}_{\mathrm{tot}}$ and $\dot{S}^{\mathcal{X}}_{\mathrm{tot}}$ be the subsystem entropy production rate associated with the kinesin and the cargo, respectively.
Each entropy production rate consists of the time derivative of the corresponding subsystem entropy and the entropy increase rate in the environment:
\begin{align}
  \dot{S}^{(\mathcal{S}, \mathcal{M})}_{\mathrm{tot}} &:= d_t S[\mathcal{S}_t, \mathcal{M}_t] + \dot{S}^{(\mathcal{S}, \mathcal{M})}_{\mathrm{env}}, \label{eq:setup:(S, M) def of the entropy pruduction rate}\\
  \dot{S}^{\mathcal{X}}_{\mathrm{tot}} &:= d_t S[\mathcal{X}_t] + \dot{S}^\mathcal{X}_{\mathrm{env}}.
\end{align}
Then, by using Eqs.~(\ref{eq:setup: Info Flow of I(S,M)}) and (\ref{eq:setup: Info Flow of I(X)}), the partial entropy production rates $\dot{\sigma}^{(\mathcal{S}, \mathcal{M})}$ and $\dot{\sigma}^\mathcal{X}$ can be expressed as 
\begin{align}
  \dot{\sigma}^{(\mathcal{S}, \mathcal{M})} &= \dot{S}^{(\mathcal{S}, \mathcal{M})}_{\mathrm{tot}} - \dot{I}^{\mathcal{X} \to (\mathcal{S}, \mathcal{M})} \geq 0 ,
  \label{eq:setup: the second law of (S,M)} \\
  \dot{\sigma}^{\mathcal{X}} &= \dot{S}^{\mathcal{X}}_{\mathrm{tot}} - \dot{I}^{(\mathcal{S}, \mathcal{M}) \to \mathcal{X}} \geq 0.
  \label{eq:setup: the second law of X}
\end{align}
These inequalities can be identified as the second law for each subsystem, which are also called the second law of information thermodynamics~\cite{horowitz2014thermodynamics,tanogami2023universal,leighton2023inferring}. 
Interestingly, the subsystem entropy production rate $\dot{S}^{(\mathcal{S}, \mathcal{M})}_{\mathrm{tot}}$ ($\dot{S}^{\mathcal{X}}_{\mathrm{tot}}$) can be negative if $\dot{I}^{\mathcal{X} \to (S,M)}$ ($\dot{I}^{(\mathcal{S}, \mathcal{M}) \to X}$) is negative.
This apparent violation of the second law of thermodynamics caused by the information flow lies at the heart of the mechanism of Maxwell's demon~\cite{parrondo2015thermodynamics}.

\subsection{Various definitions of efficiency\label{Various definitions of efficiency}}
Here, we introduce various thermodynamic efficiencies based on the thermodynamic quantities defined in Sec.~\ref{subsec: Thermodynamic properties} and \ref{subsec: Information-thermodynamic properties}.
At this point, we note that the master equation~(\ref{eq:setup: Entire Master eq.}) does not generally admit a steady-state distribution $p_{\mathrm{ss}}(s,m,x)$ because the kinesin--cargo complex exhibits unbounded biased diffusion.
To focus on universal aspects independent of the details of initial conditions or transient processes, we define these efficiencies in the long-time limit $\mathcal{T}\rightarrow\infty$ by assuming that $d_{\mathcal{T}}S[\mathcal{S}_{\mathcal{T}},\mathcal{M}_{\mathcal{T}},\mathcal{X}_{\mathcal{T}}]=d_{\mathcal{T}}S[\mathcal{S}_{\mathcal{T}},\mathcal{M}_{\mathcal{T}}]=d_{\mathcal{T}}S[\mathcal{X}_{\mathcal{T}}]=0$ in this limit.

\subsubsection{Thermodynamic efficiency}
We first define the standard thermodynamic efficiency based on the second law for the total system Eq.~(\ref{eq:setup: the second law of thermodynamics}).
By taking the long-time limit, the second law can be expressed as
\begin{align}
  \dot{S}^{(\mathcal{S}, \mathcal{M})}_{\mathrm{env}} + \dot{S}^\mathcal{X}_{\mathrm{env}} = - \frac{\dot{Q}^{(\mathcal{S}, \mathcal{M})} + \dot{Q}^\mathcal{X}}{k_{\mathrm{B}} T} \geq 0.
\end{align}
By applying the first law of thermodynamics [Eq.~\eqref{eq:setup: the first law of thermodynamics (entire system)}] with $d_{\mathcal{T}}\langle U\rangle=0$ in the limit $\mathcal{T}\rightarrow\infty$, this inequality can also be expressed in terms of the power $\dot{W}^{(\mathcal{S}, \mathcal{M})}$ and $\dot{W}^\mathcal{X}$ as
\begin{align}
    \dot{W}^{(\mathcal{S}, \mathcal{M})} + \dot{W}^\mathcal{X} \geq 0,
\end{align}
which can be regarded as another expression of the second law of thermodynamics.

Let us consider the case where the kinesin--cargo complex works as a steady-state thermodynamic engine.
In this case, the kinesin pulls the cargo against the external load, transducing the input power supplied by ATP hydrolysis, $\dot{W}^{(\mathcal{S}, \mathcal{M})}>0$, into the output power $\dot{W}^\mathcal{X} < 0$.
For such work transducers, we can define the following thermodynamic efficiency as the ratio of output to input power~\cite{leighton2025flow},
\begin{align}
    \eta := \frac{-\dot{W}^\mathcal{X}}{\dot{W}^{(\mathcal{S}, \mathcal{M})}}, \label{eq:setup: def. thermodynamics efficiency}
\end{align}
which satisfies $0\le\eta\le1$.

\subsubsection{Information-thermodynamic efficiency}
From the second law of information thermodynamics, we can define thermodynamic efficiencies at the subsystem level.
By taking the long-time limit, the second law of information thermodynamics [Eqs.~(\ref{eq:setup: the second law of (S,M)}) and (\ref{eq:setup: the second law of X})] can be expressed as 
\begin{align}
  -\dot{Q}^{(\mathcal{S}, \mathcal{M})} - k_{\mathrm{B}} T\dot{I}^{\mathcal{X} \to (\mathcal{S}, \mathcal{M})} \geq 0,
  \label{eq:setup: (S,M) Steady state the second law of info.} \\
  -\dot{Q}^{\mathcal{X}} - k_{\mathrm{B}} T\dot{I}^{(\mathcal{S}, \mathcal{M}) \to \mathcal{X}} \geq 0.
  \label{eq:setup: X Steady state the second law of info.}
\end{align}
By using the first law for the subsystem [Eqs.~\eqref{eq:setup: the first law of the kinesin} and \eqref{eq:setup: the first law of the cargo}], Eqs.~\eqref{eq:setup: (S,M) Steady state the second law of info.} and \eqref{eq:setup: X Steady state the second law of info.} can be expressed in terms of the power as
\begin{align}
  \dot{W}^{(\mathcal{S}, \mathcal{M})} - \dot{W}^{(\mathcal{S}, \mathcal{M}) \to \mathcal{X}} - k_{\mathrm{B}} T \dot{I}^{\mathcal{X} \to (\mathcal{S}, \mathcal{M})} \geq 0,
  \label{eq:setup: (S,M) the second law of info in terms of work}\\
  \dot{W}^\mathcal{X} - \dot{W}^{\mathcal{X} \to (\mathcal{S}, \mathcal{M})} - k_{\mathrm{B}} T \dot{I}^{(\mathcal{S}, \mathcal{M}) \to \mathcal{X}}\geq 0.
  \label{eq:setup: X the second law of info in terms of work}
\end{align}
Here, we note that $\dot{W}^{(\mathcal{S}, \mathcal{M}) \to \mathcal{X}} + k_{\mathrm{B}} T \dot{I}^{\mathcal{X} \to (\mathcal{S}, \mathcal{M})}=-\dot{W}^{\mathcal{X} \to (\mathcal{S}, \mathcal{M})} - k_{\mathrm{B}} T \dot{I}^{(\mathcal{S}, \mathcal{M}) \to \mathcal{X}}$ from the steady-state condition.

Now, suppose that the kinesin--cargo complex works as a steady-state thermodynamic engine, i.e., $\dot{W}^{\mathcal{X}} < 0$ and $\dot{W}^{(\mathcal{S}, \mathcal{M})} > 0$.
In this case, Eqs.~(\ref{eq:setup: (S,M) the second law of info in terms of work}) and (\ref{eq:setup: X the second law of info in terms of work}) can be simply expressed as
\begin{align}
\dot{W}^{(\mathcal{S}, \mathcal{M})} \ge \dot{W}^{(\mathcal{S}, \mathcal{M}) \to \mathcal{X}} + k_{\mathrm{B}} T \dot{I}^{\mathcal{X} \to (\mathcal{S}, \mathcal{M})} \ge -\dot{W}^\mathcal{X}>0.
\end{align}
Here, the sum of the energy and information flows, $\dot{W}^{(\mathcal{S}, \mathcal{M}) \to \mathcal{X}} + k_{\mathrm{B}} T \dot{I}^{\mathcal{X} \to (\mathcal{S}, \mathcal{M})}$, is called the transduced capacity because it acts as a bottleneck between the input and output power~\cite{leighton2023inferring,ehrich2023energy,leighton2025flow}.
In analogy with conventional thermodynamics, the transduced capacity can be interpreted as the total free energy available to the system, including the contribution from information flow.
Then, we can define the following information-thermodynamic efficiencies, which quantify the efficiency of free-energy transduction for each subsystem~\cite{ehrich2023energy,tanogami2023universal,leighton2023inferring}:
\begin{align}
  \eta^{(\mathcal{S}, \mathcal{M})}_{\mathrm{info}} &:= \frac{\dot{W}^{(\mathcal{S}, \mathcal{M}) \to \mathcal{X}} +  k_{\mathrm{B}} T \dot{I}^{\mathcal{X} \to (\mathcal{S}, \mathcal{M})}}{\dot{W}^{(\mathcal{S}, \mathcal{M})}},
  \label{eq:setup: Definition of Info. Efficiency} \\
  \eta^{\mathcal{X}}_{\mathrm{info}} &:= \frac{-\dot{W}^{\mathcal{X}}}{\dot{W}^{(\mathcal{S}, \mathcal{M}) \to \mathcal{X}} +  k_{\mathrm{B}} T \dot{I}^{\mathcal{X} \to (\mathcal{S}, \mathcal{M})}}, \label{eq:setup: Def of Info Eff of X}
\end{align}
which satisfy $0 \leq \eta^{(\mathcal{S}, \mathcal{M})}_{\mathrm{info}} \leq 1$ and $0 \leq \eta^{\mathcal{X}}_{\mathrm{info}} \leq 1$. 
Note that their product equals the thermodynamic efficiency of the whole system, $\eta^{(\mathcal{S}, \mathcal{M})}_{\mathrm{info}} \eta^{\mathcal{X}}_{\mathrm{info}} = \eta$.

\subsubsection{TUR efficiency}
The thermodynamic uncertainty relation (TUR) provides a tighter lower bound for entropy production than the conventional second law of thermodynamics~\cite{barato2015thermodynamic,gingrich2016dissipation,horowitz2020thermodynamic,shiraishi2021optimal}.
In its standard form, the TUR states that suppressing the relative fluctuations of an arbitrary time-integrated current $\hat{\mathcal{J}}$ necessarily involves a thermodynamic cost.
In this paper, we focus on the current associated with the transitions of kinesin, expressed in the following form:
\begin{align}
    \hat{\mathcal{J}}:= \sum_{(s, m)} \sum_{\substack{(s', m') \\ (\neq (s, m))}} \hat{n}^{ss'}_{mm'} d^{ss'}_{mm'}.
  \label{eq:setup: def of current in bipartite}
\end{align}
Here, $\hat{n}^{ss'}_{mm'}$ denotes the number of transitions from state $(s', m')$ to state $(s, m)$ during the time interval $[0, \mathcal{T}]$, and $d^{ss'}_{mm'}$ is an arbitrary antisymmetric weight that satisfies $d^{ss'}_{mm'}=-d^{s's}_{m'm}$. 
For example, the choice of $d^{ss'}_{mm'}=d_{mm'}$ with
\begin{align}
d_{mm'}:=
\begin{cases}
\delta\hspace{5pt} &\text{if} \hspace{5pt} m' = m - \delta \\
-\delta\hspace{5pt} &\text{if} \hspace{5pt} m' = m + \delta \\
\end{cases}
\end{align}
yields the displacement of the kinesin during the time interval $[0,\mathcal{T}]$, which can be expressed as
\begin{align}
  \Delta \hat{m} &:= \sum_{(s, m)} \sum_{\substack{(s', m') \\ (\neq (s, m))}} \hat{n}^{ss'}_{mm'} d_{mm'}\notag\\
  &= \sum_{m} (\hat{n}_{m + \delta, m} - \hat{n}_{m - \delta, m}) \delta,
    \label{eq:def of M hat}
\end{align}
where $\hat{n}_{m,m'}$ denotes the number of transitions from state $m'$ to state $m$ during the time interval.
Then, in the long-time limit, the TUR can be expressed as
\begin{align}
\dfrac{D_{\mathcal{J}}}{J^2_\mathcal{J}}\ge\dfrac{1}{\dot{\sigma}},
\end{align}
where $D_\mathcal{J} := \lim_{\mathcal{T} \to \infty}\mathrm{Var}[\hat{\mathcal{J}}] / 2 \mathcal{T}$ denotes the fluctuation of $\hat{\mathcal{J}}$, and $J_\mathcal{J} := \lim_{\mathcal{T} \to \infty} \langle \hat{\mathcal{J}} \rangle / \mathcal{T}$ denotes the mean current.
From this expression, we can define another type of efficiency for the whole system,
\begin{align}
    \eta_{\mathrm{TUR}} := \frac{J^2_\mathcal{J}}{D_\mathcal{J}} \frac{1}{\dot{\sigma}},
    \label{eq:setup: Definition of TUR Efficiency}
\end{align}
which satisfies $0\le \eta_{\mathrm{TUR}}\le 1$.
While $\eta_{\mathrm{TUR}}$ is sometimes referred to as transport efficiency~\cite{dechant2018current, hwang2018energetic}, we shall refer to it as TUR efficiency because the choice of the current $\hat{\mathcal{J}}$ is not limited to the displacement.

For a bipartite system, TUR can be extended to a subsystem~\cite{tanogami2023universal}.
Specifically, for any current $\hat{\mathcal{J}}$ of the form given by Eq.~(\ref{eq:setup: def of current in bipartite}), the following \textit{bipartite} TUR holds in the long-time limit~\cite{tanogami2023universal}:
\begin{align}
  \frac{D_\mathcal{J}}{J^2_\mathcal{J}}\ge\frac{(1 + \delta_\mathcal{J})^2}{\dot{\sigma}^{(\mathcal{S}, \mathcal{M})}},
\end{align}
where the additional term $\delta_{\mathcal{J}}$ denotes the correction arising from the interaction between the kinesin and the cargo~\cite{tanogami2023universal}. 
In analogy with the standard TUR, we can define the bipartite TUR efficiency for the subsystem $(\mathcal{S}, \mathcal{M})$:
\begin{align}
  \eta_{\mathrm{BTUR}} := \frac{J^2_\mathcal{J}}{D_\mathcal{J}} \frac{(1 + \delta_\mathcal{J})^2}{\dot{\sigma}^{(\mathcal{S}, \mathcal{M})}}.
  \label{eq:setup: bipartite TUR Efficiency}
\end{align}
When the cargo $\mathcal{X}$ relaxes sufficiently fast compared to the kinesin $(\mathcal{S}, \mathcal{M})$, the additional term $\delta_{\mathcal{J}}$ vanishes~\cite{tanogami2023universal}. 
Under this time-scale separation, the bipartite TUR efficiency is given by
\begin{align}
  \eta_{\mathrm{BTUR}} = \frac{J^2_\mathcal{J}}{D_\mathcal{J}} \frac{1}{\dot{\sigma}^{(\mathcal{S}, \mathcal{M})}}.
  \label{eq:setup: bipartite TUR Efficiency under time-scale separation}
\end{align}
In this case, because $\dot{\sigma}^{(\mathcal{S},\mathcal{M})}\le \dot{\sigma}$, it follows that $\eta_{\mathrm{BTUR}}$ is larger than the conventional TUR efficiency, i.e., $\eta_{\mathrm{BTUR}} \ge \eta_{\mathrm{TUR}}$.
Below, we use Eq.~(\ref{eq:setup: bipartite TUR Efficiency}) for the numerical calculations, evaluating $\delta_\mathcal{J}$ explicitly, and Eq.~(\ref{eq:setup: bipartite TUR Efficiency under time-scale separation}) for the analytical calculations.

\section{Perturbation expansion\label{sec:Perturbation expansion}}
When a separation of time scales exists between the kinesin and the cargo, the various efficiencies defined in the previous section can be evaluated analytically in a unified manner through a perturbative expansion. 
As shown in Sec.~\ref{sec:Results on various efficiencies}, such a separation of time scales is considered valid, at least in the typical parameter regimes corresponding to the \textit{in vitro} condition. 
The analytical results obtained through the perturbative expansion remain useful even in situations where time-scale separation is not strictly satisfied, because they provide a clear benchmark for comparison. 
In this section, we first derive the effective slow dynamics of kinesin using a perturbative expansion (Sec.~\ref{subsec:Time scale separation}) and then outline the parameter estimation procedure based on the slow dynamics (Sec.~\ref{subsec:Parameter estimation}).

\subsection{Time scale separation\label{subsec:Time scale separation}}
We first compare the characteristic time scales of the kinesin and the cargo.
For the cargo, the relevant time scale is the relaxation time $\tau^{\mathcal{X}} := \gamma / \kappa$.
The characteristic time scale for the kinesin is determined by the transition rate matrix $W^{ss'}_{mm'}(x)$ and depends on the cargo position $x$.
For simplicity, we use the bare model parameters $\tau_f, \tau_b, k_c^{-1}, (k_c^\dag)^{-1}$ introduced in Sec.~\ref{sec:model} rather than the $x$-dependent transition rates, to define a representative time scale for the kinesin dynamics. 
We thus set $\tau^{(\mathcal{S},\mathcal{M})} := \min \{\tau_f, \tau_b, k_c^{-1}, (k_c^\dag)^{-1}\}$.
In the typical experimental setup~\cite{ariga2018nonequilibrium}, the ratio of these time scales is estimated as $\epsilon :=\tau^{\mathcal{X}}/\tau^{(\mathcal{S},\mathcal{M})}\sim10^{-2}$, implying that the cargo relaxes much faster than the kinesin. 
(The parameter values used in this estimate are specified in Sec.~\ref{sec:Results on various efficiencies} and Appendix~\ref{sec:appendix:parameter estimation}.)
Hence, below we consider the idealized limit where these time scales are well separated.

Several remarks are in order regarding the above estimation. 
First, the actual characteristic time scale of kinesin depends on the external force $F$ through the transition rates; thus, this estimate may not remain valid depending on the magnitude of the force. 
Indeed, as discussed in Sec.~\ref{sec:Results on various efficiencies}, the assumption of time-scale separation is strictly violated away from the stall force, where the kinesin velocity vanishes. 
Second, the friction coefficient $\gamma$ estimated under the \textit{in vivo} condition is significantly larger than that measured \textit{in vitro}, suggesting that time-scale separation is likely to break down in living cells. 
Specifically, assuming that only the friction coefficient varies between \textit{in vivo} and \textit{in vitro} conditions while the spring constant $\kappa$ and the time scale associated with the kinesin $\tau^{(\mathcal{S},\mathcal{M})}$ remain unchanged, we find that $\epsilon \sim 10$.
(The parameter values used in this estimate are specified in Sec.~\ref{sec:Results on various efficiencies} and Appendix~\ref{sec:appendix:parameter estimation}.)
We numerically estimate the various efficiencies, including those under such an \textit{in vivo}-like condition, in Sec.~\ref{sec:Results on various efficiencies}.

By assuming a separation of time scales, we now carry out a perturbative expansion in terms of the small parameter $\epsilon$.
We first rewrite the master equation (\ref{eq:setup: Entire Master eq.}) in a dimensionless form by introducing a dimensionless slow time $\tau := t / \tau^{(\mathcal{S},\mathcal{M})}$:
\begin{align}
  \partial_\tau p_\tau (s, m, x) &= \sum_{s'} \sum_{m'} \widetilde{W}^{ss'}_{mm'}(x) p_\tau(s', m', x) \notag \\
  & \hspace{10pt} -  \frac{1}{\epsilon} \partial_x \widetilde{J}_\tau^{\mathcal{X}} (s, m, x),
  \label{eq:setup: dimensionless master equation}
\end{align}
where the dimensionless transition matrix and probability current are defined as
\begin{align}
    \widetilde{W}^{ss'}_{mm'}(x) &:= \tau^{(\mathcal{S},\mathcal{M})} W^{ss'}_{mm'}(x), \\
    \widetilde{J}^{\mathcal{X}}_{\tau} (s, m, x) &:= \tau^{\mathcal{X}} J^{\mathcal{X}}_{\tau} (s, m, x)
\end{align}
with the dimensionless transition rates $\widetilde{k}_\alpha:= \tau^{(\mathcal{S},\mathcal{M})} k_\alpha$ and $\widetilde{k}^\dag_\alpha:= \tau^{(\mathcal{S},\mathcal{M})} k^\dag_\alpha$ ($\alpha\in\{f,b,c\}$).

We now assume that $p_\tau (s, m, x)$ has an asymptotic expansion in terms of the asymptotic sequences $\{\epsilon^n \}_{n = 0}^{\infty}$ for $\epsilon \to 0$:
\begin{align}
  p_\tau (s, m, x) = p_\tau^{(0)}(s, m, x) + \epsilon p_\tau^{(1)}(s, m, x) + \cdots.
  \label{eq:setup: Perturbation expansion of p}
\end{align}
Here, we impose the normalization condition,
\begin{align}
    \int_{-\infty}^{\infty} dx p_\tau^{(0)}(s, m, x) = \int_{-\infty}^{\infty} dx p_\tau (s, m, x) = p_\tau(s, m).
\end{align} 
Substituting the expansion \eqref{eq:setup: Perturbation expansion of p} into the master equation \eqref{eq:setup: dimensionless master equation}, we find that the leading order $O(\epsilon^{-1})$ yields
\begin{align}
  \mathcal{L} [p_\tau^{(0)}] (s, m, x)=0,
\end{align}
where $\mathcal{L}$ is the Fokker--Planck operator~\cite{risken1996fokker},
\begin{align}
  &\mathcal{L} [p_\tau^{(0)}] (s, m, x)\notag\\
  &\quad:=- \partial_x \bigg[ \bigg( - (x - m) + \frac{F}{\kappa}\bigg) - \frac{k_{\mathrm{B}} T}{\kappa} \partial_x  \bigg] p_\tau^{(0)}(s, m, x).
  \label{eq:setup: FP operator}
\end{align}
Let $\pi_{\mathrm{ss}}(x | m)$ be the normalized right eigenfunction of $\mathcal{L}$ corresponding to the zero eigenvalue. 
From the definition of the Fokker--Planck operator~(\ref{eq:setup: FP operator}), $\pi_{\mathrm{ss}}(x | m)$ can be explicitly calculated as
\begin{align}
  \pi_{\mathrm{ss}} (x | m) = 
  \sqrt{\frac{\kappa}{2 \pi k_{\mathrm{B}} T}} \exp \bigg( - \frac{\kappa}{2 k_{\mathrm{B}} T} \bigg( x - m - \frac{F}{\kappa} \bigg)^2 \bigg).
  \label{eq:setup:(pi_ss) steady-state probability density of cargo}
\end{align}
Note that the eigenfunction is independent of the internal state $s$ because $\mathcal{L}$ does not depend on $s$.
Because the drift term of $\mathcal{L}$ has no singularities, this normalized zero eigenfunction is unique for each $m$~\cite{risken1996fokker}.
Hence, from the normalization condition, we conclude that $p_\tau^{(0)}(s, m, x)$ can be expressed as
\begin{align}
  p_\tau ^{(0)} (s, m, x) = p_\tau(s, m) \pi_{\mathrm{ss}} (x | m). 
  \label{eq:setup:decomposition of the prob distribution}
\end{align}

The subleading order $O(\epsilon^0)$ yields
\begin{align}
  \partial_\tau p^{(0)}_\tau (s, m,x) &= \sum_{s'} \sum_{m'} \widetilde{W}^{ss'}_{m m'}(x) p_\tau^{(0)}(s', m',x) \notag \\
  &\hspace{15pt} + \mathcal{L}[p_\tau^{(1)}](s, m,x).
  \label{eq:setup: subleading order eq}
\end{align}
Note that Eq.~(\ref{eq:setup: subleading order eq}) is linear with respect to $p_\tau^{(1)}(s,m,x)$ and that $\mathcal{L}$ has the left zero eigenfunction $1$ because $\mathcal{L}^\dag[1]=0$, where $\mathcal{L}^\dag$ denotes the adjoint of $\mathcal{L}$~\cite{touchette2018introduction}:
\begin{align}
    \mathcal{L}^\dag := \bigg[ - (x - m) + \frac{F}{\kappa} \bigg]\partial_x + \frac{k_{\mathrm{B}} T}{\kappa} \partial_x^2.
\end{align}
These properties guarantee that a solution $p_\tau^{(1)}(s,m,x)$ exists only under the solvability condition:
\begin{align}
  &\int_{-\infty}^{\infty}dx 1 \cdot \bigg[ \partial_\tau p_\tau^{(0)}(s, m, x) - \sum_{s'} \sum_{m'} \widetilde{W}^{ss'}_{mm'}(x) p_\tau^{(0)}(s', m',x) \bigg] \notag \\
  &\hspace{15pt} = \int_{-\infty}^{\infty} dx \mathcal{L}^\dag [1] \cdot p_\tau^{(1)}(s, m, x) = 0.
  \label{eq:setup: adjoint FP eq}
\end{align}
By substituting Eq.~\eqref{eq:setup:decomposition of the prob distribution} into Eq.~\eqref{eq:setup: adjoint FP eq}, we arrive at the effective master equation that governs the slow dynamics of the kinesin:
\begin{align}
  \partial_\tau p_\tau (s, m) = \sum_{s'} \sum_{m'} \overline{\widetilde{W}}^{ss'}_{mm'} p_\tau (s', m').
  \label{eq:setup: slow dynamics of kinesin}
\end{align}
Here, $\overline{\widetilde{W}}^{ss'}_{mm'}$ is the effective transition matrix defined as 
\begin{align}
  \overline{\widetilde{W}}^{ss'}_{mm'} &:= \int_{-\infty}^{\infty}dx \widetilde{W}^{ss'}_{mm'}(x) \pi_{\mathrm{ss}}(x | m')\notag\\
        &=
        \begin{cases}
            \overline{\widetilde{k}}_f
            &\text{for } \ (2,m-\delta) \to (1,m), \\[3pt]
            \overline{\widetilde{k}}_b
            &\text{for } \ (2,m+\delta) \to (1,m), \\[3pt]
            \overline{\widetilde{k}}_c
            &\text{for } \ (1,m) \to (2,m), \\[3pt]
            \overline{\widetilde{k}}_f^\dag
            &\text{for } \ (1,m+\delta) \to (2,m), \\[3pt]
            \overline{\widetilde{k}}_b^\dag
            &\text{for } \ (1,m-\delta) \to (2,m), \\[3pt]
            \overline{\widetilde{k}}_c^\dag
            &\text{for } \ (2,m) \to (1,m),
        \end{cases}
  \label{def: effective transition matrix}
\end{align}
for $(s,m)\neq(s',m')$, where we defined the effective transition rates as
\begin{align}
  \overline{\widetilde{k}}_f &= \int_{-\infty}^{\infty} dx \widetilde{k}_f(m, x) \pi_{\mathrm{ss}}(x|m) \notag \\
  &= \frac{\tau^{(\mathcal{S},\mathcal{M})}}{\tau_f} \exp \bigg( \frac{\theta_f}{k_{\mathrm{B}} T} \bigg[ (\theta_f - 1) \frac{ \kappa}{2} \delta^2 +  \Delta \mu_{\mathrm{mech}}+ F \delta \bigg] \bigg), \label{eq:effective rates kf}\\
  \overline{\widetilde{k}}_f^\dag &= \int_{-\infty}^{\infty} dx \widetilde{k}_f^\dag(m, x) \pi_{\mathrm{ss}}(x|m) \notag \\
  &= \frac{\tau^{(\mathcal{S},\mathcal{M})}}{\tau_f} \exp \bigg( \frac{\theta_f - 1}{k_{\mathrm{B}} T} \bigg[ \theta_f \frac{\kappa}{2} \delta^2 + \Delta \mu_{\mathrm{mech}}+ F \delta \bigg] \bigg), \label{eq:effective rates kf_dag}\\
  \overline{\widetilde{k}}_b &= \int_{-\infty}^{\infty} dx \widetilde{k}_b(m, x) \pi_{\mathrm{ss}}(x|m) \notag \\
  &= \frac{\tau^{(\mathcal{S},\mathcal{M})}}{\tau_b} \exp \bigg( \frac{\theta_b}{k_{\mathrm{B}} T}\bigg[ (\theta_b - 1) \frac{\kappa}{2} \delta^2   +  \Delta \mu_{\mathrm{mech}}  - F \delta \bigg] \bigg), \label{eq:effective rates kb}\\
  \overline{\widetilde{k}}_b^\dag &= \int_{-\infty}^{\infty} dx \widetilde{k}_b^\dag(m, x) \pi_{\mathrm{ss}}(x|m) \notag \\
  &= \frac{\tau^{(\mathcal{S},\mathcal{M})}}{\tau_b} \exp \bigg[ \frac{\theta_b - 1}{k_{\mathrm{B}} T} \bigg( \theta_b \frac{\kappa}{2} \delta^2  + \Delta \mu_{\mathrm{mech}}- F \delta \bigg)\bigg] \label{eq:effective rates kb_dag}.
\end{align}
Note that these transition rates are constants independent of $(s,m,x)$. 
Although $\widetilde{k}_c$ and $\widetilde{k}_c^\dag$ are unchanged by the averaging procedure, we use the notation $\overline{\widetilde{k}}_c = \widetilde{k}_c$ and $\overline{\widetilde{k}}_c^\dag = \widetilde{k}_c^\dag$ for notational consistency.

Summing over $m$ in Eq.~(\ref{eq:setup: slow dynamics of kinesin}) yields the effective dynamics for the internal state transition:
\begin{align}
  \partial_\tau p_\tau (s) = \sum_{s'} \overline{\widetilde{W}}^{ss'} p_\tau (s')
  \label{eq:setup: slow dynamics for internal state transition}
\end{align}
with the effective transition matrix defined as
\begin{align}
  \overline{\widetilde{W}}^{ss'} &:= \sum_m\overline{\widetilde{W}}^{ss'}_{mm'} \notag\\
        &=
        \begin{cases}
            \overline{\widetilde{k}}_f + \overline{\widetilde{k}}_c^\dag + \overline{\widetilde{k}}_b \hspace{5pt} &\text{for} \hspace{5pt} (s,s')=(1,2), \\
            \overline{\widetilde{k}}_f^\dag + \overline{\widetilde{k}}_c + \overline{\widetilde{k}}_b^\dag \hspace{5pt} &\text{for} \hspace{5pt} (s,s')=(2,1),
        \end{cases}
\end{align}
for $s\neq s'$.
Because $\overline{\widetilde{W}}^{ss'}$ is a finite irreducible transition-rate matrix, there exists a unique steady-state distribution $p_{\mathrm{ss}}(s)$, which is given by (see, e.g., Ref.~\cite{schnakenberg1976network})
\begin{align}
  p_{\mathrm{ss}}(1) &= \frac{\overline{\widetilde{k}}_f  + \overline{\widetilde{k}}_c^\dag + \overline{\widetilde{k}}_b }{\overline{\widetilde{k}}_f + \overline{\widetilde{k}}_c^\dag + \overline{\widetilde{k}}_b + \overline{\widetilde{k}}_f^\dag + \overline{\widetilde{k}}_c + \overline{\widetilde{k}}_b^\dag}, \label{p_ss1} \\
  p_{\mathrm{ss}}(2) &= \frac{\overline{\widetilde{k}}_f^\dag  + \overline{\widetilde{k}}_c + \overline{\widetilde{k}}_b^\dag }{ \overline{\widetilde{k}}_f + \overline{\widetilde{k}}_c^\dag + \overline{\widetilde{k}}_b + \overline{\widetilde{k}}_f^\dag + \overline{\widetilde{k}}_c + \overline{\widetilde{k}}_b^\dag }.\label{p_ss2}
\end{align}

\subsection{Parameter estimation\label{subsec:Parameter estimation}}
Under the assumption of a time-scale separation, we can explicitly calculate the steady-state average velocity of the kinesin--cargo complex.
Indeed, the average of the displacement of the kinesin, defined in Eq.~(\ref{eq:def of M hat}), is given by (see, e.g., Ref.~\cite{peliti2021stochastic})
\begin{align}
  \langle \Delta \hat{m} \rangle&= \int_{0}^{\mathcal{T}} d\tau \sum_{m} \bigg( \overline{\widetilde{k}}_f p_\tau (2, m) +  \overline{\widetilde{k}}_b^\dag p_\tau(1, m) \notag \\
  & \hspace{100pt} -  \overline{\widetilde{k}}_f^\dag p_\tau(1, m) -  \overline{\widetilde{k}}_b p_\tau(2, m) \bigg) \notag \\
  &= \mathcal{T} \bigg[ \left( \overline{\widetilde{k}}_f -  \overline{\widetilde{k}}_b\right) p_{\mathrm{ss}} (2) + \left( \overline{\widetilde{k}}_b ^\dag -  \overline{\widetilde{k}}_f^\dag\right) p_{\mathrm{ss}} (1) \bigg] \delta.
\end{align}
Here, we note that $\mathcal{T}$ is an arbitrary dimensionless time in units of $\tau^{(\mathcal{S},\mathcal{M})}$.
Then, the average velocity of the kinesin in the steady state is given by
\begin{align}
  v &:= \frac{1}{\tau^{(\mathcal{S},\mathcal{M})}}\lim_{\mathcal{T}\rightarrow\infty}\frac{\langle \Delta \hat{m} \rangle}{\mathcal{T}} \notag \\
  &= \dfrac{1}{\tau^{(\mathcal{S},\mathcal{M})}}\bigg[ \left( \overline{\widetilde{k}}_f -  \overline{\widetilde{k}}_b\right) p_{\mathrm{ss}} (2) + \left( \overline{\widetilde{k}}_b ^\dag -  \overline{\widetilde{k}}_f^\dag\right) p_{\mathrm{ss}} (1) \bigg] \delta. 
  \label{eq:setup:steady-state velocity}
\end{align}
Note that this velocity can also be regarded as the average velocity of the cargo.

From the force--velocity relationship, we can determine the six parameters $\tau_f, \tau_b, \theta_f, \theta_b, k_c, \Delta \mu_{\mathrm{chem}}$ by fitting the theoretical prediction (\ref{eq:setup:steady-state velocity}) to the experimental data in Ref.~\cite{ariga2018nonequilibrium}.
For the details of the fitting procedure and the specific values of the estimated parameters, see Appendix~\ref{sec:appendix:parameter estimation}.

\begin{figure}[t]
    \centering
    \includegraphics[width=8.5cm]{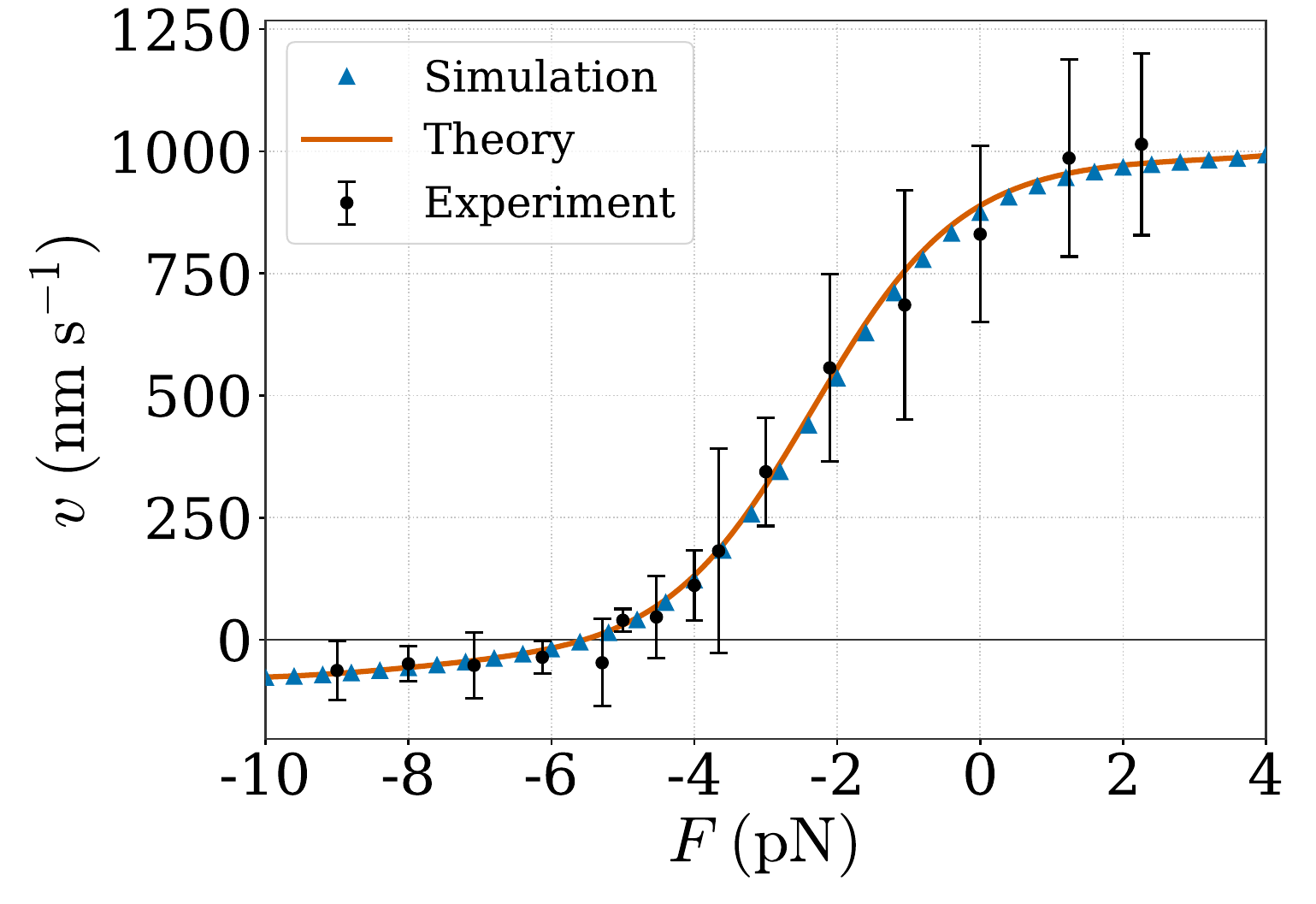}
    \caption{Force--velocity relationship of single-molecule kinesin motion under the \textit{in vitro} high-ATP condition.
    The experimental data are cited from Ref.~\cite{ariga2018nonequilibrium}.
    For the simulation, we use the parameter set given in Appendix~\ref{sec:appendix:parameter estimation}.
    }
    \label{fig:velocity vs Force}
\end{figure}

Figure~\ref{fig:velocity vs Force} shows the force--velocity relation obtained from Eq.~(\ref{eq:setup:steady-state velocity}) using the fitted high-ATP parameter set, together with the experimental results.
The qualitative behavior of the force--velocity relation is determined by the competition between the force-dependent forward and backward transitions. 
As the external force becomes more negative, the forward motion of kinesin is increasingly hindered and the mean velocity decreases. 
We define the stall force $F_{\mathrm{stall}}$ as the external force at which the steady-state velocity vanishes. 
For the fitted high-ATP parameter set, Eq.~(\ref{eq:setup:steady-state velocity}) gives $F_{\mathrm{stall}}\simeq-5.5\,\mathrm{pN}$. 
Thus, kinesin moves forward for $F>F_{\mathrm{stall}}$ and backward for $F<F_{\mathrm{stall}}$.

To examine whether the analytical expression accurately represents the full kinesin–cargo dynamics, we also perform numerical simulations using the parameters obtained from the fit.
The Fokker–Planck dynamics of the cargo is discretized as a continuous-time Markov jump process, which is coupled to the kinesin transitions and simulated using the Gillespie algorithm~\cite{gillespie1977exact}.
Details of the simulation procedures are provided in Appendix~\ref{sec:appendix:simulation detail}.
As shown in Fig.~\ref{fig:velocity vs Force}, the numerical results agree well with both the analytical prediction and the experimental data, confirming that the analytical force–velocity relation provides an accurate description of the model.


\section{Results on various efficiencies\label{sec:Results on various efficiencies}}
In this section, we evaluate the efficiencies introduced above using the perturbation expansion developed in the previous section.
As noted earlier, the perturbative approach based on time-scale separation may fail to fully capture the system's behavior when the external force \(F\) significantly deviates from the stall force or when the friction coefficient \(\gamma\) is large, as expected in intracellular environments.
We therefore compare the analytical predictions with numerical simulations, particularly in regimes where the time-scale separation approximation is expected to become less accurate.

In both the theoretical analysis and numerical simulations, the parameters \(\tau_f, \tau_b, \theta_f, \theta_b, k_c\), and \(\Delta \mu_{\mathrm{chem}}\) were set to the values estimated in the previous section; the specific values are given in Appendix~\ref{sec:appendix:parameter estimation}.
Following the experimental results in Ref.~\cite{ariga2018nonequilibrium}, the total free energy of ATP hydrolysis, the spring constant, and the friction coefficient were set to $\Delta \mu = 85\,\mathrm{pN\,nm} \simeq 21\,k_{\mathrm{B}}T $ at room temperature $T=25 \, ^\circ \mathrm{C}$, $\kappa = 0.075\,\mathrm{pN/nm}$, and $ \gamma=\gamma_{\mathrm{vitro}}:=3.09 \times 10^{-5}\, \mathrm{pN \ s / nm} $, respectively. 
Although $\kappa$ may depend on the external force $F$ in general, we neglect this possible dependence and use the above constant value for simplicity.
In this case, the time-scale ratio is estimated to be $\epsilon\sim10^{-2}$, ensuring the validity of the perturbative approach, at least in the vicinity of the stall force.
For comparison, we also performed calculations mimicking an intracellular environment by adopting a larger friction coefficient, $\gamma=\gamma_{\mathrm{vivo}}:= 1.0 \times 10^{-2}\,\mathrm{pN \ s / nm} $, which was estimated from the data in Ref.~\cite{shtridelman2008force} via the Einstein--Stokes relation, while keeping all other parameters unchanged. 
Under this condition, the time-scale ratio reaches the order of $\epsilon \sim 10$, indicating that the assumption of time-scale separation no longer holds.
We emphasize that, in the present model, the \textit{in vivo} condition is introduced only through the friction coefficient; actual intracellular environments may also alter parameters and noise statistics.
The parameter values used in the main text correspond to a high-ATP condition (\(1\,\mathrm{mM}\) ATP, \(0.1\,\mathrm{mM}\) ADP, and \(1\,\mathrm{mM}\,\mathrm{P}_\mathrm{i}\)), which reflects physiological ATP concentrations~\cite{ariga2018nonequilibrium}.
We also performed both analytical and numerical calculations using parameter values estimated under low-ATP conditions ($10\,\mu\mathrm{M}$ ATP, $1\,\mu\mathrm{M}$ ADP, $1\,\mathrm{mM}\,\mathrm{P}_\mathrm{i}$).
The results were not qualitatively altered; therefore, for brevity, we relegate the low-ATP results to Appendix~\ref{sec:appendix:results in Low ATP}.

\begin{table*}[t]
  \centering
  \setlength{\tabcolsep}{14pt}
  \caption{Comparison between the experimental, theoretical, and simulation values of the thermodynamic quantities at a constant external force of $F = -2\,\mathrm{pN}$ under the high-ATP \textit{in vitro} condition ($\gamma = \gamma_{\mathrm{vitro}}$).
  The experimental data are taken from Ref.~\cite{ariga2018nonequilibrium}.
  Here, $-F_{0}\overline{v}$, $J_{x}$, $\Delta\mu/\tau$, and $J_{\mathrm{all\ others}}$ in Ref.~\cite{ariga2018nonequilibrium} correspond to $-\dot{W}^{\mathcal{X}}$, $-\dot{Q}^{\mathcal{X}}$, $\dot{W}^{(\mathcal{S},\mathcal{M})}$, and $-\dot{Q}^{(\mathcal{S},\mathcal{M})}$, respectively.}
  \begin{ruledtabular}
  \begin{tabular}{l@{\,}llll}
  \multicolumn{2}{l}{Thermodynamic quantity}
    & Experiment & Simulation & Theory \\
  \hline
  $-\dot{W}^{\mathcal{X}}$
    & $(\mathrm{pN}\,\mathrm{nm}\,\mathrm{s}^{-1})$
    & $1150\pm120$ & $1077\pm7$ & $1082$ \\
  $-\dot{Q}^{\mathcal{X}}$
    & $(\mathrm{pN}\,\mathrm{nm}\,\mathrm{s}^{-1})$
    & $63.9\pm41.5$ & $10.3\pm16.7$ & $7.4$ \\
  $\dot{W}^{(\mathcal{S},\mathcal{M})}$
    & $(\mathrm{pN}\,\mathrm{nm}\,\mathrm{s}^{-1})$
    & $6160\pm560$ & $6061\pm34$ & $6081$ \\
  $-\dot{Q}^{(\mathcal{S},\mathcal{M})}$
    & $(\mathrm{pN}\,\mathrm{nm}\,\mathrm{s}^{-1})$
    & $4946\pm574$ & $4968\pm28$ & $4991$ \\
  \hline
  $\eta$
    &
    & $0.187\pm0.026$ & $0.178\pm0.002$ & $0.179$ \\
  \end{tabular}
  \end{ruledtabular}
  \label{tab:result: comparison of exp. and solution}
\end{table*}

\begin{figure*}[t]
    \centering
    \includegraphics[width=0.9\textwidth]{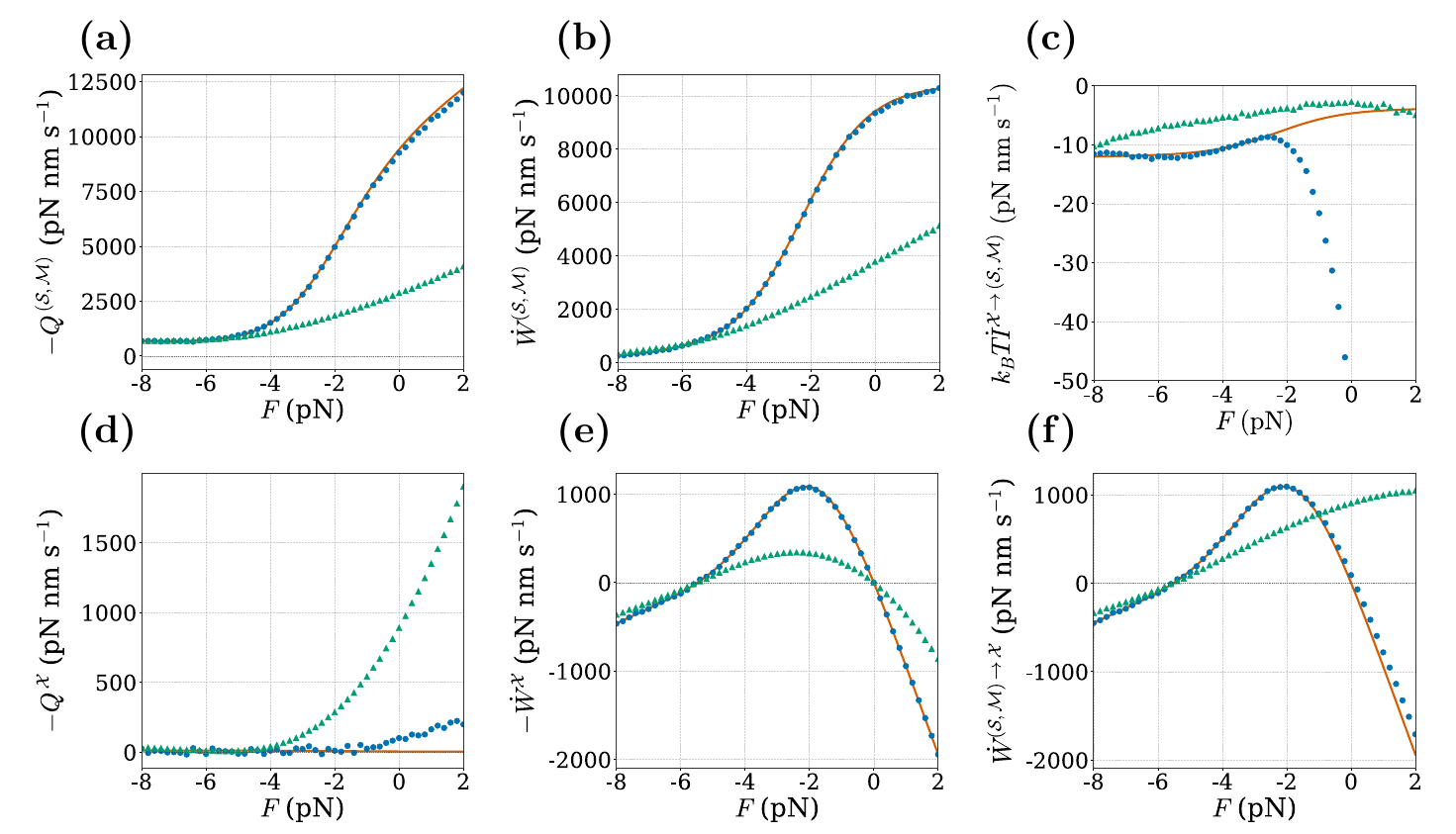}
    \caption{$F$-dependence of the various thermodynamic quantities under the high-ATP condition: (a) $- \dot{Q}^{(\mathcal{S}, \mathcal{M})}$, (b) $\dot{W}^{(\mathcal{S}, \mathcal{M})}$, (c) $k_\mathrm{B} T \dot{I}^{\mathcal{X} \to (\mathcal{S}, \mathcal{M})}$, (d) $-\dot{Q}^{\mathcal{X}}$, (e) $- \dot{W}^{\mathcal{X}}$, and (f) $\dot{W}^{(\mathcal{S}, \mathcal{M}) \to \mathcal{X}}$. 
    The orange solid line shows the result of the perturbative expansion, while the blue circles and green triangles indicate the numerical results for $\gamma=\gamma_{\mathrm{vitro}}$ and $\gamma_{\mathrm{vivo}}$, respectively.
    The error bars are smaller than the marker size.
    }
    \label{fig:thermodynamics properties}
\end{figure*}

\subsection{Various thermodynamic quantities\label{subsec:Various thermodynamic quantities}}
Before presenting the results for the various efficiencies, we first validate our analytical and numerical evaluations of the thermodynamic quantities by comparing them with experimental data.
Figures~\ref{fig:thermodynamics properties}(a)--(f) show the force dependence of \(-\dot{Q}^{(\mathcal{S}, \mathcal{M})}\), \(\dot{W}^{(\mathcal{S}, \mathcal{M})}\), $k_{\mathrm{B}}T\dot{I}^{\mathcal{X} \to (\mathcal{S}, \mathcal{M})}$, \(-\dot{Q}^{\mathcal{X}}\), \(-\dot{W}^{\mathcal{X}}\), and $\dot{W}^{(\mathcal{S}, \mathcal{M}) \to \mathcal{X}}$, respectively.
The orange solid lines represent the results of the perturbation expansion up to $O(\epsilon^0)$. 
For detailed calculations and explicit analytical expressions of these quantities, see Appendix~\ref{sec:appendix: calc detail}. 
The numerical simulation results for \(\gamma = \gamma_{\mathrm{vitro}}\) and \(\gamma = \gamma_{\mathrm{vivo}}\) are shown by blue circles and green triangles, respectively.
For the \textit{in vitro} condition ($\gamma = \gamma_{\mathrm{vitro}}$), the analytical predictions show good agreement with the numerical results over a wide range of the external force, except for the heat flux $\dot{Q}^{\mathcal{X}}$ and the information flow $k_{\mathrm{B}}T\dot{I}^{\mathcal{X} \to (\mathcal{S}, \mathcal{M})}$. 
For these two quantities, discrepancies between the analytical and numerical results become significant when $F \gtrsim -2\,\mathrm{pN}$. 
This deviation arises because their absolute values are relatively small, making them more sensitive to the breakdown of time-scale separation as the transition rates (e.g., $k_f$) increase with the external force. 
This suggests that the time-scale separation assumption is most accurate near the stall force $F_{\mathrm{stall}} \simeq -5.5\,\mathrm{pN}$.

To assess the consistency of our theory with the experimental results reported in Ref.~\cite{ariga2018nonequilibrium}, we compare the experimental and theoretical values of these quantities at a fixed external force \(F = -2\,\mathrm{pN}\) under the \textit{in vitro} condition with $\gamma=\gamma_{\mathrm{vitro}}$, as summarized in Table~\ref{tab:result: comparison of exp. and solution}.
For the analytical expressions, all quantities except \(\dot{Q}^{\mathcal{X}}\) agree with the experimental values within error bars.
For the numerical simulations, all quantities are consistent with the experimental values within error bars.
These comparisons demonstrate that our theoretical framework, despite the simplified time-scale separation assumption, effectively captures the essential energetics of kinesin under the \textit{in vitro} condition.

Next, we compare the analytical predictions with the numerical results for the \textit{in vivo}-like condition (see green triangles in Fig.~\ref{fig:thermodynamics properties}).
Although the time-scale separation assumption is not justified in this regime, our numerical analysis indicates that its breakdown remains relatively modest near the stall force.
Accordingly, the leading-order predictions remain close to the numerical results for most thermodynamic quantities in the near-stall regime.
The information flow is a notable exception: a visible discrepancy remains even in the near stall force regime.
Unlike the other quantities, the information flow depends on a logarithmic ratio of probability distributions and is relatively small in magnitude.
It is therefore more sensitive to modest deviations in the stationary joint distribution and probability current, which may produce a substantial relative error even when the corresponding errors in the energetic quantities remain barely visible.
Away from the stall force, discrepancies also emerge in the other quantities, reflecting the clear breakdown of the time-scale separation assumption.

Finally, we remark on the sign of the information flow. 
As shown in Fig.~\ref{fig:thermodynamics properties}(c), the information flow remains negative across the entire range of $F$ examined here, regardless of whether $\gamma$ corresponds to the \textit{in vitro} or \textit{in vivo} condition. 
Importantly, with our sign convention, this negative value indicates that information propagates from the kinesin to the cargo, rather than vice versa.

\subsection{Thermodynamic efficiency}
\begin{figure}[t]
    \centering
    \includegraphics[width=8.6cm]{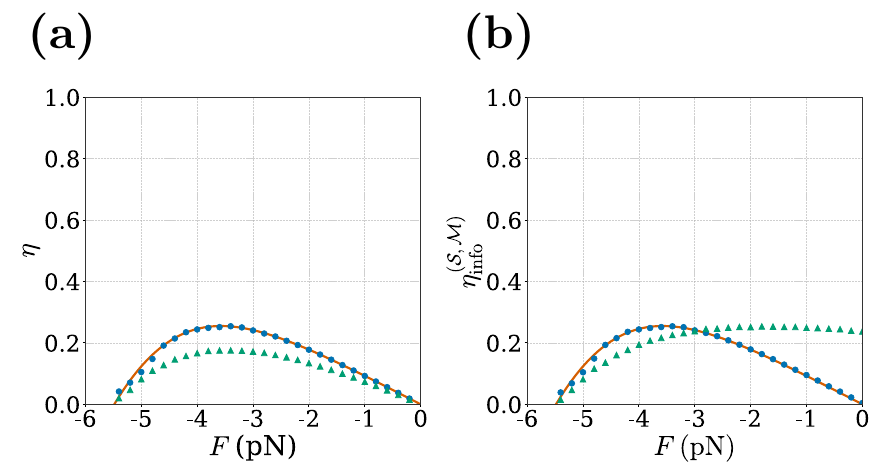}
    \caption{$F$-dependence of (a) the thermodynamic efficiency $\eta$, (b) the information-thermodynamic efficiency $\eta_\mathrm{info}^{(\mathcal{S}, \mathcal{M})}$.
    The orange solid line shows the result of the perturbative expansion, while the blue circles and green triangles indicate the numerical results for $\gamma=\gamma_{\mathrm{vitro}}$ and $\gamma_{\mathrm{vivo}}$, respectively.
    The error bars are smaller than the marker size.
    These efficiencies are calculated under the high-ATP condition.}
    \label{fig:thermoeff and info eff}
\end{figure}

We first discuss the conventional thermodynamic efficiency defined in Eq.~(\ref{eq:setup: def. thermodynamics efficiency}) as a baseline for comparison.
Figure~\ref{fig:thermoeff and info eff}(a) shows the force dependence of the thermodynamic efficiency.
Because the thermodynamic efficiency is defined only when \(\dot{W}^{\mathcal{X}} < 0\), it is plotted only in the range \(F_{\mathrm{stall}} \leq F \leq 0\).

Under the \textit{in vitro} condition, the analytical prediction agrees well with the numerical results within numerical uncertainty.
This agreement indicates that the thermodynamic efficiency is well captured by the \(O(\epsilon^0)\) approximation based on time-scale separation.
The maximum thermodynamic efficiency under the \textit{in vitro} condition is approximately \(26\%\).

We next examine the thermodynamic efficiency under the \textit{in vivo}-like condition.
The maximum value decreases to approximately \(18\%\), which is lower than the corresponding \textit{in vitro} value.
These results indicate that the thermodynamic efficiency of kinesin remains low even under intracellular-like conditions.
Taken together, these results suggest that kinesin does not appear to be optimized for efficient mechanical work output.

\begin{figure*}[t]
    \centering
    \includegraphics[width=0.95\textwidth]{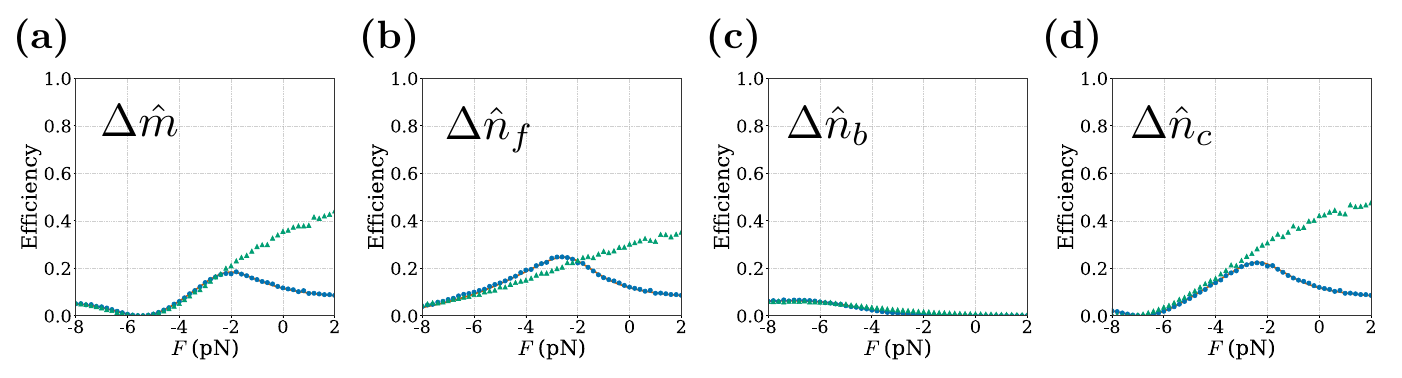}
    \caption{$F$-dependence of the TUR efficiency for each current: (a) $\Delta\hat{m}$, (b) $\Delta\hat{n}_f$, (c) $\Delta\hat{n}_b$, and (d) $\Delta\hat{n}_c$.
    The orange solid line shows the result of the perturbative expansion, while the blue circles and green triangles indicate the numerical results for $\gamma=\gamma_{\mathrm{vitro}}$ and $\gamma_{\mathrm{vivo}}$, respectively.
    The error bars are smaller than the marker size.
    These efficiencies are calculated under the high-ATP condition.}
    \label{fig:TUR eff.}
\end{figure*}

\subsection{Information-thermodynamic efficiency}

We next ask whether accounting for information flow between the motor and cargo changes the conclusion obtained from the standard thermodynamic efficiency.
To this end, we examine the information-thermodynamic efficiency
\(\eta_\mathrm{info}^{(\mathcal{S}, \mathcal{M})}\) defined in Eq.~(\ref{eq:setup: Definition of Info. Efficiency}).
Figure~\ref{fig:thermoeff and info eff}(b) shows the $F$-dependence of the information-thermodynamic efficiency. 
Note that this efficiency is defined only when the transduced capacity is positive: $\dot{W}^{(\mathcal{S}, \mathcal{M}) \to \mathcal{X}} +  k_{\mathrm{B}} T \dot{I}^{\mathcal{X} \to (\mathcal{S}, \mathcal{M})} > 0$. 
Even for $F>-2\,\mathrm{pN}$, where the time-scale separation may break down, the analytical prediction is still in good agreement with the numerical simulation.
Interestingly, under the \textit{in vitro} condition, the curve in Fig.~\ref{fig:thermoeff and info eff}(b) is quite similar to that in Fig.~\ref{fig:thermoeff and info eff}(a), suggesting that the information-thermodynamic efficiency and the thermodynamic efficiency exhibit nearly identical behavior.

This behavior can be understood from the time-scale separation between the cargo and kinesin.
Because the cargo relaxes much faster than the kinesin subsystem, the cargo probability current satisfies
\(\widetilde{J}^{\mathcal{X}}(s,m,x)=O(\epsilon)\).
In fact, using $\widetilde{J}^{\mathcal{X}}(s,m,x)=O(\epsilon)$, which follows from Eq.~(\ref{eq:setup:decomposition of the prob distribution}), it follows that the partial entropy production rate $\dot{\sigma}^{\mathcal{X}}$ is negligible: $\dot{\sigma}^{\mathcal{X}}=O(\epsilon)$ in units of $\tau^{(\mathcal{S},\mathcal{M})}$.
Consequently, we can show that (see Appendix~\ref{sec:appendix: calc detail} for the derivation)
\begin{align}
    \dot{Q}^{\mathcal{X}} = k_{\mathrm{B}} T \dot{I}^{\mathcal{X} \to (\mathcal{S}, \mathcal{M})} + O(\epsilon).
\end{align}
From this relation and the first law for the subsystem (\ref{eq:setup: the first law of the cargo}), we can rewrite the transduced capacity as
\begin{align}
    \dot{W}^{(\mathcal{S}, \mathcal{M}) \to \mathcal{X}} + k_{\mathrm{B}} T \dot{I}^{\mathcal{X} \to (\mathcal{S}, \mathcal{M})} = -\dot{W}^{\mathcal{X}} + O(\epsilon),
    \label{eq:tranceduced capacity in steady state}
\end{align} 
and thus we have 
\begin{align}
    \eta_\mathrm{info}^{(\mathcal{S}, \mathcal{M})} &:= \frac{\dot{W}^{(\mathcal{S}, \mathcal{M}) \to \mathcal{X}} + k_{\mathrm{B}} T \dot{I}^{\mathcal{X} \to (\mathcal{S}, \mathcal{M})}}{\dot{W}^{(\mathcal{S}, \mathcal{M})}} \notag \\ 
    &= \frac{-\dot{W}^{\mathcal{X}}}{\dot{W}^{(\mathcal{S}, \mathcal{M})}} + O(\epsilon)\notag\\
    &\simeq \eta.
\end{align}
We remark that this result is equivalent to $\eta^{\mathcal{X}}_{\mathrm{info}}\simeq1$ because $\eta^{(\mathcal{S}, \mathcal{M})}_{\mathrm{info}} \eta^{\mathcal{X}}_{\mathrm{info}} = \eta$.

We finally remark on the special case of zero external force $F=0$, discussed in Ref.~\cite{leighton2023inferring}.
Under the \textit{in vitro} condition, Fig.~\ref{fig:thermoeff and info eff}(b) indicates that $\eta^{(\mathcal{S}, \mathcal{M})}_{\mathrm{info}}=0$.
Indeed, Eq.~\eqref{eq:tranceduced capacity in steady state} immediately implies $\eta^{(\mathcal{S}, \mathcal{M})}_{\mathrm{info}}=O(\epsilon)$ at $F=0$.
We emphasize that this result is universal within our model and holds irrespective of the specific parameter values.
In contrast, under the \textit{in vivo}-like condition, the efficiency at $F=0$ is close to its maximum value, $\eta^{(\mathcal{S}, \mathcal{M})}_{\mathrm{info}} \simeq 0.24$.
Notably, even at $F=0$---the condition considered in previous work~\cite{leighton2023inferring}---the information-thermodynamic efficiency is at most comparable to the maximum value attained under the \textit{in vitro} condition, and its magnitude remains modest.
Thus, kinesin does not appear to be optimized with respect to the information-thermodynamic efficiency.

\subsection{TUR efficiency}
\begin{figure*}[t]
    \centering
    \includegraphics[width=0.95\textwidth]{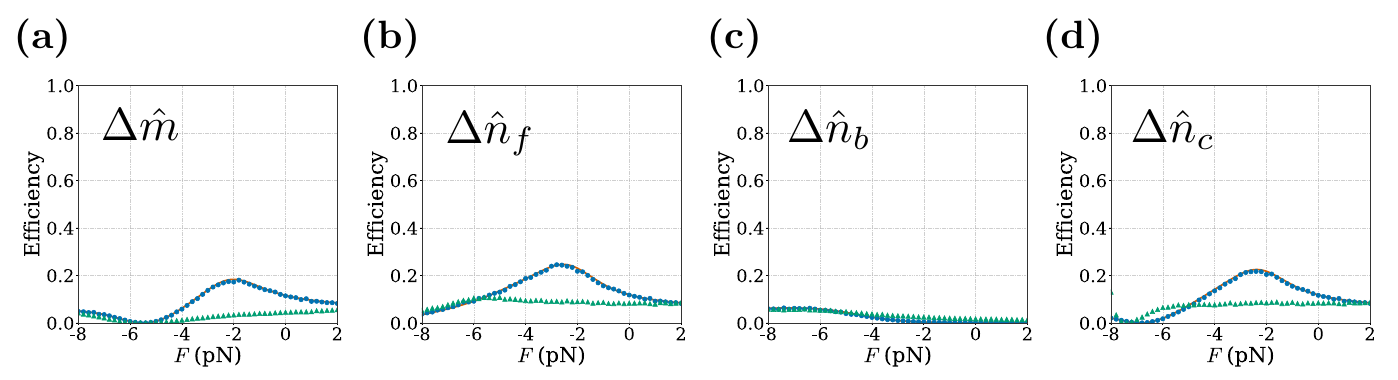}
    \caption{$F$-dependence of the bipartite TUR efficiency for each current: (a) $\Delta\hat{m}$, (b) $\Delta\hat{n}_f$, (c) $\Delta\hat{n}_b$, and (d) $\Delta\hat{n}_c$.
    The orange solid line shows the result of the perturbative expansion, while the blue circles and green triangles indicate the numerical results for $\gamma=\gamma_{\mathrm{vitro}}$ and $\gamma_{\mathrm{vivo}}$, respectively.
    The error bars are smaller than the marker size.
    These efficiencies are calculated under the high-ATP condition.}
    \label{fig:Bipartite TUR eff.}
\end{figure*}

We now examine the TUR efficiency \(\eta_{\mathrm{TUR}}\) introduced in Sec.~\ref{Various definitions of efficiency}.
As a time-integrated current $\hat{\mathcal{J}}$, we choose the displacement of the kinesin $\Delta\hat{m}$, defined in Eq.~(\ref{eq:def of M hat}), and the net numbers of occurrences of the following transitions:
\begin{align}
    \Delta\hat{n}_f &:= \sum_m \left(\hat{n}^{12}_{m + \delta, m} - \hat{n}^{21}_{m - \delta, m}\right), \label{eq: def of sf hat} \\
    \Delta\hat{n}_b &:= \sum_m \left(\hat{n}^{12}_{m - \delta, m} - \hat{n}^{21}_{m + \delta, m}\right), \label{eq: def of sb hat}\\
    \Delta\hat{n}_c &:= \sum_m \left(\hat{n}^{21}_{m,m} - \hat{n}^{12}_{m,m}\right).
    \label{eq: def of sc hat}
\end{align}
Here, $\Delta\hat{n}_f$, $\Delta\hat{n}_b$, and $\Delta\hat{n}_c$ denote the net numbers of mechanical forward, mechanical backward, and internal chemical transitions, respectively, during the time interval $[0,\mathcal{T}]$.

For each current, under the \textit{in vitro} condition, we analytically calculate the mean $J_\mathcal{J} := \lim_{\mathcal{T} \to \infty} \langle \hat{\mathcal{J}} \rangle / \mathcal{T}$ and the fluctuation $D_\mathcal{J} := \lim_{\mathcal{T} \to \infty}\mathrm{Var}[\hat{\mathcal{J}}] / 2 \mathcal{T}$ in the limit $\epsilon\rightarrow0$, as shown in Appendix~\ref{sec:appendix:TUR calculation}.
Figure~\ref{fig:TUR eff.}(a)-(d) plots the resulting TUR efficiency as a function of $F$ for $\Delta\hat{m}$, $\Delta\hat{n}_f$, $\Delta\hat{n}_b$, and $\Delta\hat{n}_c$, respectively.
The analytical predictions are in good agreement with the numerical simulations over the force range considered.

Note that the standard TUR efficiency for $\Delta\hat{m}$ was previously investigated for a six-state model in Ref.~\cite{hwang2018energetic}.
Although our model is constructed differently, Fig.~\ref{fig:TUR eff.}(a) shows that our result is consistent with the value reported in Ref.~\cite{hwang2018energetic}.
Moreover, as shown in Fig.~\ref{fig:TUR eff.}(b)--(d), choosing alternative currents $\Delta\hat{n}_f$, $\Delta\hat{n}_b$, and $\Delta\hat{n}_c$ does not substantially increase the typical magnitude of the efficiency.

Under the \textit{in vivo}-like condition, the efficiencies tend to increase compared with the \textit{in vitro} case.
However, even their maximal values do not approach unity.
Taken together, these results suggest that kinesin is not strongly optimized for suppressing current fluctuations relative to the associated thermodynamic cost, at least for the currents considered here.

\subsection{Bipartite TUR efficiency}
Finally, we ask whether focusing on the motor subsystem changes the conclusion obtained from the standard TUR efficiency.
To this end, we examine the bipartite TUR efficiency \(\eta_{\mathrm{BTUR}}\) introduced in Sec.~\ref{Various definitions of efficiency}.
As discussed in Sec.~\ref{subsec:Various thermodynamic quantities}, the typical magnitudes of the information flow \(k_{\mathrm{B}}T\,\dot{I}^{\mathcal{X}\to(\mathcal{S},\mathcal{M})}\) and the heat flow \(\dot{Q}^{\mathcal{X}}\) are much smaller than that of \(\dot{Q}^{(\mathcal{S},\mathcal{M})}\).
Therefore, within the time-scale separation approximation, the theoretical prediction for the bipartite TUR efficiency becomes nearly identical to that for the standard TUR efficiency, i.e., \(\eta_{\mathrm{BTUR}} \simeq \eta_{\mathrm{TUR}}\).

Figure~\ref{fig:Bipartite TUR eff.}(a)--(d) shows \(\eta_{\mathrm{BTUR}}\) as a function of \(F\) for the currents \(\Delta\hat{m}\), \(\Delta\hat{n}_f\), \(\Delta\hat{n}_b\), and \(\Delta\hat{n}_c\), respectively.
As in the case of the standard TUR efficiency, the analytical predictions based on the time-scale separation approximation agree well with the numerical simulations.
Because \(\eta_{\mathrm{BTUR}}\) is nearly identical to \(\eta_{\mathrm{TUR}}\) within the time-scale separation approximation, the maximum value of \(\eta_{\mathrm{BTUR}}\) for each current is also nearly the same as that of the corresponding standard TUR efficiency under the \textit{in vitro} condition.
Under the \textit{in vivo}-like condition, the maximum value of \(\eta_{\mathrm{BTUR}}\) for each current is lower than the corresponding \textit{in vitro} value.
This behavior contrasts with the standard TUR efficiency, which tends to increase under the \textit{in vivo}-like condition.
Taken together, these results suggest that, even when focusing on the subsystem, kinesin is not optimized for suppressing current fluctuations, at least for the currents considered here.

\section{Concluding remarks \label{sec:conclusion}}

In this study, we systematically compared various thermodynamic efficiencies within a single kinesin--cargo model.
By constructing a thermodynamically consistent kinesin--cargo model, we evaluated four efficiencies on an equal footing: the thermodynamic efficiency, the information-thermodynamic efficiency, the TUR efficiency, and the bipartite TUR efficiency.
Assuming time-scale separation between the motor and the cargo, we derived analytical expressions for these efficiencies using a perturbative expansion and compared them with numerical simulation results.
Our results suggest that kinesin may not be optimized for any efficiency considered here.
However, it should be emphasized that this conclusion does not imply that kinesin is an inefficient molecular machine.
Rather, our results suggest that the functional performance of kinesin may not be fully captured by thermodynamic efficiencies alone.
This observation motivates further investigation of other biological or kinetic requirements, such as directionality, robustness, processivity, or reliable cargo transport under fluctuating intracellular conditions.

We close with several remarks and limitations. 
We first comment on the origin of the discrepancy between our estimate of the information-thermodynamic efficiency and that reported by Leighton and Sivak in Ref.~\cite{leighton2023inferring}. 
In that study, upper and lower bounds on \(\eta^{(\mathcal{S}, \mathcal{M})}_{\mathrm{info}}\) were derived using a Langevin description of the kinesin--cargo complex. 
Because the upper bound relies on the Langevin description, it is not directly applicable to our present model, whereas the lower bound can still be applied. 
Under no-load conditions \((F=0)\), this lower bound reads
\begin{align}
    \frac{\gamma v^2}{\dot{W}^{(\mathcal{S},\mathcal{M})}}
    \le
    \eta^{(\mathcal{S}, \mathcal{M})}_{\mathrm{info}}
\end{align}
in our notation. 
Using the values of the physical quantities obtained from our model, we find that the lower bound is approximately 0.0021 under the \textit{in vitro} condition and 0.23 under the \textit{in vivo}-like condition.
These values are consistent with the corresponding information-thermodynamic efficiencies computed directly from our model, $\eta_{\mathrm{info}}^{(\mathcal{S}, \mathcal{M})}=0.0040$ for the \textit{in vitro} condition and $0.24$ for the \textit{in vivo}-like condition, and hence the lower-bound relation is satisfied in both cases.
By contrast, Leighton and Sivak estimated this lower bound to be approximately 0.7 under \textit{in vivo} conditions.
This larger value appears to arise mainly from the use of a relatively small ATP free-energy change, $\Delta\mu=15k_{\mathrm{B}}T$, together with the relatively large \textit{in vivo} velocity measured at $37^\circ\mathrm{C}$.
Indeed, if the chemical free-energy input rate is estimated using the tight-coupling assumption as $\dot{W}^{(\mathcal{S},\mathcal{M})}=v\Delta\mu/\delta$, the lower bound becomes
\begin{align}
\frac{\gamma v^2}{\dot{W}^{(\mathcal{S},\mathcal{M})}}
=
\frac{\gamma v\delta}{\Delta\mu},
\end{align}
showing that the estimate increases linearly with the velocity and inversely with $\Delta\mu$.
Thus, the discrepancy with Ref.~\cite{leighton2023inferring} is likely attributable not to the lower-bound relation itself, but primarily to the parameter values used to evaluate this Stokes-efficiency-type quantity under intracellular conditions.

It should be noted that the intracellular kinesin velocity parameter used in Ref.~\cite{leighton2023inferring} was taken from intracellular vesicle velocities \cite{shtridelman2008force}, rather than from single-molecule measurements of kinesin. Such vesicle motion may involve multiple motors acting on the same cargo and may also include contributions from other kinesin family members. In addition,
the use of velocity data obtained at $37^\circ\mathrm{C}$, whereas our estimates are based on parameters calibrated at $25 \, ^\circ \mathrm{C}$, suggests that the information-thermodynamic efficiency may have a nontrivial temperature dependence through the motor velocity, transition kinetics, and possibly other parameters.
A definitive quantitative assessment of $\eta^{(\mathcal{S}, \mathcal{M})}_{\mathrm{info}}$ under \textit{in vivo} conditions therefore requires more systematic measurements and modeling of kinesin dynamics under controlled temperature and intracellular-like conditions.

This point is closely related to a broader limitation of our model in comparing \textit{in vitro} and \textit{in vivo} conditions.
As mentioned above, our \textit{in vivo}-like condition is implemented solely by increasing the friction coefficient, while keeping all kinetic parameters unchanged.
In actual living cells, however, the intracellular environment can modify not only the effective friction but also the transition kinetics and the noise statistics of molecular motors.
Living cells actively generate non-thermal fluctuations~\cite{guo2014probing,ariga2024nonthermal}, and such fluctuations may affect kinesin motility~\cite{ariga2021noise,feng2023unraveling}.
Because these fluctuations can violate local detailed balance as well as the fluctuation--dissipation relation of the second kind~\cite{tanogami2022violation}, the applicability of both our model and the Langevin framework of Ref.~\cite{leighton2023inferring} under intracellular conditions remains unclear.
In addition, limited experimental information on intracellular kinesin dynamics currently precludes a rigorous quantitative assessment of existing \textit{in vivo} estimates.
We defer these open questions to future research.

\begin{acknowledgements}
    We thank Kazuma Nishimura for sharing his master's thesis.
    T.A.~was supported by JSPS KAKENHI Grant Number JP24K00600 and JST CREST Grant Number JPMJCR24T2, Japan.
    T.T.~was supported by JSPS KAKENHI Grant Number JP25K17315 and JST PRESTO Grant Number JPMJPR23O6, Japan.
\end{acknowledgements}

\section*{Data Availability}
The numerical data and source code that support the findings of this article are available from the authors upon reasonable request.

\appendix
\section{Comparison with Bell's equation \label{sec:appendix: comparison with bell's equation}}
In this section, we clarify the difference between the functional form of the transition rates employed in this paper [Eqs.~(\ref{eq:setup:form of kf})-(\ref{eq:setup: Specific transition rate of kinesin (kb dag)})] and that used in previous studies~\cite{taniguchi2005entropy, nishiyama2002chemomechanical,ariga2018nonequilibrium,ariga2021noise}.
In previous studies, the transition rates $k_{f/b}$ were described by Bell's equation~\cite{bell1978models}:
\begin{align}
    k_f &= k_f^0 \exp \bigg(\frac{d_fF}{k_{\mathrm{B}} T}\bigg), \label{eq:appendix:kf}\\
    k_f^\dag  &= k_f^{0\dag} \exp \bigg(\frac{d_f^\dag F}{ k_{\mathrm{B}} T} \bigg), \label{eq:appendix:kf dag}\\
    k_b &= k_b^0 \exp \bigg(\frac{d_bF}{k_{\mathrm{B}} T}\bigg), \label{eq:appendix:kb}\\
    k_b^\dag &= k_b^{0\dag} \exp \bigg(\frac{d_b^\dag F}{k_{\mathrm{B}} T} \bigg), \label{eq:appendix:kb dag}
\end{align}
where $d_{f/b}$ and $d_{f/b}^\dag$ denote the characteristic distances satisfying $d_f - d_f^\dag = \delta$ and $d_b - d_b^\dag = - \delta$, and $k_{f/b}^0$ and $k_{f/b}^{0\dag}$ denote the rate constants at zero load satisfying
\begin{align}
    \frac{k_{f/b}^{0}}{k_{f/b}^{0\dag}} = \exp\left[\frac{\Delta\mu_{\mathrm{mech}}}{k_{\mathrm{B}} T}\right].
\end{align}
Note that these transition rates are independent of $m$ and $x$ and do not satisfy the local detailed balance condition in Eqs.~(\ref{eq:setup: LDB of the kinesin (kf)}) and (\ref{eq:setup: LDB of the kinesin (kb)}), but instead satisfy
\begin{align}
  \frac{k_f}{k_f^\dag} &= \exp \bigg[ \frac{F\delta + \Delta\mu_{\mathrm{mech}}}{k_{\mathrm{B}} T} \bigg],\\
  \frac{k_b}{k_b^\dag} &= \exp \bigg[ \frac{-F\delta + \Delta\mu_{\mathrm{mech}}}{k_{\mathrm{B}} T} \bigg].
\end{align}
Thus, from the viewpoint of the local detailed balance conditions in Eqs.~(\ref{eq:setup: LDB of the kinesin (kf)}) and (\ref{eq:setup: LDB of the kinesin (kb)}), these transition rates are not thermodynamically consistent in general.
In other words, these transition rates become thermodynamically consistent only under the condition of force balance, $F=\kappa(x-m)$.

Although the transition rates (\ref{eq:setup:form of kf})-(\ref{eq:setup: Specific transition rate of kinesin (kb dag)}) have a form different from Bell's equation, the effective transition rates obtained under the assumption of time-scale separation, shown again below [Eqs.~(\ref{eq:effective rates kf})-(\ref{eq:effective rates kb_dag})], take the form of Bell's equation:
\begin{align}
  \overline{\widetilde{k}}_f &= \frac{\tau^{(\mathcal{S},\mathcal{M})}}{\tau_f} \exp \bigg( \frac{\theta_f}{k_{\mathrm{B}} T} \bigg[(\theta_f - 1) \frac{\kappa}{2} \delta^2 +  \Delta\mu_{\mathrm{mech}} + F \delta \bigg] \bigg), \\
  \overline{\widetilde{k}}_f^\dag &= \frac{\tau^{(\mathcal{S},\mathcal{M})}}{\tau_f} \exp \bigg( \frac{\theta_f - 1}{k_{\mathrm{B}} T} \bigg[ \theta_f \frac{\kappa}{2} \delta^2 + \Delta\mu_{\mathrm{mech}} + F \delta \bigg] \bigg), \\
  \overline{\widetilde{k}}_b &= \frac{\tau^{(\mathcal{S},\mathcal{M})}}{\tau_b} \exp \bigg( \frac{\theta_b}{k_{\mathrm{B}} T}\bigg[ (\theta_b - 1) \frac{\kappa}{2} \delta^2   +  \Delta\mu_{\mathrm{mech}}  - F \delta \bigg] \bigg), \\
  \overline{\widetilde{k}}_b^\dag &= \frac{\tau^{(\mathcal{S},\mathcal{M})}}{\tau_b} \exp \bigg[ \frac{\theta_b - 1}{k_{\mathrm{B}} T} \bigg( \theta_b \frac{\kappa}{2} \delta^2  + \Delta\mu_{\mathrm{mech}} - F \delta \bigg)\bigg].
\end{align}
Indeed, if we identify $d_f=\theta_f\delta$, $d^\dag_f=(\theta_f-1)\delta$, $d_b=-\theta_b\delta$, and $d^\dag_b=-(\theta_b-1)\delta$, then these transition rates can be rewritten in the form of Bell's equation.
As a consequence, under the assumption of time-scale separation, the six parameters $\tau_f, \tau_b, \theta_f, \theta_b, k_c, \Delta \mu_{\mathrm{chem}}$ in our model are in one-to-one correspondence with the six parameters $k_f^0, k_b^0, d_f, d_b, k_c, \Delta \mu_{\mathrm{chem}}$ in Bell's equation.

\section{Details of the parameter estimation\label{sec:appendix:parameter estimation}}
There are six parameters to be estimated in our model: $\tau_f, \tau_b, \theta_f, \theta_b, k_c, \Delta \mu_{\mathrm{chem}}$.
We determine these parameters by fitting the theoretical prediction of the force--velocity relationship derived from the model [Eq.~(\ref{eq:setup:steady-state velocity})] to the experimental data in Ref.~\cite{ariga2018nonequilibrium} (see also Fig.~\ref{fig:velocity vs Force}).
As we have mentioned in Appendix~\ref{sec:appendix: comparison with bell's equation}, under time-scale separation, the six parameters are in one-to-one correspondence with $k_f^0, k_b^0, d_f, d_b, k_c, \Delta \mu_{\mathrm{chem}}$ in Bell's equation.
We assume that these six parameters do not depend on the time-scale separation.
Among these, $\Delta \mu_{\mathrm{chem}}$ is the only parameter that cannot be directly measured experimentally with current techniques.
To estimate a plausible value of $\Delta \mu_{\mathrm{chem}}$, we perform a series of fits.
Specifically, we first fix $\Delta \mu_{\mathrm{chem}}$ to a range of different values and estimate the remaining five parameters by fitting Eq.~(\ref{eq:setup:steady-state velocity}) to the experimental data.
Then, we compare the estimated five-parameter values with their typical experimental values to narrow down the range of plausible values of $\Delta \mu_{\mathrm{chem}}$.

\begin{table}[b]
  \centering
  \caption{Estimated values of the parameters in our model.}
  \begin{ruledtabular}
  \begin{tabular}{lll}
  Parameter   &   High ATP & Low ATP                  \\ \hline
  $\tau_f \, \mathrm{(s)}$ & $ 0.1445 \pm 0.130 $ & $ 0.255 \pm 0.398 $ \\
  $\tau_b \, \mathrm{(s)}$ & $0.194 \pm 1.476$ & $0.476 \pm 5.740$ \\ 
  $k_c \, \mathrm{(s^{-1})}$ & $123.4 \pm 18.50$ & $32.60 \pm 4.900$ \\
  $\Delta \mu_{\mathrm{chem}}  \, \mathrm{(pN \, nm)}$ (Fixed) & $42.4$ & $38.3$ \\
  $\theta_f$ & $0.493 \pm 0.184$ & $ 0.472 \pm 0.221$ \\
  $\theta_b$  & $0.000 \pm 0.312$ & $ 0.000 \pm 0.464 $ \\
  \end{tabular}
  \end{ruledtabular}
  \label{tab:appendix:params our model}
  \centering
  \caption{Estimated values of the experimentally measurable parameters.}
  \begin{ruledtabular}
  \begin{tabular}{lll}
  Parameter   &   High ATP & Low ATP                  \\ \hline
  $k_f^0 \, \mathrm{(s^{-1})}$ & $1136.1 \pm 2485.2 $ & $ 675.7 \pm 1974.9$ \\
  $k_b^0 \, \mathrm{(s^{-1})}$ & $5.20 \pm 39.3 $ & $2.1 \pm 25.4 $ \\ 
  $k_c \, \mathrm{(s^{-1})}$ & $123.4 \pm 18.50$ & $32.60 \pm 4.900$ \\
  $\Delta \mu_{\mathrm{chem}}  \, \mathrm{(pN \, nm)} $ (Fixed)  & $42.4$ & $38.3$ \\
  $d_f \, \mathrm{(nm)}$ & $3.944 \pm 1.472$ & $ 3.776 \pm 1.768 $ \\
  $d_b \, \mathrm{(nm)} $ & $0.000 \pm 2.496$ & $ 0.000 \pm 3.712 $ \\
  \end{tabular}
  \end{ruledtabular}
  \label{tab:appendix:params exp setup}
\end{table}

\begin{figure}[t]
    \centering
    \includegraphics[width=0.43\textwidth]{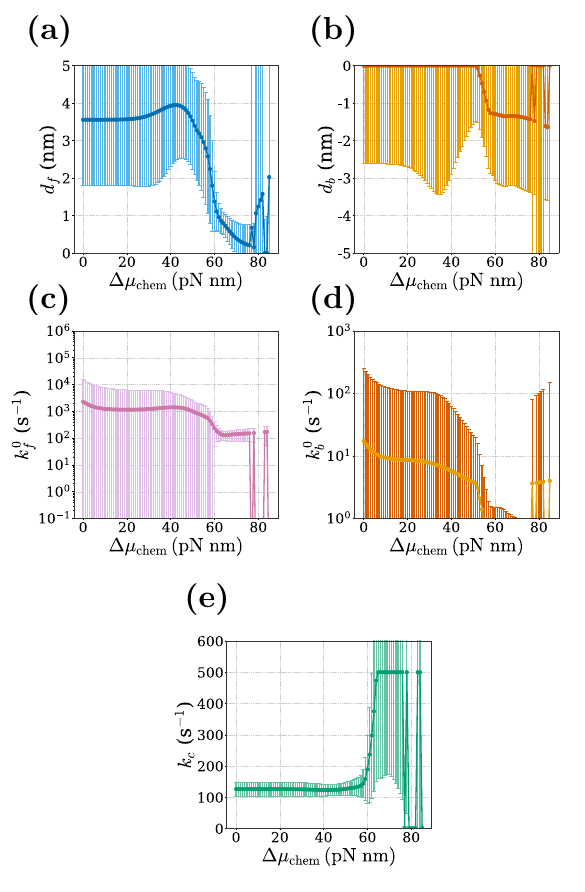}
    \caption{$\Delta \mu_{\mathrm{chem}}$-dependence of the estimated parameters: (a) the characteristic distance of the forward step $d_f$, (b) the characteristic distance of the backward step $d_b$, (c) the forward rate constant at zero load $k^0_f$, (d) the backward rate constant at zero load $k^0_b$, and (e) the internal transition rate $k_c$.}
    \label{fig:appendix:parameter estimation}
\end{figure}

First, we investigate how the estimated five parameters $k_f^0, k_b^0, d_f, d_b, k_c$ change as a function of $\Delta \mu_{\mathrm{chem}}$. 
The total free energy of ATP hydrolysis was fixed at $\Delta \mu = 85\,\mathrm{pN\,nm} \simeq 21\,k_{\mathrm{B}}T$, while $\Delta \mu_{\mathrm{chem}}$ was varied from $0$ to $85\,\mathrm{pN\,nm}$ in increments of $1\,\mathrm{pN\,nm}$. 
We set $\kappa = 0.075\,\mathrm{pN\,nm^{-1}}$, which is estimated at $F = -2\,\mathrm{pN}$ and $\gamma = 3.09 \times 10^{-5} \, \mathrm{pN \ s/ nm}$~\cite{ariga2018nonequilibrium}, and $k_{\mathrm{B}}T = 4.12\,\mathrm{pN\,nm}$ at room temperature $25 \, ^\circ \mathrm{C}$. 
Figure~\ref{fig:appendix:parameter estimation} shows the resulting estimated five parameters as a function of $\Delta \mu_{\mathrm{chem}}$.
The fitting was performed using a weighted least squares method.

\begin{figure*}[!t]
    \centering
    \hspace*{-8mm}
    \vspace*{-10mm}
    \includegraphics[width=0.85\textwidth]{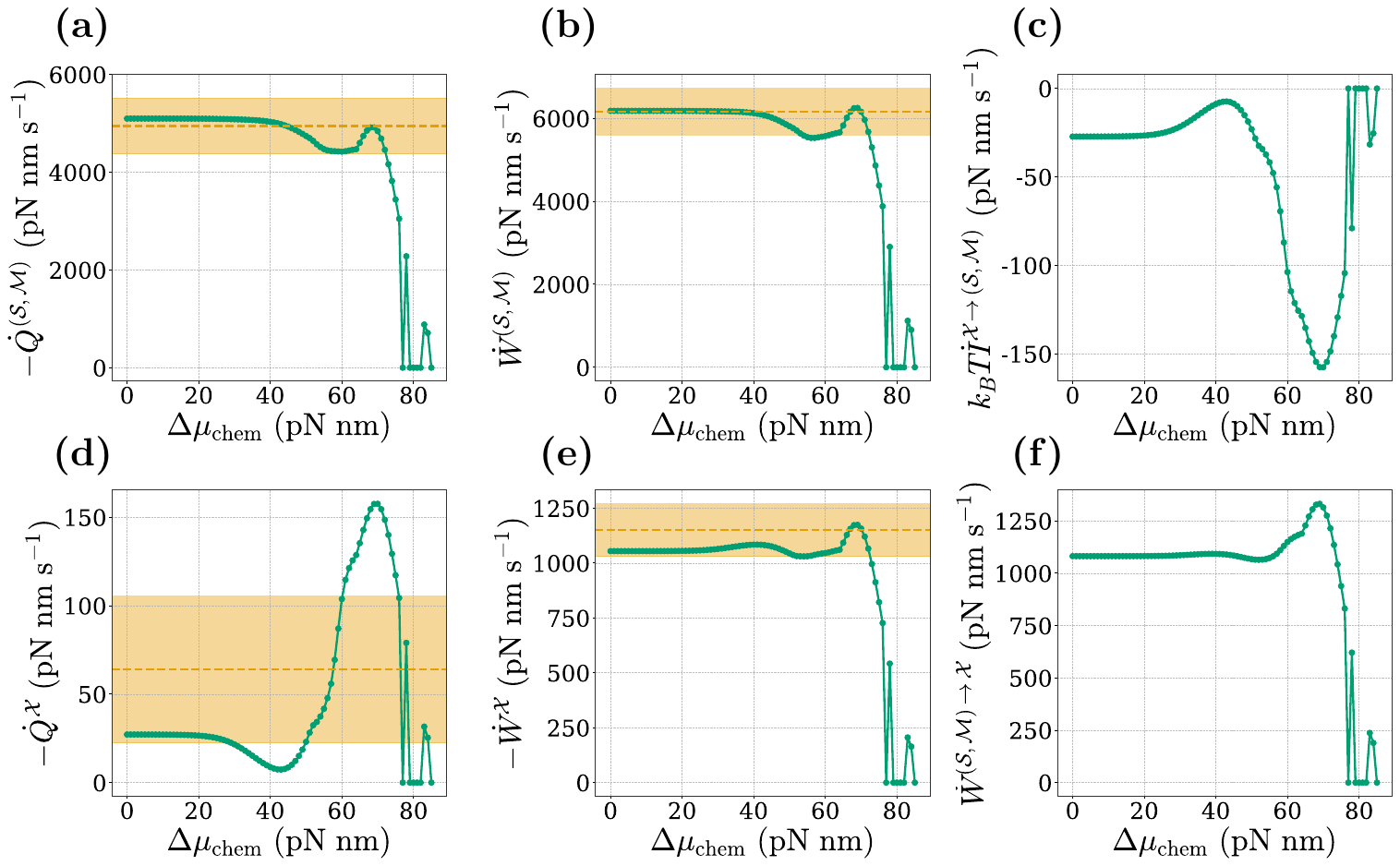}
    \caption{Dependence of the thermodynamic quantities on $\Delta \mu_{\mathrm{chem}}$: (a) $- \dot{Q}^{(\mathcal{S}, \mathcal{M})}$, (b) $\dot{W}^{(\mathcal{S}, \mathcal{M})}$, (c) $k_{\mathrm{B}}T \dot{I}^{\mathcal{X} \to (\mathcal{S}, \mathcal{M})}$, (d) $-\dot{Q}^{\mathcal{X}}$, (e) $- \dot{W}^{\mathcal{X}}$, and (f) $\dot{W}^{(\mathcal{S}, \mathcal{M}) \to \mathcal{X}}$.
    In all panels, the constant external force is set to $F = -2\,\mathrm{pN}$.
    The dashed lines and shaded regions in panels (a), (b), (d), and (e) denote the experimental values and their standard errors, respectively; these values are listed in Table~\ref{tab:result: comparison of exp. and solution}.}
    \label{fig:appendix:parameter estimation of thermo quantities}
\end{figure*}

These plots show that the five parameters are nearly independent of $\Delta \mu_{\mathrm{chem}}$ for $\Delta \mu_{\mathrm{chem}}\lesssim50\,\mathrm{pN\,nm}$, whereas they change significantly for $\Delta \mu_{\mathrm{chem}}\gtrsim50\,\mathrm{pN\,nm}$.
When $\Delta \mu_{\mathrm{chem}}\gtrsim50\,\mathrm{pN\,nm}$, the estimated parameter values, especially $k_f^0$ and $k_c$, deviate significantly from the experimentally estimated values reported in previous studies~\cite{ariga2018nonequilibrium,taniguchi2005entropy,nishiyama2002chemomechanical}.
Therefore, it is reasonable to suppose that the plausible range for $\Delta \mu_{\mathrm{chem}}$ is $0$--$50\,\mathrm{pN\,nm}$.
Because various properties of our model, such as the force--velocity relationship and the thermodynamic properties, do not change significantly within this range, we can say that our model is robust against the specific choice of $\Delta \mu_{\mathrm{chem}}$ (see also Fig.~\ref{fig:appendix:parameter estimation of thermo quantities}).
In other words, our main results remain unchanged regardless of the specific value of $\Delta \mu_{\mathrm{chem}}$ in the range $0$--$50\,\mathrm{pN\,nm}$.
Nevertheless, we can still estimate a plausible value of $\Delta \mu_{\mathrm{chem}}$ by using a relation between the stall force $F_{\mathrm{stall}}$ and $\Delta\mu_{\mathrm{mech}}$ reported in a previous study~\cite{hyeon2009kinesin}, which states that $\Delta\mu_{\mathrm{mech}}$ is equal to the work required to stall the motor, $W_{\mathrm{stall}} := - F_{\mathrm{stall}} \times \delta$. 
In our model, the stall force can be estimated as $F_{\mathrm{stall}} \simeq -5.5 \hspace{5pt} \mathrm{pN}$ (high ATP) and $F_{\mathrm{stall}} \simeq -6.0 \hspace{5pt} \mathrm{pN}$ (low ATP) from the force--velocity relationship, which is insensitive to the specific value of $\Delta \mu_{\mathrm{chem}}$. 
Then, $\Delta\mu_{\mathrm{mech}}$ can be estimated as $\Delta\mu_{\mathrm{mech}} \simeq 10.7\,k_{\mathrm{B}}T$ and $\Delta\mu_{\mathrm{mech}} \simeq 11.7\,k_{\mathrm{B}}T$ for the high- and low-ATP conditions, respectively. 
By noting that the total free energy of ATP hydrolysis is $\Delta \mu = 21\,k_{\mathrm{B}}T$, the free energy change $\Delta \mu_{\mathrm{chem}}$ can thus be estimated as $\Delta \mu_{\mathrm{chem}} \simeq 10.3\,k_{\mathrm{B}}T$ ($\simeq 42.4\,\mathrm{pN\,nm}$) for the high-ATP condition and $\Delta \mu_{\mathrm{chem}} \simeq 9.3\,k_{\mathrm{B}}T$ ($\simeq 38.3\,\mathrm{pN\,nm}$) for the low-ATP condition.
The resulting parameter values $\tau_f, \tau_b, \theta_f, \theta_b, k_c, \Delta \mu_{\mathrm{chem}}$ (equivalently, $k_f^0, k_b^0, d_f, d_b, k_c, \Delta \mu_{\mathrm{chem}}$) are summarized in Table~\ref{tab:appendix:params our model} (Table~\ref{tab:appendix:params exp setup}).

\section{Details of the numerical simulations}
\label{sec:appendix:simulation detail}

This section describes the numerical scheme used to simulate the kinesin--cargo dynamics considered in the main text.
The system consists of a Markov jump process for the discrete kinesin states and overdamped Langevin dynamics for the continuous cargo coordinate.

A direct hybrid simulation would combine an event-driven update of the kinesin states with a time-discretized integration of the cargo dynamics.
However, because the kinesin transition rates depend on the continuously evolving cargo coordinate, the escape rate of the kinesin subsystem generally varies during each waiting interval.

For a fixed cargo position $x$, we define the total escape rate from the kinesin state $(s,m)$ as
\begin{align}
    \Gamma(s,m;x)
    :=
    \sum_{(s',m')\neq(s,m)}
    W_{m'm}^{s's}(x),
\end{align}
where the sum runs over all allowed kinesin transitions out of $(s,m)$.
If $x$ remained fixed, the waiting time to the next kinesin event would be exponentially distributed with rate $\Gamma(s,m;x)$.
In the coupled dynamics, however, $x=x_t$ evolves continuously, and the instantaneous escape rate becomes $\Gamma(s,m;x_t)$.
Conditional on a realization of the cargo trajectory, the probability that no kinesin transition occurs during the interval $[t,t+\Delta t]$ is therefore
\begin{align}
    \exp\left[
    -\int_t^{t+ \Delta t}
    \Gamma(s,m;x_{t'})\,dt'
    \right],
\end{align}
rather than $\exp[-\Gamma(s,m;x_t) \Delta t]$ with the rate fixed at the beginning of the interval.
An exact simulation would thus require the integrated escape rate to be tracked simultaneously with the stochastic cargo dynamics, making the event-detection procedure substantially more involved than the standard Gillespie algorithm.


We therefore construct a spatially discretized Markov jump process whose continuum limit reproduces the original Langevin dynamics.
The kinesin and cargo degrees of freedom can then be treated within a single continuous-time Markov jump framework, and the full coupled system can be simulated using the Gillespie algorithm.
This approach eliminates the need for a fixed temporal discretization and places all kinesin and cargo transitions on the same event-driven time axis.

Consider the overdamped Langevin equation for the cargo coordinate,
\begin{equation}
    \gamma \frac{dx_t}{dt} 
    = 
    -\partial_x V(m,x_t)
    + \sqrt{2\gamma k_B T}\,\xi_t,
\end{equation}
where
\begin{equation}
    V(m,x):=U(m,x)-Fx
\end{equation}
is the effective potential including the constant external force $F$.
For simplicity, we describe the lattice implementation of the cargo dynamics for a fixed kinesin position $m$. 
In the full coupled simulation, the kinesin state $(s,m)$ is updated whenever a kinesin transition occurs, and the cargo transition rates are recalculated using the resulting value of $m$.
Then, the corresponding Fokker--Planck equation for the cargo coordinate is
\begin{equation}
    \frac{\partial}{\partial t}p_t(x)
    =
    \frac{1}{\gamma}
    \frac{\partial}{\partial x}
    \left[
        \partial_x V(m,x)\,p_t(x)
    \right]
    +
    \frac{k_B T}{\gamma}
    \frac{\partial^2}{\partial x^2}p_t(x).
    \label{eq:appendix:FP of cargo}
\end{equation}
We discretize space with lattice spacing $\Delta x$ and consider a one-dimensional Markov jump process.
Let $k_+(m,x)$ denote the transition rate from $x$ to $x+\Delta x$, and let $k_-(m,x)$ denote the transition rate from $x$ to $x-\Delta x$. 
Let $P_i(t)$ denote the probability of finding the cargo at the lattice site $x_i=i\Delta x$. 
In the continuum limit, it is related to the probability density by
\begin{align}
    P_i(t) \simeq p_t(x_i)\Delta x.
\end{align}
The lattice probability evolves according to the master equation
\begin{align}
    \frac{d}{dt}P_i(t)
    ={}&
    k_+(m,x_{i-1})P_{i-1}(t)
    +
    k_-(m,x_{i+1})P_{i+1}(t)
    \notag\\
    &-
    \bigl[
        k_+(m,x_i)+k_-(m,x_i)
    \bigr]P_i(t).
    \label{eq:appendix:master eq. of discrete cargo}
\end{align}
To ensure that the discrete process is physically consistent with the underlying Langevin dynamics, we impose local detailed balance on the transition rates:
\begin{equation}
    \frac{k_+(m,x)}{k_-(m,x+\Delta x)}
    = 
    \exp\left(\frac{V(m,x)-V(m, x+\Delta x)}{k_B T}\right).
\end{equation}
We define the transition rates by
\begin{align}
    k_+(m,x) &:= \frac{1}{\tau_{\Delta x}}
    \exp 
    \left[
        \frac{\theta}{k_B T} 
        \bigl( 
            V(m,x)-V(m, x+\Delta x)
        \bigr)
    \right], \label{eq:appendix:transition rate of k_+}\\
    k_-(m,x) 
    &:= \frac{1}{\tau_{\Delta x}}
    \exp
    \left[
        \frac{\theta-1}{k_B T}
        \bigl(
            V(m, x-\Delta x)-V(m,x)
        \bigr)
    \right],
    \label{eq:appendix:transition rate of k_-}
\end{align}
where $\theta$ is a dimensionless parameter and $\tau_{\Delta x}$ is a characteristic time scale.
Assuming that $\Delta x$ is sufficiently small compared with the characteristic length scale over which the potential varies, we expand all terms up to second order in $\Delta x$.

The transition rates in Eqs.~\eqref{eq:appendix:transition rate of k_+} and \eqref{eq:appendix:transition rate of k_-} are expanded to second order as
\begin{align}
    &k_+(m,x) \notag \\
    &\simeq
    \frac{1}{\tau_{\Delta x}}
    \Biggl[
        1
        - \frac{\theta}{k_B T}
        \partial_x V(m,x)\,\Delta x
        \notag\\
    &\quad
        +
        \left\{
            \frac{\theta^2}{2(k_B T)^2}
            \bigl(\partial_x V(m,x)\bigr)^2
            - \frac{\theta}{2k_B T}
            \partial_x^2 V(m,x)
        \right\}
        (\Delta x)^2
    \Biggr],
    \\
    &k_-(m,x) \notag \\
    &\simeq
    \frac{1}{\tau_{\Delta x}}
    \Biggl[
        1
        - \frac{\theta-1}{k_B T}
        \partial_x V(m,x)\,\Delta x
        \notag\\
    &\quad
        +
        \left\{
            \frac{(\theta-1)^2}{2(k_B T)^2}
            \bigl(\partial_x V(m,x)\bigr)^2
            + \frac{\theta-1}{2k_B T}
            \partial_x^2 V(m,x)
        \right\}
        (\Delta x)^2
    \Biggr].
\end{align}
Substituting $P_i(t)\simeq p_t(x_i)\Delta x$ into the master equation and expanding $p_t(x_i\pm\Delta x)$ and the transition rates in powers of $\Delta x$, we obtain, to leading order in the continuum limit,
\begin{equation}
    \frac{\partial}{\partial t} p_t(x)
    \simeq 
    \frac{(\Delta x)^2}{\tau_{\Delta x}}
    \left[
        \partial_x^2 p_t(x)
        + 
        \frac{1}{k_B T}\partial_x 
        \bigl( 
            \partial_x V(m,x)\,p_t(x)
        \bigr)
    \right].
    \label{eq:appendix:reduced master eq. of cargo}
\end{equation}
Comparing Eq.~\eqref{eq:appendix:reduced master eq. of cargo} with Eq.~\eqref{eq:appendix:FP of cargo}, we find that the same dynamics is reproduced if
\begin{equation}
    \tau_{\Delta x} = \frac{\gamma}{k_B T}(\Delta x)^2.
    \label{eq:appendix:matching condition of cargo}
\end{equation}

In all numerical results reported in this work, the continuous coordinate is discretized with lattice spacing $\Delta x$ and evolved as the Markov jump process defined above. 
We set $\tau_{\Delta x}$ according to the matching condition in Eq.~\eqref{eq:appendix:matching condition of cargo}, so that, for sufficiently small $\Delta x$, the resulting master equation provides an accurate approximation to the Fokker--Planck dynamics in Eq.~\eqref{eq:appendix:FP of cargo}. 
The parameter $\theta$ is not fixed by the leading-order continuum matching condition; different choices affect only higher-order finite-lattice-spacing corrections. 
We adopt the symmetric convention $\theta=1/2$, which distributes the energy difference equally between the forward and backward transition rates.
Throughout this work, we use the lattice spacing $\Delta x=0.05\,\mathrm{nm}$, which is much smaller than the kinesin step size $\delta=8\,\mathrm{nm}$ and the thermal width $\sqrt{k_{\mathrm{B}}T/\kappa}\simeq7.4\,\mathrm{nm}$ of the cargo distribution.
We confirmed that decreasing the lattice spacing further does not appreciably change the observables reported in the main text.

For a given state $(s,m,x_i)$, the possible events consist of kinesin transitions $(s,m)\to(s',m')$ and cargo transitions $x_i\to x_{i\pm1}$. 
The total escape rate of the combined process is
\begin{align}
    &\Lambda(s,m,x_i)\notag\\
    &=
    \sum_{(s',m')\neq(s,m)} W_{m'm}^{s's}(x_i)
    +
    k_+(m,x_i)
    +
    k_-(m,x_i).
\end{align}
At each simulation step, the waiting time to the next event is sampled as
\begin{align}
    \Delta t_{\mathrm{event}}
    =
    -\frac{\ln r_1}{\Lambda(s,m,x_i)},
\end{align}
where $r_1$ is uniformly distributed on $(0,1)$. 
The type of event is then selected with probability proportional to its transition rate using a second independent uniform random number.
Because the cargo transition rates scale as
$k_\pm=O((\Delta x)^{-2})$, the computational cost increases as the lattice spacing is reduced. 
The value of $\Delta x$ was therefore chosen to balance continuum-limit accuracy and computational efficiency.

All physical observables reported in the main text were evaluated after the statistics of the relative coordinate had relaxed to stationarity.
Before data collection, each independent trajectory was evolved for a duration of $3\tau_{\mathcal X}$ to allow the cargo to relax to the stationary state.
No observables were accumulated during this initial relaxation period.
Except for the information flow, the observables were then measured for $3.0\,\mathrm{s}$ along each of $10000$ independent trajectories for each value of the external force.
The corresponding error bars represent standard errors of the mean calculated from the trajectory-to-trajectory fluctuations.
For the information flow, stationary residence-time data with a total observation time of $100000\,\mathrm{s}$ were divided into $30$ equal-duration blocks.
The error bar was estimated as the standard error of the $30$ blockwise estimates.

\begin{widetext}

\section{Detailed calculation of thermodynamic and information-theoretic quantities \label{sec:appendix: calc detail}}
In this section, we calculate explicit expressions for the thermodynamic and information-theoretic quantities $\dot{W}^{(\mathcal{S}, \mathcal{M})}$, $\dot{W}^{\mathcal{X}}$, $\dot{Q}^{(\mathcal{S}, \mathcal{M})}$, $\dot{Q}^{\mathcal{X}}$, $\dot{W}^{(\mathcal{S}, \mathcal{M}) \to \mathcal{X}}$, and $\dot{I}^{\mathcal{X}\rightarrow(\mathcal{S}, \mathcal{M})}$ in the limit $\epsilon\rightarrow0$.
First, we calculate the chemical input power $\dot{W}^{(\mathcal{S}, \mathcal{M})}$ defined in Eq.~(\ref{eq:setup: work of (S,M)}).
Using Eq.~(\ref{eq:setup:decomposition of the prob distribution}) and the definition of the effective transition matrix in Eq.~(\ref{def: effective transition matrix}), the leading-order contribution in the limit $\epsilon\rightarrow0$ reads
\begin{align}
  \dot{W}^{(\mathcal{S}, \mathcal{M})} &= \int_{-\infty}^{\infty} dx \sum_{s, s', m, m'} W_{mm'}^{ss'}(x) p_t (s',m',x) \Delta\mu_{mm'}^{ss'}\notag\\
  &= \dfrac{1}{\tau^{(\mathcal{S},\mathcal{M})}} \sum_{s, s', m, m'} \overline{\widetilde{W}}_{mm'}^{ss'} p_\tau(s',m') \Delta\mu_{mm'}^{ss'}\notag\\
  &=\dfrac{1}{\tau^{(\mathcal{S},\mathcal{M})}}\sum_m \bigg[ \bigg( \overline{\widetilde{k}}_f p_\tau(2, m) - \overline{\widetilde{k}}_f^\dag p_\tau(1, m + \delta) \bigg) \Delta\mu_{\mathrm{mech}} \notag \\
  &\qquad\qquad + \bigg( \overline{\widetilde{k}}_c p_\tau(1, m) - \overline{\widetilde{k}}_c^\dag p_\tau(2, m) \bigg) \Delta \mu_{\mathrm{chem}} \notag \\
  &\qquad\qquad + \bigg( \overline{\widetilde{k}}_b p_\tau(2, m) - \overline{\widetilde{k}}_b^\dag p_\tau(1, m - \delta) \bigg) \Delta\mu_{\mathrm{mech}} \bigg]\notag\\
  &=\dfrac{1}{\tau^{(\mathcal{S},\mathcal{M})}} \bigg[ \bigg( \overline{\widetilde{k}}_f p_{\mathrm{ss}}(2) - \overline{\widetilde{k}}_f^\dag p_{\mathrm{ss}}(1) \bigg) \Delta\mu_{\mathrm{mech}} \notag \\
  &\qquad\qquad + \bigg( \overline{\widetilde{k}}_c p_{\mathrm{ss}}(1) - \overline{\widetilde{k}}_c^\dag p_{\mathrm{ss}}(2) \bigg) \Delta \mu_{\mathrm{chem}} \notag \\
  &\qquad\qquad + \bigg( \overline{\widetilde{k}}_b p_{\mathrm{ss}}(2) - \overline{\widetilde{k}}_b^\dag p_{\mathrm{ss}}(1) \bigg) \Delta\mu_{\mathrm{mech}} \bigg].
\end{align}
By substituting Eqs.~(\ref{eq:effective rates kf})-(\ref{eq:effective rates kb_dag}) into this expression, we obtain
\begin{align}
  \dot{W}^{(\mathcal{S}, \mathcal{M})} &= \frac{1}{\tau^{(\mathcal{S},\mathcal{M})}} \bigg[ \frac{\tau^{(\mathcal{S},\mathcal{M})}}{\tau_f} \Delta\mu_{\mathrm{mech}}  \bigg( p_{\mathrm{ss}}(2) - p_{\mathrm{ss}}(1) e^{- \frac{\Delta\mu_{\mathrm{mech}} + F \delta}{k_{\mathrm{B}} T}} \bigg) e^{\frac{\theta_f}{k_{\mathrm{B}} T} \big[ \frac{\kappa}{2} \delta^2 (\theta_f - 1) + \Delta\mu_{\mathrm{mech}} + F \delta \big]} \notag \\
  &\hspace{30pt}+ \frac{\tau^{(\mathcal{S},\mathcal{M})}}{\tau_b} \Delta\mu_{\mathrm{mech}}  \bigg( p_{\mathrm{ss}}(2) - p_{\mathrm{ss}}(1) e^{- \frac{\Delta\mu_{\mathrm{mech}} - F \delta}{k_{\mathrm{B}} T}} \bigg) e^{\frac{\theta_b}{k_{\mathrm{B}} T} \big[ \frac{\kappa}{2} \delta^2 (\theta_b - 1) + \Delta\mu_{\mathrm{mech}} - F \delta \big]} \notag \\
  &\hspace{45pt} - \overline{\widetilde{k}}_c \Delta \mu_{\mathrm{chem}} \bigg( p_{\mathrm{ss}}(2) e^{- \frac{\Delta \mu_{\mathrm{chem}}}{k_{\mathrm{B}} T}}- p_{\mathrm{ss}}(1)  \bigg) \bigg].
\end{align}

The input or output power exerted by the constant external force on the cargo, $\dot{W}^\mathcal{X}$, can be obtained using the explicit form of the steady-state mean velocity in Eq.~(\ref{eq:setup:steady-state velocity}):
\begin{align}
  \dot{W}^\mathcal{X} &= F \langle \dot{x}_t \rangle \notag \\
  &= \dfrac{1}{\tau^{(\mathcal{S},\mathcal{M})}}\bigg[ \left( \overline{\widetilde{k}}_f -  \overline{\widetilde{k}}_b\right) p_{\mathrm{ss}} (2) + \left( \overline{\widetilde{k}}_b ^\dag -  \overline{\widetilde{k}}_f^\dag\right) p_{\mathrm{ss}} (1) \bigg] F\delta\notag\\
  &= \frac{1}{\tau^{(\mathcal{S},\mathcal{M})}} \bigg[ \frac{\tau^{(\mathcal{S},\mathcal{M})}}{\tau_f}   \bigg( p_{\mathrm{ss}}(2) - p_{\mathrm{ss}}(1) e^{- \frac{\Delta\mu_{\mathrm{mech}} + F \delta}{k_{\mathrm{B}} T}} \bigg) e^{\frac{\theta_f}{k_{\mathrm{B}} T} \big[ \frac{\kappa}{2} \delta^2 (\theta_f - 1) + \Delta\mu_{\mathrm{mech}} + F \delta \big]} \notag \\
  &\hspace{30pt} - \frac{\tau^{(\mathcal{S},\mathcal{M})}}{\tau_b}  \bigg( p_{\mathrm{ss}}(2) - p_{\mathrm{ss}}(1) e^{- \frac{\Delta\mu_{\mathrm{mech}} - F \delta}{k_{\mathrm{B}} T}} \bigg) e^{\frac{\theta_b}{k_{\mathrm{B}} T} \big[ \frac{\kappa}{2} \delta^2 (\theta_b - 1) + \Delta\mu_{\mathrm{mech}} - F \delta \big]} \bigg] F \delta.
\end{align}

Next, we calculate the power delivered from the kinesin to the cargo, $\dot{W}^{(\mathcal{S}, \mathcal{M}) \to \mathcal{X}}$, which can be expressed as
\begin{align}
    \dot{W}^{(\mathcal{S}, \mathcal{M}) \to \mathcal{X}} &= \int_{-\infty}^{\infty} dx \sum_{s, s', m, m'} W^{ss'}_{mm'} (x) p_t(s', m', x) \Big( U(m, x) - U(m', x) \Big) \notag\\
    &= \int_{-\infty}^{\infty} dx \sum_m \bigg[ \left(k_f(m, x) p_t(2, m, x) - k_f^\dag (m + \delta, x) p_t(1, m + \delta, x)\right) \Big(U(m + \delta, x) - U(m, x)\Big) \notag \\
    &\qquad\qquad + \left(k_b(m, x) p_t(2, m, x) - k_b^\dag (m - \delta, x) p_t(1, m - \delta, x)\right) \Big(U(m - \delta, x) - U(m, x)\Big) \bigg].
\end{align}
Using Eq.~(\ref{eq:setup:decomposition of the prob distribution}) and changing variables to $\chi = x - m$, we obtain
\begin{align}
  \dot{W}^{(\mathcal{S}, \mathcal{M}) \to \mathcal{X}} &= \dfrac{1}{\tau^{(\mathcal{S},\mathcal{M})}}\int_{-\infty}^{\infty} d\chi \bigg[ \widetilde{k}_f(0, \chi)p_\mathrm{ss}(2) \pi_{\mathrm{ss}}(\chi | 0) - \widetilde{k}_b(0, \chi - \delta) p_\mathrm{ss}(2) \pi_{\mathrm{ss}}(\chi - \delta | 0) \notag \\
  &\qquad- \widetilde{k}_f^\dag(0, \chi - \delta) p_\mathrm{ss}(1) \pi_{\mathrm{ss}} (\chi - \delta | 0) + \widetilde{k}_b^\dag(0, \chi) p_\mathrm{ss}(1) \pi_{\mathrm{ss}} (\chi | 0) \bigg] \Big(U(0, \chi - \delta) - U(0, \chi)\Big).
\end{align}
By substituting Eqs.~(\ref{eq:setup:(pi_ss) steady-state probability density of cargo}), (\ref{eq:effective rates kf})-(\ref{eq:effective rates kb_dag}) into this expression, we obtain
\begin{align}
  \dot{W}^{(\mathcal{S}, \mathcal{M}) \to \mathcal{X}} & = \frac{1}{\tau^{(\mathcal{S},\mathcal{M})}} \kappa \delta^2 \bigg[ \frac{\tau^{(\mathcal{S},\mathcal{M})}}{\tau_f} \bigg(p_{\mathrm{ss}} (2) - p_{\mathrm{ss}}(1) e^{- \frac{\Delta\mu_{\mathrm{mech}} + F \delta}{k_{\mathrm{B}} T}} \bigg)\bigg(- \theta_f + \frac{1}{2} - \frac{F}{\kappa \delta} \bigg) e^{\frac{\theta_f}{k_{\mathrm{B}} T} \big[ \frac{\kappa}{2} \delta^2 (\theta_f - 1) + \Delta\mu_{\mathrm{mech}} + F \delta \big]} \notag \\
  &\qquad\qquad + \frac{\tau^{(\mathcal{S},\mathcal{M})}}{\tau_b} \bigg(p_{\mathrm{ss}} (2) - p_{\mathrm{ss}}(1) e^{- \frac{\Delta\mu_{\mathrm{mech}} - F \delta}{k_{\mathrm{B}} T}} \bigg) \bigg(- \theta_b + \frac{1}{2} + \frac{F}{\kappa \delta} \bigg) e^{\frac{\theta_b}{k_{\mathrm{B}} T} \big[ \frac{\kappa}{2} \delta^2 (\theta_b - 1) + \Delta\mu_{\mathrm{mech}} - F \delta \big]} \bigg] .
\end{align}

Although the heat flows $\dot{Q}^{(\mathcal{S}, \mathcal{M})}$ and $\dot{Q}^{\mathcal{X}}$ can similarly be calculated from their definitions, it is simpler to use the first laws for the subsystems [Eqs.~(\ref{eq:setup: the first law of the kinesin}) and (\ref{eq:setup: the first law of the cargo})]:
\begin{align}
  \dot{Q}^{(\mathcal{S}, \mathcal{M})} &= \dot{W}^{(\mathcal{S}, \mathcal{M}) \to \mathcal{X}} - \dot{W}^{(\mathcal{S}, \mathcal{M})}\notag\\
  &= -\frac{1}{\tau^{(\mathcal{S},\mathcal{M})}} \kappa \delta^2 \bigg[  \frac{\tau^{(\mathcal{S},\mathcal{M})}}{\tau_f} \bigg( p_{\mathrm{ss}}(2) - p_{\mathrm{ss}}(1) e^{- \frac{\Delta\mu_{\mathrm{mech}} + F \delta }{k_{\mathrm{B}} T}} \bigg) \bigg(\theta_f - \frac{1}{2} + \frac{\Delta\mu_{\mathrm{mech}}}{\kappa \delta^2} + \frac{F}{\kappa \delta} \bigg) e^{\frac{\theta_f}{k_{\mathrm{B}} T} \big[ \frac{\kappa}{2} \delta^2 (\theta_f - 1) + \Delta\mu_{\mathrm{mech}}  + F \delta \big] } \notag \\
  &\qquad\qquad\hspace{15pt} + \frac{\tau^{(\mathcal{S},\mathcal{M})}}{\tau_b} \bigg( p_{\mathrm{ss}}(2) - p_{\mathrm{ss}}(1) e^{- \frac{\Delta\mu_{\mathrm{mech}} - F \delta }{k_{\mathrm{B}} T}} \bigg) \bigg(\theta_b - \frac{1}{2} + \frac{\Delta\mu_{\mathrm{mech}}}{\kappa \delta^2} - \frac{F}{\kappa \delta}\bigg) e^{\frac{\theta_b}{k_{\mathrm{B}} T} \big[\frac{\kappa}{2} \delta^2 (\theta_b - 1) +  \Delta\mu_{\mathrm{mech}}  - F \delta \big]}  \notag \\
  &\quad\hspace{45pt} - \overline{\widetilde{k}}_c \frac{\Delta \mu_{\mathrm{chem}}}{\kappa \delta^2}  \bigg(p_{\mathrm{ss}}(2) e^{- \frac{\Delta \mu_{\mathrm{chem}}}{k_{\mathrm{B}} T}} - p_{\mathrm{ss}}(1) \bigg)\bigg],\\
  \dot{Q}^{\mathcal{X}} &= \dot{W}^{\mathcal{X} \to (\mathcal{S}, \mathcal{M})} - \dot{W}^\mathcal{X}\notag\\
  &= \frac{1}{\tau^{(\mathcal{S},\mathcal{M})}} \kappa \delta^2\bigg[ \frac{\tau^{(\mathcal{S},\mathcal{M})}}{\tau_f}  \left(p_{\mathrm{ss}}(2) - p_{\mathrm{ss}}(1) e^{- \frac{\Delta\mu_{\mathrm{mech}} + F \delta}{k_{\mathrm{B}} T}}\right) \bigg( \theta_f - \frac{1}{2} \bigg) e^{\frac{\theta_f}{k_{\mathrm{B}} T} \big[ \frac{\kappa}{2} \delta^2 (\theta_f - 1) +  \Delta\mu_{\mathrm{mech}}  + F \delta \big]} \notag \\
  &\qquad + \frac{\tau^{(\mathcal{S},\mathcal{M})}}{\tau_b} \left(p_{\mathrm{ss}}(2) - p_{\mathrm{ss}}(1) e^{- \frac{\Delta\mu_{\mathrm{mech}} - F \delta}{k_{\mathrm{B}} T}}\right) \bigg( \theta_b - \frac{1}{2}\bigg) e^{\frac{\theta_b}{k_{\mathrm{B}} T} \big[ \frac{\kappa}{2} \delta^2 (\theta_b - 1) + \Delta\mu_{\mathrm{mech}}  - F \delta \big]} \bigg].
\end{align}

Finally, we calculate $\dot{I}^{\mathcal{X}\rightarrow(\mathcal{S}, \mathcal{M})}$. 
We first note that, using $\sum_{s',m'}W^{s's}_{m'm}(x)=0$, this information flow can be rewritten as
\begin{align}
  \dot{I}^{\mathcal{X}\rightarrow(\mathcal{S}, \mathcal{M})} &= \int_{-\infty}^{\infty} dx \sum_{s, s', m, m'} W^{ss'}_{mm'} (x) p_t (s', m', x) \ln \frac{p_t (s, m, x)}{p_t (s, m) p_t(x)}\notag\\
  &=\int^\infty_{-\infty} dx\,\sum_{s, s', m, m'}W^{ss'}_{mm'}(x)p_t(s',m',x)\ln\dfrac{p_t(x|s,m)}{p_t(x|s',m')}.
\end{align}
Using Eq.~(\ref{eq:setup:decomposition of the prob distribution}) and changing variables to $\chi = x - m$, we obtain
\begin{align}
  \dot{I}^{\mathcal{X}\rightarrow(\mathcal{S}, \mathcal{M})} &= \frac{1}{\tau^{(\mathcal{S},\mathcal{M})}}\int_{-\infty}^{\infty} d\chi \bigg[\widetilde{k}_f(0, \chi) p_{\mathrm{ss}}(2) \pi_{\mathrm{ss}}(\chi | 0) - \widetilde{k}_f^\dag(0, \chi - \delta) p_{\mathrm{ss}}(1) \pi_{\mathrm{ss}}(\chi - \delta | 0) \notag \\
  &\qquad\qquad - \widetilde{k}_b(0, \chi - \delta) p_{\mathrm{ss}}(2) \pi_{\mathrm{ss}}(\chi - \delta | 0) + \widetilde{k}_b^\dag(0, \chi) p_{\mathrm{ss}}(1) \pi_{\mathrm{ss}}(\chi | 0) \bigg] \ln \frac{\pi_{\mathrm{ss}}(\chi - \delta | 0)}{\pi_{\mathrm{ss}}(\chi | 0)}.
\end{align}
By substituting Eqs.~(\ref{eq:setup:(pi_ss) steady-state probability density of cargo}), (\ref{eq:effective rates kf})-(\ref{eq:effective rates kb_dag}) into this expression, we obtain
\begin{align}
  \dot{I}^{\mathcal{X}\rightarrow(\mathcal{S}, \mathcal{M})} &= \frac{1}{\tau^{(\mathcal{S},\mathcal{M})}} \frac{\kappa \delta^2}{k_{\mathrm{B}} T} \bigg[ \frac{\tau^{(\mathcal{S},\mathcal{M})}}{\tau_f}  \left(p_{\mathrm{ss}}(2) - p_{\mathrm{ss}}(1) e^{- \frac{\Delta\mu_{\mathrm{mech}} + F \delta}{k_{\mathrm{B}} T}}\right) \bigg( \theta_f - \frac{1}{2} \bigg) e^{\frac{\theta_f}{k_{\mathrm{B}} T} \big[ \frac{\kappa}{2} \delta^2 (\theta_f - 1) +  \Delta\mu_{\mathrm{mech}}  + F \delta \big]} \notag \\
  &\qquad + \frac{\tau^{(\mathcal{S},\mathcal{M})}}{\tau_b} \left(p_{\mathrm{ss}}(2) - p_{\mathrm{ss}}(1) e^{- \frac{\Delta\mu_{\mathrm{mech}} - F \delta}{k_{\mathrm{B}} T}}\right) \bigg( \theta_b - \frac{1}{2}\bigg) e^{\frac{\theta_b}{k_{\mathrm{B}} T} \big[ \frac{\kappa}{2} \delta^2 (\theta_b - 1) + \Delta\mu_{\mathrm{mech}}  - F \delta \big]} \bigg].
\end{align}
Note that $\dot{Q}^{\mathcal{X}} = k_{\mathrm{B}} T \dot{I}^{\mathcal{X} \to (\mathcal{S}, \mathcal{M})}$ in the limit $\epsilon\rightarrow0$.

\section{Detailed calculation of TUR efficiency \label{sec:appendix:TUR calculation}}
In this section, we analytically calculate the mean $J_\mathcal{J} := \lim_{\mathcal{T} \to \infty} \langle \hat{\mathcal{J}} \rangle / \mathcal{T}$ and the fluctuation $D_\mathcal{J} := \lim_{\mathcal{T} \to \infty}\mathrm{Var}[\hat{\mathcal{J}}] / 2 \mathcal{T}$ for the time-integrated currents $\hat{\mathcal{J}}=\Delta\hat{m}$, $\Delta\hat{n}_f$, $\Delta\hat{n}_b$, and $\Delta\hat{n}_c$ in the limit $\epsilon\rightarrow0$.
Hereafter, we use dimensionless time measured in units of $\tau^{(\mathcal{S},\mathcal{M})}$.
To calculate $J_\mathcal{J}$ and $D_\mathcal{J}$, we consider the scaled cumulant generating function for $\hat{\mathcal{J}}$ defined by
\begin{align}
    \mu_{\mathcal{J}} (\lambda) := \lim_{\mathcal{T} \to \infty} \frac{1}{\mathcal{T}} \ln \langle e^{\lambda \hat{\mathcal{J}}} \rangle,
    \label{eq:appendix:def of cumulant generating function}
\end{align}
where $\lambda\in\mathbb{R}$ is the conjugate field associated with $\hat{\mathcal{J}}$.
From $\mu_{\mathcal{J}} (\lambda)$, the mean $J_\mathcal{J}$ and the fluctuation $D_\mathcal{J}$ can be obtained as
\begin{align}
    J_\mathcal{J} &= \left. \frac{\partial}{\partial \lambda} \mu_{\mathcal{J}}(\lambda) \right|_{\lambda = 0}, \\
    D_\mathcal{J} &= \left. \frac{1}{2} \frac{\partial^2}{\partial \lambda^2} \mu_{\mathcal{J}}(\lambda) \right|_{\lambda = 0}.
\end{align}

To compute $\mu_{\mathcal{J}} (\lambda)$, we note that it can be expressed as
\begin{align}
    \mu_{\mathcal{J}} (\lambda) := \lim_{\mathcal{T} \to \infty} \frac{1}{\mathcal{T}} \ln \sum_s G^\mathcal{J}_{\mathcal{T}}(s),
\end{align}
where $G^\mathcal{J}_{\mathcal{T}}(s)$ denotes the generating function for the effective slow dynamics of the kinesin [Eq.~\eqref{eq:setup: slow dynamics for internal state transition}], defined as 
\begin{align}
    G^\mathcal{J}_\mathcal{T}(s) := \sum_\mathcal{J} p_\mathcal{T} (s, \mathcal{J}) e^{\lambda \mathcal{J}},
    \label{eq:appendix:def of generating function}
\end{align}
where $p_\mathcal{T} (s, \mathcal{J})$ denotes the joint probability that the internal state of the system and the time-integrated current at time $\mathcal{T}$ are $s$ and $\mathcal{J}$, respectively, and $\sum_\mathcal{J}$ denotes the summation over the possible values of $\hat{\mathcal{J}}$.
The time-evolution equation of $G^\mathcal{J}_\mathcal{T}(s)$ can be obtained from the effective dynamics (\ref{eq:setup: slow dynamics for internal state transition}).
For $\hat{\mathcal{J}}=\Delta\hat{m}$, for example, we first note that the time-evolution equation of $p_\mathcal{T} (s, \Delta m)$ reads
\begin{align}
\partial_\mathcal{T} p_\mathcal{T}(1,\Delta m) &= -\left(\overline{\widetilde{k}}_f^\dag + \overline{\widetilde{k}}_c + \overline{\widetilde{k}}_b^\dag\right)p_\mathcal{T}(1,\Delta m) + \overline{\widetilde{k}}_fp_\mathcal{T}(2,\Delta m-\delta) + \overline{\widetilde{k}}_c^\dag p_\mathcal{T}(2,\Delta m) + \overline{\widetilde{k}}_bp_\mathcal{T}(2,\Delta m+\delta),\\
\partial_\mathcal{T} p_\mathcal{T}(2,\Delta m) &= -\left(\overline{\widetilde{k}}_f + \overline{\widetilde{k}}_c^\dag + \overline{\widetilde{k}}_b\right)p_\mathcal{T}(2,\Delta m) + \overline{\widetilde{k}}^\dag_fp_\mathcal{T}(1,\Delta m+\delta) + \overline{\widetilde{k}}_c p_\mathcal{T}(1,\Delta m) + \overline{\widetilde{k}}^\dag_bp_\mathcal{T}(1,\Delta m-\delta).
\end{align}
From this equation, the time-evolution equation of $G^{\Delta m}_\mathcal{T}(s)$ can be obtained as
\begin{align}
\partial_\mathcal{T} G^{\Delta m}_\mathcal{T}(1) &= -\left(\overline{\widetilde{k}}_f^\dag + \overline{\widetilde{k}}_c + \overline{\widetilde{k}}_b^\dag\right)G^{\Delta m}_\mathcal{T}(1) + \left(\overline{\widetilde{k}}_fe^{\lambda\delta} + \overline{\widetilde{k}}_c^\dag + \overline{\widetilde{k}}_be^{-\lambda\delta}\right) G^{\Delta m}_\mathcal{T}(2),\\
\partial_\mathcal{T} G^{\Delta m}_\mathcal{T}(2) &= -\left(\overline{\widetilde{k}}_f + \overline{\widetilde{k}}_c^\dag + \overline{\widetilde{k}}_b\right)G^{\Delta m}_\mathcal{T}(2) + \left(\overline{\widetilde{k}}^\dag_fe^{-\lambda\delta} + \overline{\widetilde{k}}_c + \overline{\widetilde{k}}^\dag_be^{\lambda\delta}\right) G^{\Delta m}_\mathcal{T}(1).
\end{align}
These equations can be concisely expressed as
\begin{align}
    \partial_\mathcal{T} G^{\Delta m}_\mathcal{T}(s) = \sum_{s'}\mathcal{L}^{\Delta m}_{ss'}(\lambda) G^{\Delta m}_\mathcal{T}(s'),
    \label{eq:appendix:time evolution of G delta m}
\end{align}
where $\mathcal{L}^{\Delta m}(\lambda)$ denotes the tilted generator defined by
\begin{align}
  \mathcal{L}^{\Delta m}(\lambda) &:=
  \begin{pmatrix}
    -\left(\overline{\widetilde{k}}_f^\dag +  \overline{\widetilde{k}}_c + \overline{\widetilde{k}}_b^\dag\right) & e^{\lambda \delta} \overline{\widetilde{k}}_f  + \overline{\widetilde{k}}_c^\dag  + e^{- \lambda \delta} \overline{\widetilde{k}}_b  \\[2pt]
    e^{- \lambda \delta} \overline{\widetilde{k}}_f^\dag  + \overline{\widetilde{k}}_c  +  e^{\lambda \delta} \overline{\widetilde{k}}_b^\dag  & -\left(\overline{\widetilde{k}}_f + \overline{\widetilde{k}}_c^\dag + \overline{\widetilde{k}}_b\right) 
  \end{pmatrix}.
  \label{eq:appendix:def of L M}
\end{align}
Similarly, the time-evolution equation of $G^\mathcal{J}_\mathcal{T}$ for the other currents $\hat{\mathcal{J}}=\Delta\hat{n}_f$, $\Delta\hat{n}_b$, and $\Delta\hat{n}_c$ can be expressed as
\begin{align}
    \partial_\mathcal{T} G^{\mathcal{J}}_\mathcal{T}(s) = \sum_{s'}\mathcal{L}^{\mathcal{J}}_{ss'}(\lambda) G^{\mathcal{J}}_\mathcal{T}(s')
    \label{eq:appendix:time evolution of G}
\end{align}
with
\begin{align}
  \mathcal{L}^{\Delta n_f}(\lambda) &:= 
  \begin{pmatrix}
    -\left(\overline{\widetilde{k}}_f^\dag +  \overline{\widetilde{k}}_c + \overline{\widetilde{k}}_b^\dag\right) & e^{\lambda} \overline{\widetilde{k}}_f  + \overline{\widetilde{k}}_c^\dag  +  \overline{\widetilde{k}}_b \\[2pt]
    e^{- \lambda} \overline{\widetilde{k}}_f^\dag  + \overline{\widetilde{k}}_c  +  \overline{\widetilde{k}}_b^\dag & -\left(\overline{\widetilde{k}}_f + \overline{\widetilde{k}}_c^\dag + \overline{\widetilde{k}}_b\right) 
  \end{pmatrix},
  \label{eq:appendix:def of L f} \\
  \mathcal{L}^{\Delta n_b}(\lambda) &:= 
  \begin{pmatrix}
    -\left(\overline{\widetilde{k}}_f^\dag +  \overline{\widetilde{k}}_c + \overline{\widetilde{k}}_b^\dag\right) & \overline{\widetilde{k}}_f  + \overline{\widetilde{k}}_c^\dag +  e^{\lambda}  \overline{\widetilde{k}}_b \\[2pt]
     \overline{\widetilde{k}}_f^\dag  + \overline{\widetilde{k}}_c +  e^{- \lambda} \overline{\widetilde{k}}_b^\dag  & -\left(\overline{\widetilde{k}}_f + \overline{\widetilde{k}}_c^\dag + \overline{\widetilde{k}}_b\right) 
  \end{pmatrix},
  \label{eq:appendix:def of L b} \\
  \mathcal{L}^{\Delta n_c}(\lambda) &:= 
  \begin{pmatrix}
    -\left(\overline{\widetilde{k}}_f^\dag +  \overline{\widetilde{k}}_c + \overline{\widetilde{k}}_b^\dag\right) & \overline{\widetilde{k}}_f  + e^{- \lambda} \overline{\widetilde{k}}_c^\dag  + \overline{\widetilde{k}}_b  \\[2pt]
     \overline{\widetilde{k}}_f^\dag  + e^{\lambda}\overline{\widetilde{k}}_c  +  \overline{\widetilde{k}}_b^\dag  & -\left(\overline{\widetilde{k}}_f + \overline{\widetilde{k}}_c^\dag + \overline{\widetilde{k}}_b\right) 
  \end{pmatrix}.
  \label{eq:appendix:def of L c}
\end{align}

The general solution of Eq.~(\ref{eq:appendix:time evolution of G}) can be expressed as
\begin{align}
G^{\mathcal{J}}_\mathcal{T}(s) = \sum_i e^{\theta^{\mathcal{J}}_i(\lambda)\mathcal{T}} \phi^{\mathcal{J}}_{i,\mathcal{T}}(s;\lambda),
\end{align}
where $\theta^{\mathcal{J}}_i(\lambda)$ denotes the $i$th eigenvalue of $\mathcal{L}^{\mathcal{J}}(\lambda)$ ($i=1,2$), and $\phi^{\mathcal{J}}_{i,\mathcal{T}}(s;\lambda)$ is a polynomial in $\mathcal{T}$.
From the Perron--Frobenius theorem, the largest eigenvalue, denoted by $\theta^{\mathcal{J}}_{\mathrm{max}}(\lambda)$, satisfies $\theta^{\mathcal{J}}_{\mathrm{max}}(0)=0$.
Hence, for $\lambda\rightarrow0$, the scaled cumulant generating function $\mu_{\mathcal{J}} (\lambda)$ is related to the largest eigenvalue as
\begin{align}
    \mu_{\mathcal{J}} (\lambda) &= \lim_{\mathcal{T} \to \infty} \frac{1}{\mathcal{T}} \ln \sum_s\sum_i e^{\theta^{\mathcal{J}}_i(\lambda)\mathcal{T}} \phi^{\mathcal{J}}_{i,\mathcal{T}}(s;\lambda)\notag\\
    &= \theta^{\mathcal{J}}_{\mathrm{max}}(\lambda).
\end{align}
Thus, in order to calculate $J_\mathcal{J}$ and $D_\mathcal{J}$, we need to calculate the largest eigenvalue of the tilted generator $\mathcal{L}^{\mathcal{J}}(\lambda)$.
We first note that $\mathcal{L}^{\mathcal{J}}(\lambda)$ has the following form:
\begin{align}
    \mathcal{L}^{\mathcal{J}}(\lambda) := \begin{pmatrix}
    \mathcal{L}^{\mathcal{J}}_{11} & \mathcal{L}^{\mathcal{J}}_{12}(\lambda) \\
    \mathcal{L}^{\mathcal{J}}_{21}(\lambda) & \mathcal{L}^{\mathcal{J}}_{22}
  \end{pmatrix}.
\end{align}
Then, the largest eigenvalue for this matrix is calculated as 
\begin{align}
    \theta^{\mathcal{J}}_{\mathrm{max}}(\lambda) = \frac{1}{2} \bigg[ \mathcal{L}^{\mathcal{J}}_{11} + \mathcal{L}^{\mathcal{J}}_{22} +  \sqrt{(\mathcal{L}^{\mathcal{J}}_{11} - \mathcal{L}^{\mathcal{J}}_{22})^2 + 4 \mathcal{L}^{\mathcal{J}}_{12}(\lambda) \mathcal{L}^{\mathcal{J}}_{21}(\lambda)} \bigg].
    \label{eq:appendix:largest eigenvalue for L}
\end{align}
From this expression, $J_\mathcal{J}$ and $D_\mathcal{J}$ can be calculated as
\begin{align}
  J_\mathcal{J} &= \left. \frac{\partial}{\partial \lambda} \mu_{\mathcal{J}}(\lambda) \right|_{\lambda = 0}\notag\\
  &= \left. \frac{\partial}{\partial \lambda} \theta^{\mathcal{J}}_{\mathrm{max}}(\lambda)\right|_{\lambda = 0}\notag\\
  &= \left.\dfrac{\partial_\lambda \mathcal{L}^{\mathcal{J}}_{12}(\lambda) \mathcal{L}^{\mathcal{J}}_{21}(\lambda) + \mathcal{L}^{\mathcal{J}}_{12}(\lambda) \partial_\lambda \mathcal{L}^{\mathcal{J}}_{21}(\lambda)}{\Big[ (\mathcal{L}^{\mathcal{J}}_{11} - \mathcal{L}^{\mathcal{J}}_{22})^2 + 4 \mathcal{L}^{\mathcal{J}}_{12}(\lambda) \mathcal{L}^{\mathcal{J}}_{21}(\lambda) \Big]^{\frac{1}{2}}} \right|_{\lambda = 0},
  \label{eq:appendix:first derivative of theta max}
\end{align}
and
\begin{align}
  D_\mathcal{J} &= \left. \frac{1}{2} \frac{\partial^2}{\partial \lambda^2} \mu_{\mathcal{J}}(\lambda) \right|_{\lambda = 0}\notag\\
  &= \left. \dfrac{1}{2}\frac{\partial^2}{\partial \lambda^2} \theta^{\mathcal{J}}_{\mathrm{max}}(\lambda) \right|_{\lambda = 0} \notag\\
  &= \left.\dfrac{1}{2}\dfrac{\partial_\lambda^2 \mathcal{L}^{\mathcal{J}}_{12}(\lambda) \mathcal{L}^{\mathcal{J}}_{21}(\lambda) + 2 \partial_\lambda \mathcal{L}^{\mathcal{J}}_{12}(\lambda) \partial_\lambda \mathcal{L}^{\mathcal{J}}_{21}(\lambda) + \mathcal{L}^{\mathcal{J}}_{12}(\lambda)\partial_\lambda^2 \mathcal{L}^{\mathcal{J}}_{21}(\lambda)}{\Big[ (\mathcal{L}^{\mathcal{J}}_{11} - \mathcal{L}^{\mathcal{J}}_{22})^2 + 4 \mathcal{L}^{\mathcal{J}}_{12}(\lambda) \mathcal{L}^{\mathcal{J}}_{21}(\lambda) \Big]^{\frac{1}{2}}} \right|_{\lambda = 0} \notag \\
  &\qquad \left. - \dfrac{\Big[ \partial_\lambda \mathcal{L}^{\mathcal{J}}_{12}(\lambda) \mathcal{L}^{\mathcal{J}}_{21}(\lambda) + \mathcal{L}^{\mathcal{J}}_{12}(\lambda) \partial_\lambda \mathcal{L}^{\mathcal{J}}_{21}(\lambda) \Big]^2}{\Big[ (\mathcal{L}^{\mathcal{J}}_{11} - \mathcal{L}^{\mathcal{J}}_{22})^2 + 4 \mathcal{L}^{\mathcal{J}}_{12}(\lambda) \mathcal{L}^{\mathcal{J}}_{21}(\lambda) \Big]^{\frac{3}{2}}} \right|_{\lambda = 0}.
  \label{eq:appendix:second derivative of theta max}
\end{align}
Using Eqs.~\eqref{eq:appendix:first derivative of theta max} and \eqref{eq:appendix:second derivative of theta max}, we can obtain the explicit form of $J_\mathcal{J}$ and $D_\mathcal{J}$ for $\hat{\mathcal{J}}=\Delta\hat{m}$, $\Delta\hat{n}_f$, $\Delta\hat{n}_b$, and $\Delta\hat{n}_c$ in the limit $\epsilon\rightarrow0$:
\begin{align}
  J_{\Delta m} &= \tau^{(\mathcal{S},\mathcal{M})}v \notag\\
  &= \delta\frac{\Big(\overline{\widetilde{k}}_f - \overline{\widetilde{k}}_b\Big) \Big(\overline{\widetilde{k}}_f^\dag + \overline{\widetilde{k}}_c + \overline{\widetilde{k}}_b^\dag \Big) + \Big(\overline{\widetilde{k}}_b^\dag - \overline{\widetilde{k}}_f^\dag\Big) \Big(\overline{\widetilde{k}}_f + \overline{\widetilde{k}}_c^\dag + \overline{\widetilde{k}}_b\Big)}{\overline{\widetilde{k}}_f + \overline{\widetilde{k}}_b + \overline{\widetilde{k}}_c + \overline{\widetilde{k}}_f^\dag + \overline{\widetilde{k}}_b^\dag + \overline{\widetilde{k}}_c^\dag} \notag \\
  &= \delta \bigg[ \Big(\overline{\widetilde{k}}_f - \overline{\widetilde{k}}_b\Big) p_{\mathrm{ss}}(2) + \Big(\overline{\widetilde{k}}_b^\dag - \overline{\widetilde{k}}_f^\dag\Big) p_{\mathrm{ss}}(1) \bigg], \\
  D_{\Delta m} &= \frac{\delta^2}{2} \frac{ \Big(\overline{\widetilde{k}}_f + \overline{\widetilde{k}}_b\Big) \overline{\widetilde{k}}_c + \Big(\overline{\widetilde{k}}_f^\dag + \overline{\widetilde{k}}_b^\dag\Big) \overline{\widetilde{k}}_c^\dag + 4 \Big(\overline{\widetilde{k}}_f \overline{\widetilde{k}}_b^\dag + \overline{\widetilde{k}}_f^\dag \overline{\widetilde{k}}_b\Big)}{\overline{\widetilde{k}}_f + \overline{\widetilde{k}}_b + \overline{\widetilde{k}}_c + \overline{\widetilde{k}}_f^\dag + \overline{\widetilde{k}}_b^\dag + \overline{\widetilde{k}}_c^\dag} \notag \\
  &\qquad- \delta^2 \frac{ \bigg[ \Big(\overline{\widetilde{k}}_f - \overline{\widetilde{k}}_b\Big) \Big(\overline{\widetilde{k}}_f^\dag + \overline{\widetilde{k}}_c + \overline{\widetilde{k}}_b^\dag \Big) + \Big(\overline{\widetilde{k}}_b^\dag - \overline{\widetilde{k}}_f^\dag\Big) \Big(\overline{\widetilde{k}}_f + \overline{\widetilde{k}}_c^\dag + \overline{\widetilde{k}}_b\Big) \bigg]^2 }{\bigg( \overline{\widetilde{k}}_f + \overline{\widetilde{k}}_b + \overline{\widetilde{k}}_c + \overline{\widetilde{k}}_f^\dag + \overline{\widetilde{k}}_b^\dag + \overline{\widetilde{k}}_c^\dag \bigg)^3},
\end{align}
\begin{align}
  J_{\Delta n_f} &=  \frac{\overline{\widetilde{k}}_f \Big(\overline{\widetilde{k}}_c + \overline{\widetilde{k}}_b^\dag \Big) - \overline{\widetilde{k}}_f^\dag \Big(\overline{\widetilde{k}}_b + \overline{\widetilde{k}}_c^\dag \Big)}{\overline{\widetilde{k}}_f + \overline{\widetilde{k}}_b + \overline{\widetilde{k}}_c + \overline{\widetilde{k}}_f^\dag + \overline{\widetilde{k}}_b^\dag + \overline{\widetilde{k}}_c^\dag}, \\
  D_{\Delta n_f} &= \frac{1}{2} \frac{\overline{\widetilde{k}}_f \Big( \overline{\widetilde{k}}_f^\dag + \overline{\widetilde{k}}_c + \overline{\widetilde{k}}_b^\dag \Big) - 2 \overline{\widetilde{k}}_f \overline{\widetilde{k}}_f^\dag + \Big( \overline{\widetilde{k}}_f + \overline{\widetilde{k}}_c^\dag + \overline{\widetilde{k}}_b \Big) \overline{\widetilde{k}}_f^\dag}{\overline{\widetilde{k}}_f + \overline{\widetilde{k}}_b + \overline{\widetilde{k}}_c + \overline{\widetilde{k}}_f^\dag + \overline{\widetilde{k}}_b^\dag + \overline{\widetilde{k}}_c^\dag} - \frac{\bigg[ \overline{\widetilde{k}}_f \Big( \overline{\widetilde{k}}_c + \overline{\widetilde{k}}_b^\dag \Big) - \overline{\widetilde{k}}_f^\dag \Big( \overline{\widetilde{k}}_b + \overline{\widetilde{k}}_c^\dag \Big) \bigg]^2}{\bigg( \overline{\widetilde{k}}_f + \overline{\widetilde{k}}_b + \overline{\widetilde{k}}_c + \overline{\widetilde{k}}_f^\dag + \overline{\widetilde{k}}_b^\dag + \overline{\widetilde{k}}_c^\dag \bigg)^3}, 
\end{align}
\begin{align}
  J_{\Delta n_b} &=  \frac{\overline{\widetilde{k}}_b \Big( \overline{\widetilde{k}}_f^\dag + \overline{\widetilde{k}}_c \Big) - \overline{\widetilde{k}}_b^\dag \Big( \overline{\widetilde{k}}_f + \overline{\widetilde{k}}_c^\dag \Big)}{\overline{\widetilde{k}}_f + \overline{\widetilde{k}}_b + \overline{\widetilde{k}}_c + \overline{\widetilde{k}}_f^\dag + \overline{\widetilde{k}}_b^\dag + \overline{\widetilde{k}}_c^\dag}, \\
  D_{\Delta n_b} &= \frac{1}{2} \frac{\overline{\widetilde{k}}_b \Big( \overline{\widetilde{k}}_f^\dag + \overline{\widetilde{k}}_c + \overline{\widetilde{k}}_b^\dag \Big) - 2 \overline{\widetilde{k}}_b \overline{\widetilde{k}}_b^\dag + \overline{\widetilde{k}}_b^\dag \Big( \overline{\widetilde{k}}_f + \overline{\widetilde{k}}_c^\dag + \overline{\widetilde{k}}_b \Big)}{\overline{\widetilde{k}}_f + \overline{\widetilde{k}}_b + \overline{\widetilde{k}}_c + \overline{\widetilde{k}}_f^\dag + \overline{\widetilde{k}}_b^\dag + \overline{\widetilde{k}}_c^\dag} - \frac{\bigg[ \overline{\widetilde{k}}_b \Big( \overline{\widetilde{k}}_f^\dag + \overline{\widetilde{k}}_c \Big) - \overline{\widetilde{k}}_b^\dag \Big( \overline{\widetilde{k}}_f + \overline{\widetilde{k}}_c^\dag \Big) \bigg]^2}{\bigg( \overline{\widetilde{k}}_f + \overline{\widetilde{k}}_b + \overline{\widetilde{k}}_c + \overline{\widetilde{k}}_f^\dag + \overline{\widetilde{k}}_b^\dag + \overline{\widetilde{k}}_c^\dag \bigg)^3}, 
\end{align}
\begin{align}
  J_{\Delta n_c} &=  \frac{\overline{\widetilde{k}}_c \Big( \overline{\widetilde{k}}_f + \overline{\widetilde{k}}_b \Big) - \overline{\widetilde{k}}_c^\dag \Big( \overline{\widetilde{k}}_f^\dag + \overline{\widetilde{k}}_b^\dag \Big)}{\overline{\widetilde{k}}_f + \overline{\widetilde{k}}_b + \overline{\widetilde{k}}_c + \overline{\widetilde{k}}_f^\dag + \overline{\widetilde{k}}_b^\dag + \overline{\widetilde{k}}_c^\dag}, \\
  D_{\Delta n_c} &= \frac{1}{2} \frac{\overline{\widetilde{k}}_c^\dag \Big( \overline{\widetilde{k}}_f^\dag + \overline{\widetilde{k}}_c + \overline{\widetilde{k}}_b^\dag \Big) - 2 \overline{\widetilde{k}}_c \overline{\widetilde{k}}_c^\dag + \overline{\widetilde{k}}_c \Big( \overline{\widetilde{k}}_f + \overline{\widetilde{k}}_c^\dag + \overline{\widetilde{k}}_b \Big)}{\overline{\widetilde{k}}_f + \overline{\widetilde{k}}_b + \overline{\widetilde{k}}_c + \overline{\widetilde{k}}_f^\dag + \overline{\widetilde{k}}_b^\dag + \overline{\widetilde{k}}_c^\dag} - \frac{\bigg[ \overline{\widetilde{k}}_c \Big( \overline{\widetilde{k}}_f + \overline{\widetilde{k}}_b \Big) - \overline{\widetilde{k}}_c^\dag \Big( \overline{\widetilde{k}}_f^\dag + \overline{\widetilde{k}}_b^\dag \Big) \bigg]^2}{\bigg( \overline{\widetilde{k}}_f + \overline{\widetilde{k}}_b + \overline{\widetilde{k}}_c + \overline{\widetilde{k}}_f^\dag + \overline{\widetilde{k}}_b^\dag + \overline{\widetilde{k}}_c^\dag \bigg)^3}. 
\end{align}

\end{widetext}

\clearpage

\section{Results under the low-ATP condition}
\label{sec:appendix:results in Low ATP}

\begin{figure}[htb]
    \centering
    \hspace*{-6mm}
    \vspace{-3mm}
    \includegraphics[width=0.8\linewidth]{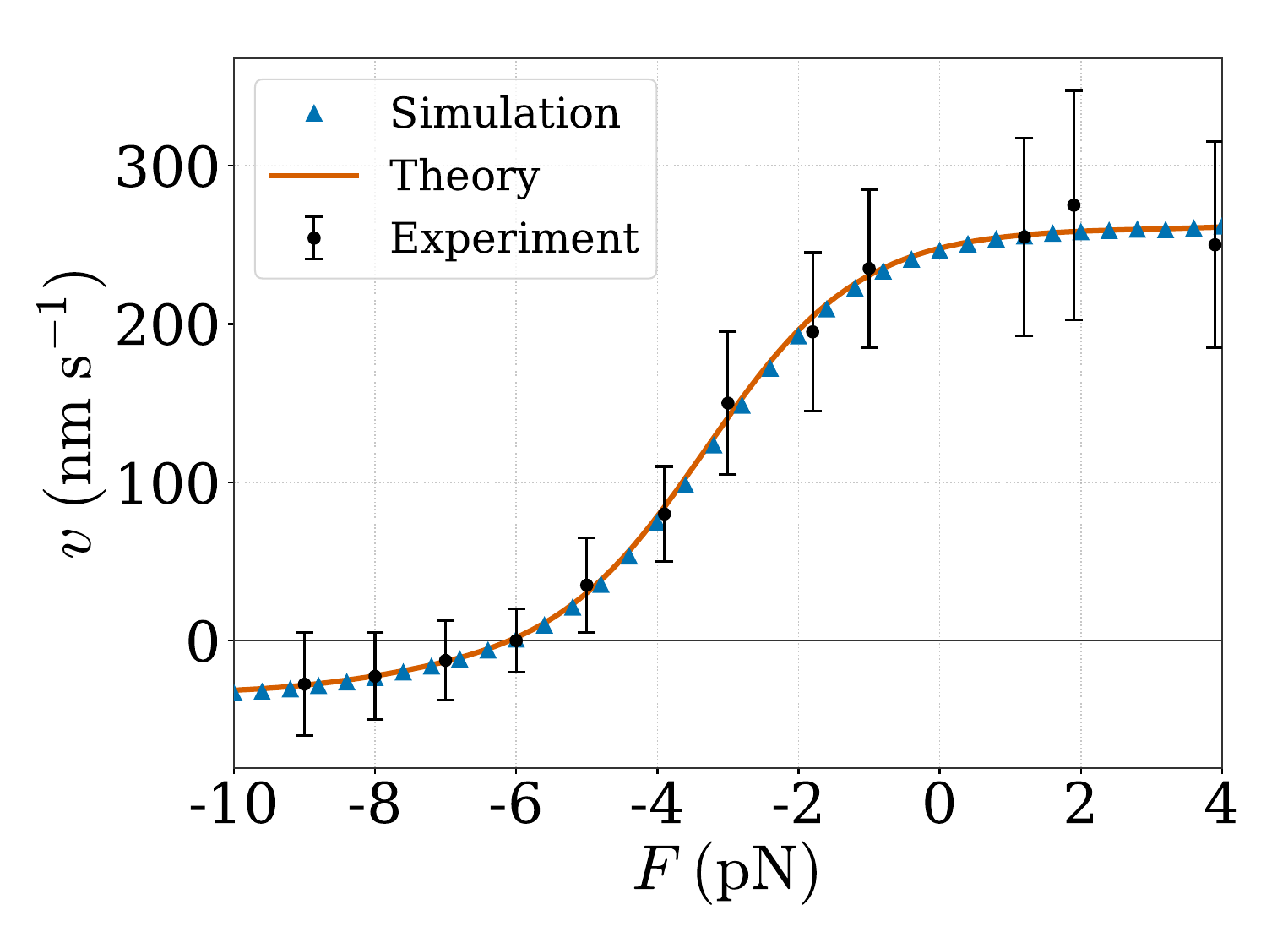}
    \caption{
    Force--velocity relation of single-molecule kinesin under the low-ATP \textit{in vitro} condition.
    The experimental data are taken from Ref.~\cite{ariga2018nonequilibrium}.
    }
    \label{fig:appendix:low ATP force curve}
\end{figure}

\begin{figure}[htb]
    \centering
    \includegraphics[width=8.5cm]{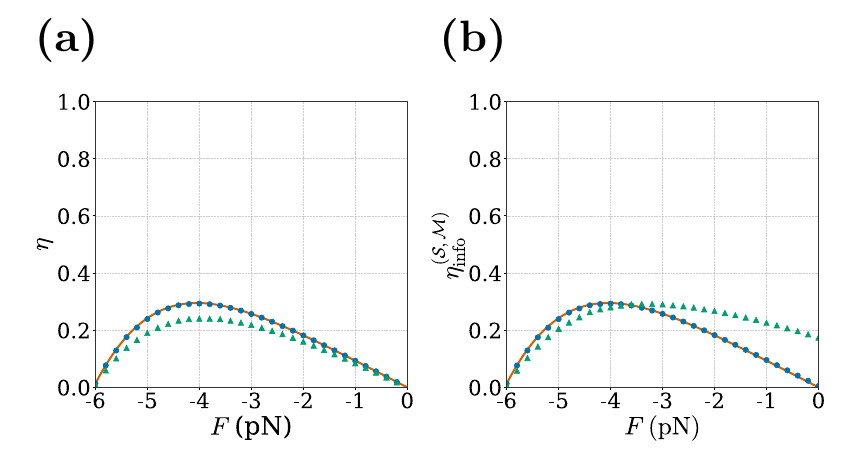}
    \caption{$F$-dependence of (a) the thermodynamic efficiency $\eta$, (b) the information-thermodynamic efficiency $\eta_\mathrm{info}^{(\mathcal{S}, \mathcal{M})}$. 
    These efficiencies are calculated under the low-ATP condition.}
    \label{fig:low ATP thermoeff and info eff}
\end{figure}

\begin{table*}[htpb]
  \centering
  \footnotesize
  \setlength{\tabcolsep}{10pt}
  \caption{Comparison between the experimental, theoretical, and simulation values of the thermodynamic quantities at a constant external force of $F = -2\,\mathrm{pN}$ under the low-ATP \textit{in vitro} condition ($\gamma = \gamma_{\mathrm{vitro}}$).
  Here, $-F_{0}\overline{v}$, $J_{x}$, $\Delta\mu/\tau$, and $J_{\mathrm{all \hspace{2pt} others}}$ in Ref.~\cite{ariga2018nonequilibrium} correspond to $-\dot{W}^{\mathcal{X}}$, $-\dot{Q}^{\mathcal{X}}$, $\dot{W}^{(\mathcal{S},\mathcal{M})}$, and $-\dot{Q}^{(\mathcal{S},\mathcal{M})}$, respectively.}
  \begin{ruledtabular}
  \begin{tabular}{l@{\,}llll}
  \multicolumn{2}{l}{Thermodynamic quantity}
    & Experiment & Simulation & Theory \\
  \hline
  $-\dot{W}^{\mathcal{X}}$
    & $(\mathrm{pN}\,\mathrm{nm}\,\mathrm{s}^{-1})$
    & $410\pm60$ & $384 \pm 0.4$ & 386 \\
  $-\dot{Q}^{\mathcal{X}}$
    & $(\mathrm{pN}\,\mathrm{nm}\,\mathrm{s}^{-1})$
    & $4.09\pm1.56$ & $10.3 \pm 1.7$ & 4.5 \\
  $\dot{W}^{(\mathcal{S},\mathcal{M})}$
    & $(\mathrm{pN}\,\mathrm{nm}\,\mathrm{s}^{-1})$
    & $2190\pm310$ & $2114 \pm 2.0$ & 2122 \\
  $-\dot{Q}^{(\mathcal{S},\mathcal{M})}$
    & $(\mathrm{pN}\,\mathrm{nm}\,\mathrm{s}^{-1})$
    & $1776 \pm 315$ & $1722 \pm 1.6$ & 1731 \\
  \hline
  $\eta$ 
    &
    & $0.187 \pm 0.038$ & $0.182 \pm 0.0002$ & 0.182 \\
  \end{tabular}
  \end{ruledtabular}
  \label{tab:appendix:low ATP comparison of exp. and solution}
\end{table*}

\begin{figure*}[htp]
  \centering
  \includegraphics[width=0.9\textwidth]{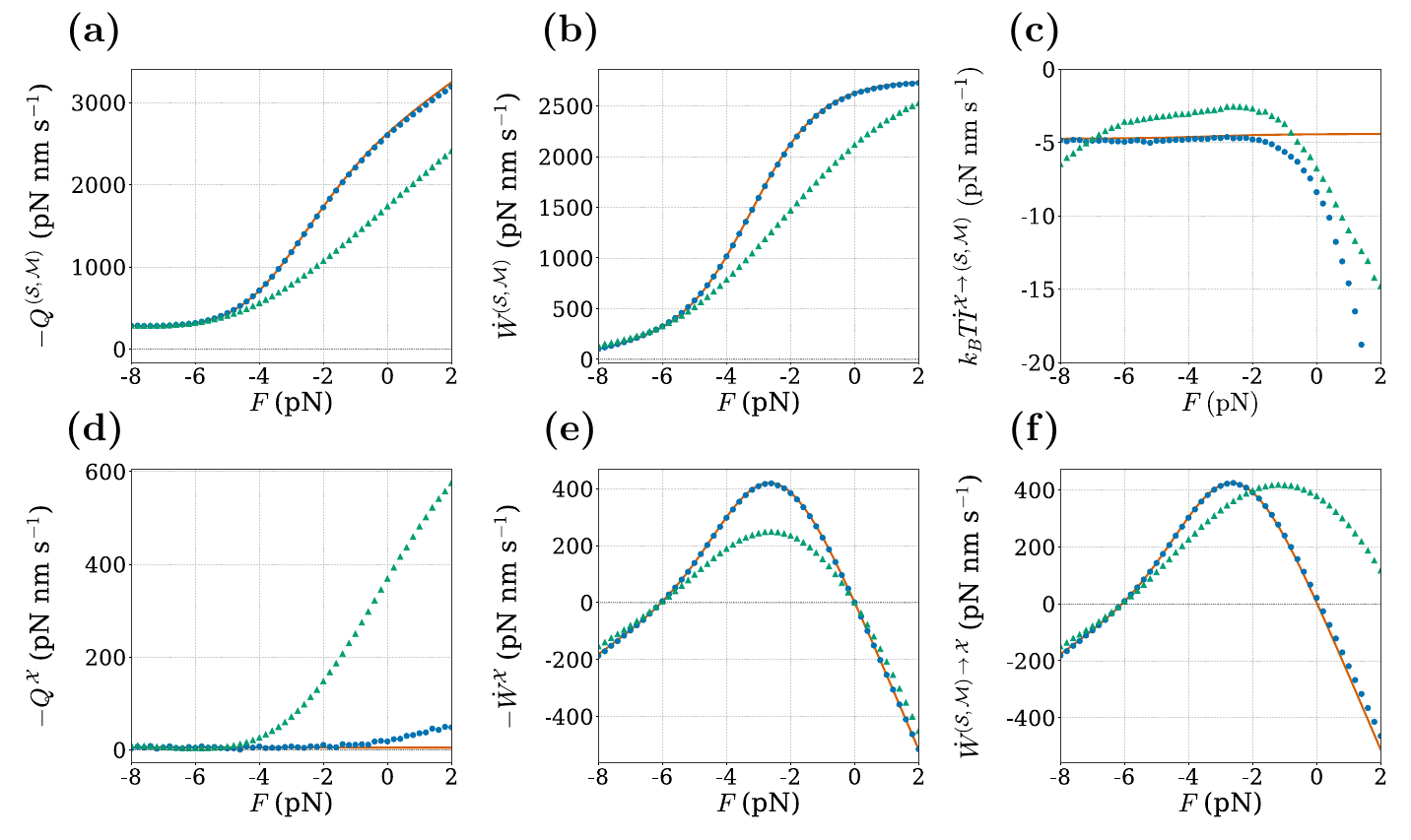}
  \caption{$F$-dependence of the various thermodynamic quantities under the low-ATP condition: (a) $-\dot{Q}^{(\mathcal{S}, \mathcal{M})}$, (b) $\dot{W}^{(\mathcal{S}, \mathcal{M})}$, (c) $k_{\mathrm{B}}T\dot{I}^{\mathcal{X}\to(\mathcal{S}, \mathcal{M})}$, (d) $-\dot{Q}^{\mathcal{X}}$, (e) $-\dot{W}^{\mathcal{X}}$, and (f) $\dot{W}^{(\mathcal{S}, \mathcal{M})\to\mathcal{X}}$.}
  \label{fig:low ATP thermodynamics properties}
\end{figure*}

In this appendix, we present the results obtained under the low-ATP condition ($10\,\mu\mathrm{M}$ ATP, $1\,\mu\mathrm{M}$ ADP, $1\,\mathrm{mM}\,\mathrm{P}_\mathrm{i}$).
The parameter values used in this section are given in Appendix~\ref{sec:appendix:parameter estimation}.
Using these parameters, we evaluate the thermodynamic quantities and efficiencies introduced in the main text.
We also validate the parameter estimation by comparing the force--velocity relation with the experimental data reported in Ref.~\cite{ariga2018nonequilibrium}, as shown in Fig.~\ref{fig:appendix:low ATP force curve}.
Overall, the qualitative trends are similar to those obtained under the high-ATP condition.
We therefore provide only a brief summary of the low-ATP results.

\paragraph{Thermodynamic and information-thermodynamic efficiencies.}

Under the low-ATP condition, we evaluate the relevant thermodynamic quantities using both analytical expressions and numerical simulations.
Assuming time-scale separation, we obtain analytical expressions from the perturbation expansion up to \(O(\epsilon^0)\).
Figures~\ref{fig:low ATP thermodynamics properties}(a)--(f) show the force dependence of
\(-\dot{Q}^{(\mathcal{S}, \mathcal{M})}\),
\(\dot{W}^{(\mathcal{S}, \mathcal{M})}\),
\(k_{\mathrm{B}}T\dot{I}^{\mathcal{X}\to(\mathcal{S},\mathcal{M})}\),
\(-\dot{Q}^{\mathcal{X}}\),
\(-\dot{W}^{\mathcal{X}}\), and
\(\dot{W}^{(\mathcal{S}, \mathcal{M})\to\mathcal{X}}\), respectively.

To assess consistency with the experimental results reported in Ref.~\cite{ariga2018nonequilibrium}, we compare the experimental, numerical, and theoretical values of these quantities at a fixed external force \(F=-2\,\mathrm{pN}\) under the low-ATP \textit{in vitro} condition, as summarized in Table~\ref{tab:appendix:low ATP comparison of exp. and solution}.
As in the high-ATP condition, the thermodynamic efficiency obtained from our model is consistent with the experimental estimate.
The mechanical output power \(-\dot{W}^{\mathcal{X}}\) and the input power \(\dot{W}^{(\mathcal{S},\mathcal{M})}\) also agree well with the experimental values.
On the other hand, the heat flow $\dot{Q}^{\mathcal{X}}$ shows larger deviations, which reflect the limitations of the assumption of time-scale separation.

We next examine the thermodynamic and information-thermodynamic efficiencies under the low-ATP condition.
Figure~\ref{fig:low ATP thermoeff and info eff}(a) shows the force dependence of the thermodynamic efficiency.
Under the \textit{in vitro} condition, the maximum value is approximately \(30\%\), whereas under the \textit{in vivo}-like condition, the maximum value is approximately \(24\%\).
Figure~\ref{fig:low ATP thermoeff and info eff}(b) shows the corresponding information-thermodynamic efficiency.
As in the high-ATP condition, the information-thermodynamic efficiency under the \textit{in vitro} condition exhibits behavior similar to that of the thermodynamic efficiency.
These results indicate that the work-conversion efficiencies are not qualitatively altered by changing the ATP concentration.

\begin{figure*}[htp]
  \centering
  \includegraphics[width=0.95\textwidth]{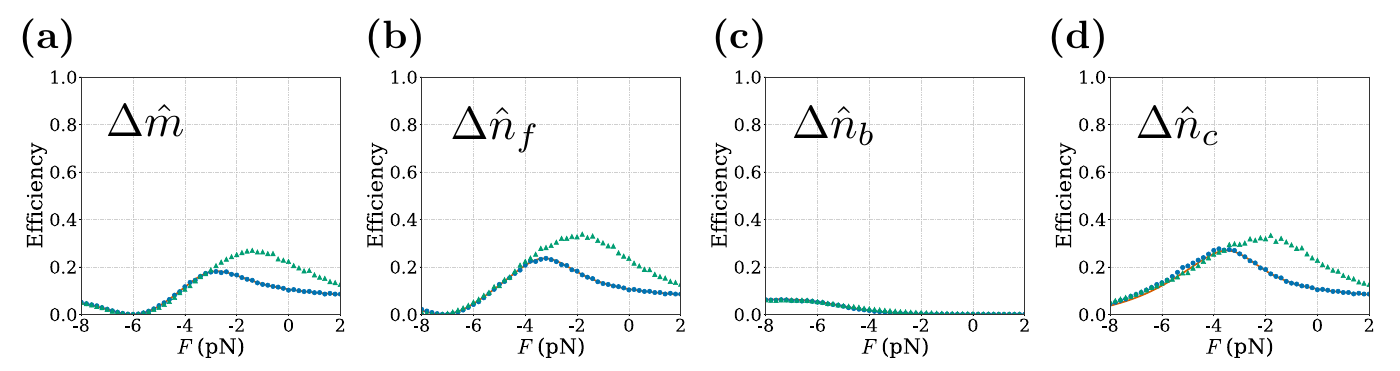}
  \caption{$F$-dependence of the TUR efficiency for each current: (a) $\Delta\hat{m}$, (b) $\Delta\hat{n}_f$, (c) $\Delta\hat{n}_b$, and (d) $\Delta\hat{n}_c$.
  These efficiencies are calculated under the low-ATP condition.}
  \label{fig:low ATP TUR eff.}
\end{figure*}

\begin{figure*}[htp]
  \centering
  \includegraphics[width=0.95\textwidth]{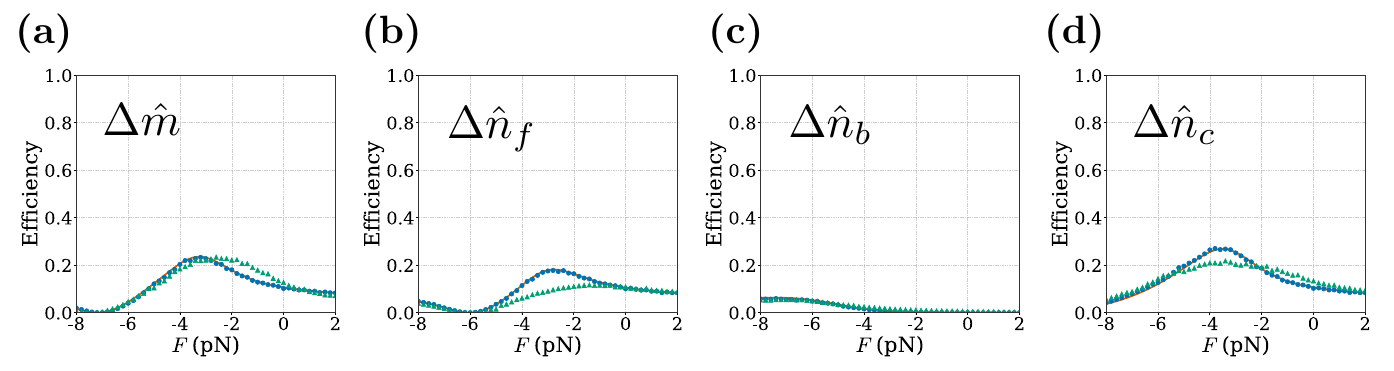}
  \caption{$F$-dependence of the bipartite TUR efficiency for each current: (a) $\Delta\hat{m}$, (b) $\Delta\hat{n}_f$, (c) $\Delta\hat{n}_b$, and (d) $\Delta\hat{n}_c$.
  These efficiencies are calculated under the low-ATP condition.}
  \label{fig:low ATP Bipartite TUR eff.}
\end{figure*}

\paragraph{TUR / bipartite TUR efficiency}

We next examine the TUR and bipartite TUR efficiencies under the low-ATP condition.
Figures~\ref{fig:low ATP TUR eff.} and \ref{fig:low ATP Bipartite TUR eff.} show the force dependence of these efficiencies, respectively.
For both efficiencies, we consider four currents:
\(\Delta\hat{m}\), \(\Delta\hat{n}_f\), \(\Delta\hat{n}_b\), and \(\Delta\hat{n}_c\).

For the TUR efficiency, the largest value among the currents considered here is approximately \(0.38\), which is obtained for the current \(\Delta\hat{n}_f\) under the \textit{in vivo}-like condition.
For the bipartite TUR efficiency, the largest value is approximately \(0.30\), which is obtained for the current \(\Delta\hat{n}_c\) under the \textit{in vitro} condition.
Although these values are somewhat larger than the work-conversion efficiencies, they remain well below unity.
Thus, even under the low-ATP condition, kinesin does not appear to be strongly optimized for suppressing current fluctuations, at least for the currents considered here.

Overall, the low-ATP results support the conclusions drawn in the main text.
Changing the ATP concentration modifies the quantitative values of some thermodynamic quantities and efficiencies, but it does not qualitatively change the thermodynamic characterization of kinesin obtained in this study.
We emphasize that this conclusion applies to the case considered here, where $\Delta\mu$ is fixed, and the ATP and ADP concentrations are varied simultaneously; it does not necessarily apply when the ATP concentration is varied while the other solution conditions are kept fixed, because $\Delta\mu$ then also changes.

\clearpage

\FloatBarrier
\bibliography{kinesin}

\end{document}